\documentclass[sigconf]{acmart}
\copyrightyear{2020}
\acmYear{2020}
\setcopyright{iw3c2w3}
\acmConference[WWW '20]{Proceedings of The Web Conference 2020}{April 20--24, 2020}{Taipei, Taiwan}
\acmBooktitle{Proceedings of The Web Conference 2020 (WWW '20), April 20--24, 2020, Taipei, Taiwan}
\acmPrice{}
\acmDOI{10.1145/3366423.3380246}
\acmISBN{978-1-4503-7023-3/20/04}

\usepackage[T1]{fontenc}
\usepackage{color}
\usepackage{url}
\usepackage{xspace}
\usepackage{balance}
\usepackage{graphicx}
\usepackage{listings}
\usepackage{array,booktabs,ragged2e}
\usepackage[lined,algoruled,noend,linesnumbered]{algorithm2e}
\usepackage{multirow}
\usepackage{colortbl}
\usepackage{natbib}
\usepackage{subcaption}
\usepackage{comment}

\includecomment{tr} \excludecomment{paper}

\begin{tr}
\newcommand{\FigOrTab}{{Table}\xspace}
\end{tr}
\begin{paper}
\newcommand{\FigOrTab}{{Figure}\xspace}
\end{paper}

\newtheorem{example}{Example}
\newtheorem{proposition}{Proposition}

\newcommand{\sys}{{Trident}\xspace}
\newcommand{\virtuoso}{\texttt{SYSTEM\_A}\xspace}

\newcommand{\fakeparagraph}[1]{\vspace{1mm}\noindent\textbf{#1.}}

\newcommand{\defeq}{\coloneqq}
\newcommand{\selectlayout}{\ensuremath{\mathsf{selectlayout}}}
\newcommand{\sizeofname}{\ensuremath{\mathsf{sizeof}}}

\newcommand{\pa}[1]{\ensuremath{\mathtt{#1}}}
\newcommand{\cR}{\mathcal{R}}
\newcommand{\cV}{\mathcal{V}}

\newcommand{\cL}{\mathcal{L}}
\newcommand{\cB}{\mathcal{B}}
\newcommand{\cI}{\mathcal{I}}
\newcommand{\cT}{\mathcal{T}}
\newcommand{\TR}{\pa{TR}\xspace}
\newcommand{\TS}{\pa{TS}\xspace}
\newcommand{\TD}{\pa{TD}\xspace}
\newcommand{\TRI}{\pa{TR'}\xspace}
\newcommand{\TSI}{\pa{TS'}\xspace}
\newcommand{\TDI}{\pa{TD'}\xspace}

\newcommand{\is}{\ensuremath{\iota_s}}
\newcommand{\ir}{\ensuremath{\iota_r}}
\newcommand{\id}{\ensuremath{\iota_d}}
\newcommand{\il}{\ensuremath{\iota_l}}
\newcommand{\dictl}{\pa{DICT_l}}
\newcommand{\dicti}{\pa{DICT_\iota}}

\newcommand{\nodemgr}{\ensuremath{\mathsf{NM}}\xspace}
\newcommand{\sizeof}[1]{\sizeofname(#1)}

\newcommand{\srd}[0]{\pa{srd}}
\newcommand{\sdr}[0]{\pa{sdr}}
\newcommand{\rsd}[0]{\pa{rsd}}
\newcommand{\rds}[0]{\pa{rds}}
\newcommand{\dsr}[0]{\pa{dsr}}
\newcommand{\drs}[0]{\pa{drs}}
\newcommand{\spa}[0]{\pa{s}}
\newcommand{\rpa}[0]{\pa{r}}
\newcommand{\dpa}[0]{\pa{d}}

\newcommand{\srsd}[0]{\pa{\{sr,sd\}}}
\newcommand{\rsrd}[0]{\pa{\{rs,rd\}}}
\newcommand{\dsdr}[0]{\pa{\{ds,dr\}}}

\newcommand{\squishlist}{
 \begin{list}{$\bullet$}
  { \setlength{\itemsep}{0pt}
     \setlength{\parsep}{3pt}
     \setlength{\topsep}{3pt}
     \setlength{\partopsep}{0pt}
     \setlength{\leftmargin}{1em}
     \setlength{\labelwidth}{1em}
     \setlength{\labelsep}{0.5em} } }

\newcommand{\squishend}{
\end{list}}

\newcommand{\isprefix}[0]{\ensuremath{\mathsf{isprefix}}\xspace}
\newcommand{\bound}[0]{\ensuremath{\mathsf{bound}}\xspace}
\newcommand{\lbl}[0]{\ensuremath{\mathsf{lbl}}\xspace}
\newcommand{\nodid}[0]{\ensuremath{\mathsf{nodid}}\xspace}
\newcommand{\edgid}[0]{\ensuremath{\mathsf{edgid}}\xspace}
\newcommand{\edg}[0]{\ensuremath{\mathsf{edg}}\xspace}
\newcommand{\grp}[0]{\ensuremath{\mathsf{grp}}\xspace}
\newcommand{\cnt}[0]{\ensuremath{\mathsf{count}}\xspace}
\newcommand{\pos}[0]{\ensuremath{\mathsf{pos}}\xspace}
\newcommand{\ans}[0]{\ensuremath{\mathsf{ans}}\xspace}

\definecolor{lightgray}{gray}{0.9}
\definecolor{gray}{gray}{0.6}

\begin{document}

\title{Adaptive Low-level Storage of Very Large Knowledge Graphs}

\author{Jacopo Urbani}
\email{jacopo@cs.vu.nl}
\affiliation{%
  \institution{Vrije Universiteit Amsterdam}
  \city{Amsterdam}
  \country{The Netherlands}
}

\author{Ceriel Jacobs}
\email{ceriel@cs.vu.nl}
\affiliation{%
  \institution{Vrije Universiteit Amsterdam}
  \city{Amsterdam}
  \country{The Netherlands}
}

\begin{abstract}

The increasing availability and usage of Knowledge Graphs (KGs) on the Web calls
for scalable and general-purpose solutions to store this type of data
structures. We propose \sys, a novel storage architecture for very large KGs on
centralized systems.  \sys uses several interlinked data structures to provide
fast access to nodes and edges, with the physical storage changing depending on
the topology of the graph to reduce the memory footprint. In contrast to single
architectures designed for single tasks, our approach offers an interface with
few low-level and general-purpose primitives that can be used to implement tasks
like SPARQL query answering, reasoning, or graph analytics. Our experiments show
that \sys can handle graphs with $10^{11}$ edges using inexpensive hardware,
delivering competitive performance on multiple workloads.

\end{abstract}

\maketitle

\section{Introduction}
\label{sec:intro}

\fakeparagraph{Motivation} Currently, a large wealth of knowledge is published
on the Web in the form of interlinked Knowledge Graphs (KGs). These KGs cover
different fields (e.g., biomedicine~\cite{bioportal,bio2rdf}, encyclopedic or
commonsense knowledge~\cite{wikidata,webchild}, etc.), and are actively used to
enhance tasks such as entity recognition~\cite{shen_entity_2015}, query
answering~\cite{kgqueryans}, or more in general Web
search~\cite{guha_semantic_2003,websearch2,websearch3}.

As the size of KGs keeps growing and their usefulness expands to new scenarios,
applications increasingly need to access large KGs for different purposes. For
instance, a search engine might need to query a KG using SPARQL~\cite{sparql},
enrich the results using embeddings of the graph ~\cite{kgemb_survey}, and then
compute some centrality metrics for ranking the answers~\cite{naga,ranking}. In
such cases, the storage engine must not only efficiently handle large KGs, but
also allow the execution of multiple types of computation so that the same KG
does not have to be loaded in multiple systems.

\fakeparagraph{Problem} In this paper, we focus on providing an efficient,
scalable, and general-purpose storage solution for large KGs on centralized
architectures. A large amount of recent research has focused on distributed
architectures~\cite{powergraph,graphx,pregel_comp,pregel,graph_mapreduce,trinity,triad,trillion},
because they offer many cores and a large storage space. However, these benefits
come at the price of higher communication cost and increased system
complexity~\cite{ringo}. Moreover, sometimes distributed solutions
cannot be used either due to financial or privacy-related constraints.

Centralized architectures, in contrast, do not have network costs, are commonly
affordable, and provide enough resources to load all-but-the-largest graphs.
Some centralized storage engines have demonstrated that they can handle large
graphs, but they focus primarily on supporting one particular type of workload
(e.g., Ringo~\cite{ringo} supports graph analytics, RDF engines like
Virtuoso~\cite{virtuoso} or RDFox~\cite{rdfox} focus on SPARQL~\cite{sparql}).
To the best of our knowledge, we still lack a single storage solution that can
handle very large KGs as well as support multiple workloads.

\fakeparagraph{Our approach} In this paper, we fill this gap presenting
\sys, a novel storage architecture that can store very
large KGs on centralized architectures, support multiple workloads, such as
SPARQL querying, reasoning, or graph analytics, and is resource-savvy.
Therefore, it meets our goal of combining scalability and general-purpose
computation.

We started the development of \sys by studying which are the most frequent
access types performed during the execution of tasks like SPARQL answering,
reasoning, etc. Some of these access types are \emph{node-centric} (i.e., access
subsets of the nodes), while others are \emph{edge-centric} (i.e., access
subsets of the edges). From this study, we distilled a small set of low-level
primitives that can be used to implement more complex tasks. Then, the research
focused on designing an architecture that supports the execution of these
primitives as efficiently as possible, resulting in \sys.

At its core, \sys uses a dedicated data structure (a B+Tree or an in-memory
array) to support fast access to the nodes, and a series of binary tables to
store subsets of the edges. Since there can be many binary tables -- possibly
billions with the largest KGs -- handling them with a relational DBMS can be
problematic. To avoid this problem, we introduce a light-weight storage scheme
where the tables are serialized on byte streams with only a little overhead per
table. In this way, tables can be quickly loaded from the secondary storage
without expensive pre-processing and offloaded in case the size of the database
exceeds the amount of available RAM.

Another important benefit of our approach is that it allows us to exploit
the topology of the graph to reduce its physical storage. To this end, we
introduce a novel procedure that analyses each binary table and decides, at
loading time, whether the table should be stored either in a row-by-row,
column-by-column, or in a cluster-based fashion. In this way, the storage
engines effectively \emph{adapts} to the input. Finally, we introduce other dynamic procedures that decide, at loading
time, whether some tables can be ignored due to their small sizes or whether the
content of some tables can be aggregated to further reduce the space.

Since \sys offers low-level primitives, we built interfaces to several engines
(RDF3X~\cite{rdf3x}, VLog~\cite{vlog}, SNAP~\cite{snap}) to evaluate the
performance of SPARQL query answering, datalog reasoning and graph analytics on
various types of graphs. Our comparison against the state-of-the-art shows that
our approach can be highly competitive in multiple scenarios.

\fakeparagraph{Contribution} We identified the followings as the main
contributions of this paper.

\squishlist{}

\item We propose a new architecture to store very large KGs on a centralized
    system. In contrast to other engines that store the KG in few data
    structures (e.g., relational tables), our architecture exhaustively
    decomposes the storage in many binary tables such that it supports both
    node- and edge-centric via a small number of primitives;

\item Our storage solution adapts to the KG as it uses different layouts to
    store the binary tables depending on its topology. Moreover, some binary
    tables are either skipped or aggregated to save further space. The
    adaptation of the physical storage is, as far as we know, a unique feature
    which is particularly useful for highly heterogeneous graphs, such as the
    KGs on the Web;

\item We present an evaluation with multiple workloads and the results
    indicate highly competitive performance while maintaining good
    scalability. In some of our largest experiments, \sys was able to load and
    process KGs with up to $10^{11}$ (100B) edges with hardware that costs less
    than \$5K.

    \squishend

\noindent The source code of \sys is freely available with an open source
license at \url{https://github.com/karmaresearch/trident}, along with links to
the datasets and instructions to replicate our experiments.
\begin{paper}
    An extended version of this paper is available at~\cite{tr}.
\end{paper}

\section{Preliminaries}

A \emph{graph} $G=(V,E,L,\phi_V,\phi_E)$ is a tuple where $V,E,L$ represent the
sets of nodes, edges and labels respectively, $\phi_V$ is a bijection that maps
each node to a label in $L$, while $\phi_E$ is a function that maps each edge to
a label in $L$. We assume that there is at most one edge with the same label
between any pair of nodes. Throughout, we use the notation $r(s,d)$ to indicate
the edge with label $r$ from node with label $s$ (source) to the node with label $d$ (destination).

We say that the graph is \emph{undirected} if $r(s,d)\in E$ implies that also
$r(d,s)\in E$. Otherwise, the graph is \emph{directed}. A graph is
\emph{unlabeled} if all edges map to the same label. In this paper, we will
mostly focus on labeled directed graphs since undirected or unlabeled graphs are
special cases of labeled directed graphs.

In practice, it is inefficient to store the graph using the raw labels as
identifiers. The most common strategy, which is the one we also follow, consists
of assigning a numerical ID to each label in $L$, and stores each edge $r(s,d)$
with the tuple $\langle \is,\ir,\id \rangle$ where $\is$, $\ir$, and $\id$ are
the IDs associated to $s$, $r$, and $d$ respectively.

The numerical IDs allow us to sort the edges and by permuting $\is,\ir,\id$ we
can define six possible ordering criteria. We use strings of three characters
over the alphabet $\{\pa{s},\pa{r},\pa{d}\}$ to identify these orderings, e.g.,
$\pa{srd}$ specifies that the edges are ordered by source, relation, and
destination. We denote with $\cR=\{\pa{srd},\pa{sdr},\ldots\}$ the collection of
six orderings while
$\cR'=\{\pa{s},\pa{r},\pa{d},\pa{sr},\pa{rs},\pa{sd},\pa{ds},\pa{dr},\pa{rd}\}$
specifies all partial orderings. We use the function \isprefix to check whether
string $a$ is a prefix of $b$, i.e., $\isprefix(a,b)=[true|false]$ and the
operator $-$ to remove all characters of one string from another one (e.g., if
$a=\pa{srd}$ and $b=\pa{sd}$, then $a-b=\pa{r}$).

Let $\mathcal{V}$ be a set of variables. A \emph{simple graph pattern} (or
\emph{triple pattern}) is an
instance of $L\cup \mathcal{V} \times L \cup \mathcal{V} \times L \cup
\mathcal{V}$ and we denote it as $(X,Y,Z)$ where $X,Y,Z\in L\cup\cV$. A
\emph{graph pattern} is a finite set of simple graph patterns. Let $\sigma:
\mathcal{V} \rightarrow L$ be a partial function from variables to labels. With
a slight abuse of notation, we also use $\sigma$ as a postfix operator that
replaces each occurrence of the variables in $\sigma$ with the corresponding
node. Given the graph $G$ and a simple graph pattern $q$, the \emph{answers} for
$q$ on $G$ correspond to the set $\ans(G,q)=\{r(s,d) \mid r(s,d) \in E \wedge
q\sigma=(s,r,d) \}$. Function $\bound(p)$ returns the positions of the labels
in the simple graph pattern $p$ left-to-right, i.e., if
$p=(X,a,b)$ where $X\in \cV$ and $a,b\in L$, then $\bound(p)=\pa{rd}$.

A \emph{Knowledge Graph (KG)} is a directed labeled graph where nodes are
entities and edges establish semantic relations between them, e.g., $\langle
Sadiq\_Khan,majorOf,London \rangle$. Usually, KGs are published on the Web using
the RDF data model~\cite{rdf}. In this model, data is represented as a set of
\emph{triples} of the form $\langle subject, predicate, object\rangle$ 
drawn from $(\cI \cup \cB) \times \cI \times (\cI \cup
\cL)$ where $\cI,\cB,\cL$ denote sets of IRIs, blank nodes and literals
respectively. Let $\cT=\cI\cup \cB \cup \cL$ be the set of all RDF terms. RDF
triples can be trivially seen as a graph where the subjects and objects are the
nodes, triples map to edges labeled with their predicate name, and $L=\cT$.

\begin{tr}
SPARQL~\cite{sparql} is a language for querying knowledge graphs which has been
standardized by W3C. It offers many SQL-like operators like \texttt{UNION},
\texttt{FILTER}, \texttt{DISTINCT} to specify complex queries and to further
process the answers. Every query contains at its core a graph pattern, which is
called \emph{Basic Graph Pattern (BGP)} in the SPARQL terminology. SPARQL graph
patterns are defined over $\cT \cup \cV$ and their answers are mappings $\sigma$
from $\cV$ to $\cT$. Therefore, answering a SPARQL graph pattern $P$ over a KG
$G$ corresponds to computing $\ans(G,p_1)\cap \ldots \cap \ans(G,p_{|P|})$ and
retrieving the corresponding labels.

\begin{example} \label{ex:1} An example of a SPARQL query is: \begin{verbatim}
SELECT ?s ?o { ?s isA ?o . ?s livesIn Rome . }\end{verbatim}
\begin{sloppypar}
If the KG contains the RDF triples $\langle
\mathtt{Eli, isA, Professor}\rangle$ and $\langle
\mathtt{Eli, livesIn, Rome}\rangle$ then one answer to the query is $\{ (?s
\rightarrow \mathtt{Eli}, ?o \rightarrow \mathtt{Professor}) \}$.
\end{sloppypar}
\end{example}
\end{tr}

\section{Graph Primitives}

We start our discussion with a description of the low-level primitives that we
wish to support. We distilled these primitives considering four types of
workloads: \emph{SPARQL~\cite{sparql} query answering}, which is the most
popular language for querying KGs; \emph{Rule-based
reasoning}~\cite{reasoningsurvey}, which is an important task in the Semantic
Web to infer new knowledge from KGs; Algorithms for \emph{graph analytics}, or
network analysis, since these are widely applied on KGs either to study
characteristics like the graph's topology or degree distribution, or within
more complex pipelines; \emph{Statistical relational models}~\cite{kgemb_survey},
which are effective techniques to make predictions using the KG as prior
evidence.

If we take a closer look at the computation performed in these tasks, we can
make a first broad distinction between \emph{edge-centric} and
\emph{node-centric} operations. The first ones can be defined as operations that
retrieve subsets of edges that satisfy some constraints. In contrast, operations
of the second type retrieve various data about the nodes, like their degree.
Some tasks, like SPARQL query answering, depend more heavily on edge-centric
operations while others depend more on node-centric operations (e.g., random
walks).

\fakeparagraph{Graph Primitives} Following a RISC-like approach, we identified a
small number of low-level primitives that can act as basic building blocks for
implementing both node- and edge-centric operations. These primitives are
reported in Table~\ref{tab:primitives} and are described below.

\begin{table}[tb]
    \centering
    \scriptsize
    \begin{tabular}{lll}
        & \emph{Name} & \emph{Output} \\
        \hline \hline
        $f_1$ & $\lbl_n(G,n)$ & Label of node $n$ (equals to
        $\phi_V(v)$).\\
        $f_2$ & $\lbl_e(G,e)$ & Label of edge $e$ (equals to $\phi_E(e)$).\\
        $f_3$ & $\nodid(G,l)$ & $\il$, i.e., the ID of node with label $l$.\\
        $f_4$ & $\edgid(G,l)$ & $\il$, i.e., the ID of edge label $l$.\\
        \hline
        $f_5$ & $\edg_{\srd}(G,p)$ & $\ans(G,p)$ sorted by $\srd$. \\
        $f_6$ & $\edg_{\sdr}(G,p)$ & $\ans(G,p)$ sorted by $\sdr$. \\
        $f_7$ & $\edg_{\drs}(G,p)$ & $\ans(G,p)$ sorted by $\drs$.  \\
        $f_8$ & $\edg_{\dsr}(G,p)$ & $\ans(G,p)$ sorted by $\dsr$. \\
        $f_9$ & $\edg_{\rsd}(G,p)$ & $\ans(G,p)$ sorted by $\rsd$. \\
        $f_{10}$ & $\edg_{\rds}(G,p)$ & $\ans(G,p)$ sorted by $\rds$. \\
        \hline
        $f_{11}$ & $\grp_{\spa}(G,p)$ & All $s$ of $\ans(G,p)$.\\
        $f_{12}$ & $\grp_{\rpa}(G,p)$ & All $r$ of $\ans(G,p)$.\\
        $f_{13}$ & $\grp_{\dpa}(G,p)$ & All $d$ of $\ans(G,p)$. \\
        $f_{14}$ & $\grp_{\srsd}(G,p)$ & Aggr.
        $(s,r)$/$(s,d)$ of $\ans(G,p)$.\\
        $f_{15}$ & $\grp_{\rsrd}(G,p)$ & Aggr.
        $(r,s)$/$(r,d)$ of $\ans(G,p)$. \\
        $f_{16}$ & $\grp_{\dsdr}(G,p)$ & Aggr.
        $(d,s)$/$(d,r)$ of $\ans(G,p)$. \\
        \hline
        $f_{17}$ & $\cnt(f_5|\ldots|f_{16})$ & Cardinality of
        $f_5,\ldots,f_{16}$.\\
        \hline
        $f_{18}$ & $\pos_{\srd}(G,p,i)$ & $i^{th}$ edge returned by
        $\edg_{\srd}(G,p)$. \\
        $f_{19}$ & $\pos_{\sdr}(G,p,i)$ & $i^{th}$ edge returned by
        $\edg_{\sdr}(G,p)$. \\
        $f_{20}$ & $\pos_{\drs}(G,p,i)$ & $i^{th}$ edge returned by
        $\edg_{\drs}(G,p)$. \\
        $f_{21}$ & $\pos_{\dsr}(G,p,i)$ & $i^{th}$ edge returned by
        $\edg_{\dsr}(G,p)$. \\
        $f_{22}$ & $\pos_{\rsd}(G,p,i)$ & $i^{th}$ edge returned by
        $\edg_{\rsd}(G,p)$.  \\
        $f_{23}$ & $\pos_{\rds}(G,p,i)$ & $i^{th}$ edge returned by
        $\edg_{\rds}(G,p)$. \\
    \end{tabular}
    \caption{Graph primitives}
    \label{tab:primitives}
\end{table}

\fakeparagraph{$\mathbf{f_1-f_4}$} These primitives retrieve the numerical IDs
associated with labels and vice-versa. The primitives $f_1$ and $f_2$ retrieve the
labels associated with nodes and edges respectively. The primitives $f_3$ and
$f_4$ retrieve the labels associated with numerical IDs.

\fakeparagraph{$\mathbf{f_5-f_{10}}$} Function $\edg_\omega(G,p)$  retrieves the
subset of the edges in $G$ that matches the simple graph pattern $p$ and returns it
sorted according to $\omega$. Primitives in this group are particularly important for
the execution of SPARQL queries since they encode the core operation of
retrieving the answers of a SPARQL triple pattern.

\fakeparagraph{$\mathbf{f_{11}-f_{16}}$} This group of primitives returns an
aggregated version of the output of $\mathbf{f_{5}-f_{10}}$. For instance,
$\grp_{s}(G,p)$ returns the list $\langle (x_1,c_1),\ldots,(x_n,c_n)\rangle$ of
all distinct sources in the edges $\ans(G,p)$ with the respective counts of the
edges that share them. Let $D$ be a set of edges, $A(x,D)=\{r(x,d) \in D\}$
and $B(D)=\{ (s,c) \mid r(s,d) \in A(s,D) \wedge c=|A(s,D)|\}$.  Then,
$\grp_{s}(G,p)$ returns the list of all tuples in $B(\ans(G,p))$ sorted by the
numerical ID of the first field. The other primitives are defined analogously.

\fakeparagraph{$\mathbf{f_{17}}$} This primitive returns the cardinality of the
output of $f_5,\ldots,f_{16}$. This computation is useful in a number of cases: For
instance, it can be used to optimize the computation of SPARQL queries by
rearranging the join ordering depending on the cardinalities of the triple
patterns or to compute the degree of nodes in the graph.

\fakeparagraph{$\mathbf{f_{18}-f_{23}}$} These primitives return the $i^{th}$
edge that would be returned by the corresponding primitives $\edg_*$. In
practice, this operation is needed in several graph
analytics algorithms or for mini-batching during the training of statistical
relational models.

\begin{tr}
    \begin{example} We show how we can use the primitives in
    Table~\ref{tab:primitives} to answer the SPARQL query of Example~\ref{ex:1},
    assuming that the KG is called $I$.
    \squishlist \item First, we retrieve the IDs of the labels \pa{isA},
    \pa{livesIn}, and \pa{Rome}. To this end, we can use the primitives $f_3$
    and $f_4$.

    \item Then, we create two single graph patterns $p_1$ and $p_2$ which map to
        the first and second triple patterns respectively. Then, we execute
        $\edg_\pa{rsd}(I,p_1)$ and $\edg_\pa{drs}(I,p_2)$ so that the edges are
        returned in a order suitable for a merge join.

    \item We invoke the primitive $f_1$ to retrieve all the labels of the nodes
    which are returned by the join algorithm. These labels are then used to
construct the answers of the query.  \squishend \end{example}
\end{tr}

\section{Architecture}

One straightforward way to implement the primitives in
Table~\ref{tab:primitives} is to store the KG in many independent data
structures that provide optimal access for each function. However, such solution
will require a large amount of space and updates will be slow. It is challenging
to design a storage engine that uses fewer data structures without excessively
compromising the performance.

Moreover, KGs are highly heterogeneous objects where some subgraphs have a
completely different topology than others. The storage engine should take
advantage of this diversity and potentially store different parts of the KGs in
different ways, effectively adapting to its structure. This adaptation lacks in
current engines, which treat the KG as a single object to store.

Our architecture addresses these two problems with a compact storage
layer that supports the execution of primitives $f_1,\ldots,f_{23}$ with a
minimal compromise in terms of performance, and in such a way that the engine
can adapt to the KG in input selecting the best strategy to store its parts.

\begin{figure}[t] \centering
    \includegraphics[width=0.95\linewidth]{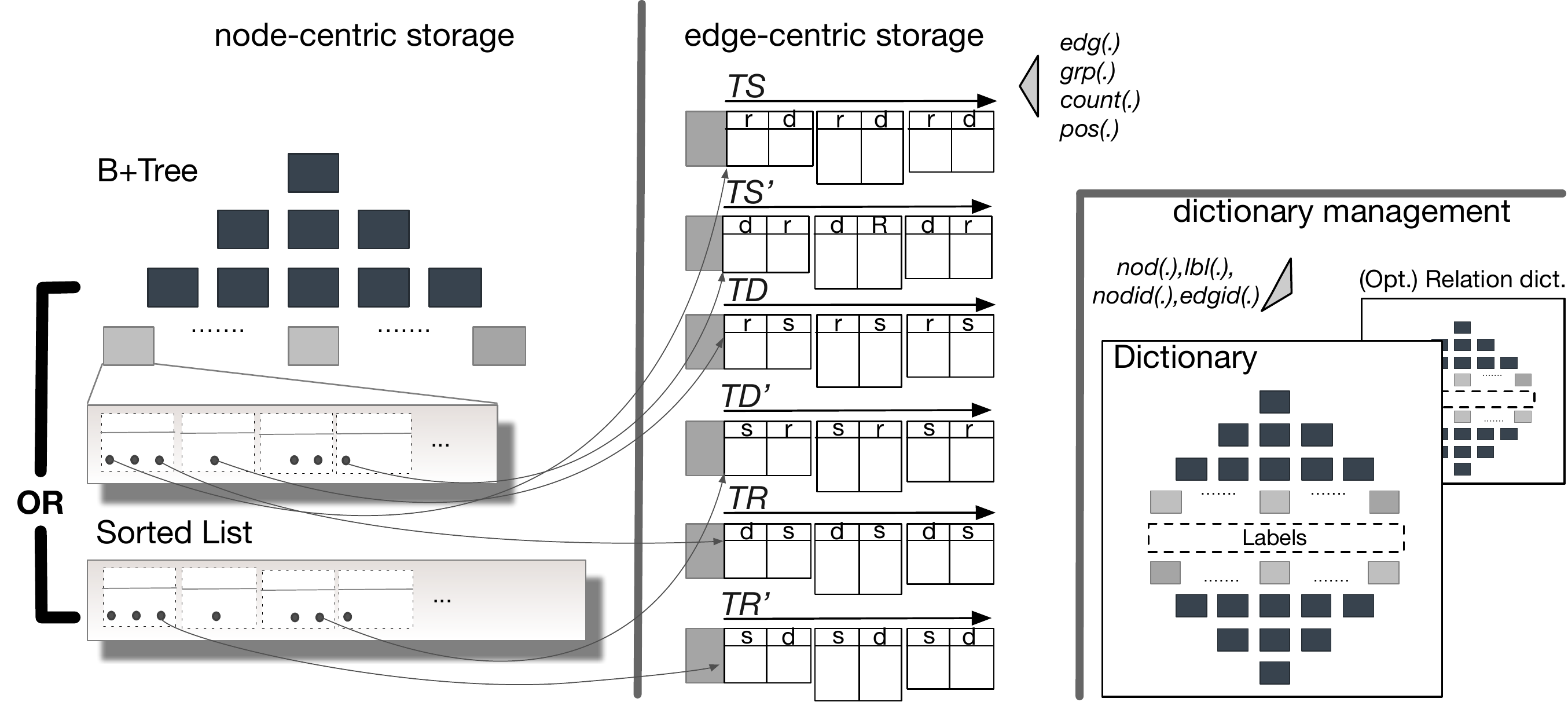} \caption{
Architectural overview of \sys} \label{fig:overview} \end{figure}

Figure~\ref{fig:overview} gives a graphical view of our approach. It uses a
series of interlinked data structures that can be grouped in \emph{three
    components}. The first one contains data structures for the mappings
    $ID\Leftrightarrow label$. The second component is called \emph{edge-centric
    storage} and contains data structures for providing fast access to the
    edges. The third one is called \emph{node-centric storage} and offers fast
    access to the nodes. Section~\ref{sec:components} describes these components
    in more detail. Section~\ref{sec:primitives} discusses how they allow an
    efficient execution of the primitives, while Section~\ref{sec:loadingdb}
    focuses on loading and updating the database.

\subsection{Architectural Components}
\label{sec:components}

\fakeparagraph{Dictionary} We store the labels on a block-based byte stream on
disk. We use one B+Tree called $\dicti$ to index the mappings $ID\Rightarrow
label$ and another one called $\dictl$ for $label\Rightarrow ID$. Using B+Trees
here is usual, so we will not discuss it further. %
\begin{tr}
    It is important to note that assigning a certain ID to a term rather
    than another one might have a significant impact on the performance. For
    instance, ~\citeauthor{kognac}~\cite{kognac} have shown that a careful
    choice of the IDs can introduce importance speedups due to the improved data
    locality.  Typically, current graph engines assign unique IDs to all labels,
    irrespectively whether a label is used as an entity or as a relation. This
    is desirable for SPARQL query answering because all data joins can operate
    on the IDs directly.  There are cases, however, where unique ID assignments
    are not optimal. For instance, most implementations of techniques for
    creating KG embeddings (e.g., TranSE~\cite{transe}) store the embeddings for
    the entities and the ones for relations in two contiguous vectors, and use
    offsets in the vectors as IDs. If the labels for the relations share the IDs
    with the entities, then the two vectors must have the same number of
    elements. This is highly inefficient because KGs have many fewer relations
    than entities which means that much space in the second vector will be
    unused. To avoid this problem, we can assign IDs to entities and
relationships in an independent manner. In this way, no space is wasted in
storing the embeddings.
\end{tr}
Note that \sys supports both global ID assignments and independent
entity/relationship assignments with an additional index specifically for the
relation labels. The first type of assignment is needed for tasks like SPARQL
query answering while the second is useful for operations like learning graph
embeddings~\cite{kgemb_survey}.

\fakeparagraph{Edge-centric storage} In order to adapt to the complex and
non-uniform topology of current KGs, we do not store all edges in a single data
structure, but store subsets of the edges independently. These subsets
correspond the edges which share a specific entity/relation. More specifically,
let us assume that we must store the graph $G=(V,E,L,\phi_V,\phi_E)$. For each
$l\in L$, we consider three types of subsets: $E_s(l)=\{ r(l,d) \in E\}$,
$E_r(l)=\{ l(s,d) \in E \}$, $E_d(l)=\{ r(s,l) \in E \}$, i.e., the subsets of
edges that have $l$ as source, edge, or destination respectively.

The choice of separating the storage of various subsets allows us to choose the
best data structure for a specific subset, but it hinders the execution of
\emph{inter-table scans}, i.e., scans where the content of multiple tables must
be taken into account. To alleviate this problem, we organize the physical
storage in such a way that all edges can still be retrieved by scanning a
contiguous memory location.

We proceed as follows: First, we compute $E_s(l)$, $E_r(l)$, $E_d(l)$ for every
$l\in L$. Let $\Omega$ be the collection of all these sets. For each $E_x(l)\in
\Omega$, we construct two sets of tuples, $F_x(l)$ and $G_x(l)$, by extracting
the free fields left-to-right and right-to-left respectively. For instance, the
set $E_s(l)$ results into the sets $F_s(l)=\{\langle r,d \rangle \mid r(l,d)\in
E\}$ and $G_s(l)=\{\langle d,r \rangle \mid r(l,d)\in E\}$. Since these sets
contains pairs of elements, we view them as binary tables. These are grouped
into the following six sets:

\squishlist
\item $T_s=\{F_s(l) \mid l \in L\}$ and $T'_s=\{ G_s(l) \mid l \in L \}$
\item $T_r=\{F_r(l) \mid l \in L\}$ and $T'_r=\{ G_r(l) \mid l \in L \}$
\item $T_d=\{F_d(l) \mid l \in L\}$ and $T'_d=\{ G_d(l) \mid l \in L \}$
\squishend

The content of these six sets is serialized on disk in corresponding byte
streams called \TS, \TSI, \TR, \TRI, \TD, and \TDI respectively (see middle
section of Figure~\ref{fig:overview}). The serialization is done by first
sorting the binary tables by their defining label IDs, and then serializing each table
one-by-one. For instance, if $F_s(l_1),F_s(l_2) \in T_s$, then $F_s(l_1)$ is
serialized before $F_s(l_2)$ iff $\iota_{l_1} < \iota_{l_2}$. At the beginning
of the byte stream, we store the list of all IDs associated to the tables,
pointers to the tables' physical location and instructions to parse them.

Since the binary tables and tuples are serialized on the byte stream with a specific order, we can
retrieve all edges sorted with any ordering in $\cR$ with a single scan of the
corresponding byte stream, using the content stored at the beginning of the
stream to decode the binary tables in it. For instance, we can scan \TS to retrieve all edges
sorted according to \srd. %
\begin{tr}
The IDs stored at the beginning of the stream specify
the sources of the edges (\spa) while the content of the tables specify the
remaining relations and destinations (\rpa and \dpa).
\end{tr}

\fakeparagraph{Node-centric storage} In order to provide fast access to the
nodes, we map each ID $\il$ (i.e., the ID assigned to label $l$) to a tuple $M_l$
that contains 15 fields:

\squishlist

\item the cardinalities $|E_s(l)|$, $|E_r(l)|$, and $|E_d(l)|$;

\item Six pointers $p_1,\ldots,p_6$ to the physical storage of $F_s(l)$,
    $G_s(l)$, $F_r(l)$, $G_r(l)$, $F_d(l)$, and $G_d(l)$;

\item Six bytes $m_1,\ldots,m_6$ that contain instructions to read the data
    structures pointed by $p_1,\ldots,p_6$. These instructions are necessary
    because the tables are stored in different ways (see
    Section~\ref{sec:dynamic}).

\squishend

We index all $M_*$ tuples by the numerical IDs using one global data structure called
\nodemgr (Node Manager), shown on the left side of Figure~\ref{fig:overview}.
This data structure is implemented either with an on-disk B+Tree or with a
in-memory sorted vector (the choice is done at loading time). The
B+Tree is preferable if the engine is used for edge-based
computation because the B+Tree does not need to load all nodes in main memory
and the nodes are accessed infrequently anyway. In contrast, the sorted vector
provides much faster access ($O(1)$ vs. $O(log|L|)$) but it requires that the
entire vector is stored in main memory. Thus, it is suitable only if the
application accesses the nodes very frequently and there are enough hardware
resources.

\begin{tr}
Note that the coordinates to the binary tables are stored both in \nodemgr and
in the meta-data in front of the byte streams. This means that the table can be
accessed either by accessing \nodemgr, or by scanning the beginning of the byte
stream. In our implementation, we consult \nodemgr when we need to answers graph
patterns with at least one constant element (e.g., for answering the query in
Example~\ref{ex:1}). In contrast, the meta-content at the beginning of the
stream is used when must perform a full scan.
\end{tr}

The way we store the binary tables in six byte streams resembles six-permutation
indexing schemes such as proposed in engines like RDF3X~\cite{rdf3x} or
Hexastore~\cite{hexastore}. There are, however, two important differences:
First, in our approach the edges are stored in multiple independent binary
tables rather than a single series of ternary tuples (as, for instance, in
RDF3X~\cite{rdf3x}). This division is important because it allows us to choose
different serialization strategies for subgraphs or to avoid storing some tables
(Section~\ref{sec:table_pruning}). The second difference is that in our case
most access patterns go through a \emph{single} B+Tree instead of six different
data structures. This allows us to save space and to store additional
information about the nodes, e.g., their degree, which is useful, for
instance, for traversal algorithms like PageRank, or random walks.

\subsection{Primitive Execution}
\label{sec:primitives}

We now discuss how we can implement the primitives in Table~\ref{tab:primitives}
with our architecture.

\fakeparagraph{$\mathbf{Primitives\;f_1,\ldots,f_4\;(\lbl_*\;\nodid,\edgid)}$}
These are executed consulting either
\dictl\;or \dicti. Thus, the time complexity follows in a straightforward
manner.

\begin{proposition} Let $G=(V,E,L,\phi_V,\phi_E)$. The time complexity of
computing $f_1,\ldots,f_4$ is $O(log(|L|))$.  \end{proposition}

\fakeparagraph{$\mathbf{Primitives\;f_5,\ldots,f_{10}\;(\edg_*)}$} Let
$\edg_\omega(G,p)$ be a generic invocation of one of $f_5,\ldots,f_{10}$. First,
we need to retrieve the numerical IDs associated to the labels in $p$ (if any).
Then, we select an ordering that allows us to 1) retrieve answers for $p$ with a
range scan, and 2) the ordering complies with $\omega$. The orderings that
satisfy 1) are \begin{equation}\label{eq:1}\Omega =\{\omega' \mid \omega' \in
        \cR \wedge \isprefix(\bound(p),\omega')=true\}\end{equation}

An ordering $\omega'\in\Omega$ which also satisfies 2) is one for which
$\omega'-\bound(p)=\omega-\bound(p)$.

\begin{example} Consider the execution of $\edg_{\pa{srd}}(G,p)$ where $p=
(X,Y,a)$. In this case, $\bound(p)=\pa{d}$, $\Omega=\{\pa{drs},\pa{dsr}\}$ and
$\omega'=\pa{dsr}$. \end{example}

The selected $\omega'$ is associated to one byte stream. If $p$ contains one or
more constants, then we can query \nodemgr to retrieve the appropriate binary
table from that binary stream and (range-)scan it to retrieve the answers of
$p$. In contrast, if $p$ only contains variables, the results can be obtained by
scanning all tables in the byte stream. Note that the cost of retrieving the
IDs for the labels in $p$ is $O(log|L|)$ since we use B+Trees for the
dictionary. This is an operation that is applied any time the input contains a
graph pattern. If we ignore this cost and look at the remaining computation,
then we can make the following observation.

\begin{proposition} Let $G=(V,E,L,\phi_V,\phi_E)$. The time complexity of
$\edg_\omega(G,p)$ is $O(|E|))$ if $p$ only contains variables, $O(log(|L|) +
|E|)$ otherwise. \end{proposition}

\fakeparagraph{$\mathbf{Primitives\;f_{11},\ldots,f_{16}\;(\grp_*)}$} Let
$\grp_\omega(G,p)$ be a general call to one of these primitives. Note that in
this case $\omega \in \cR'$, i.e., is a partial ordering. These functions can be
implemented by invoking $f_{5},\ldots,f_{10}$ and then return an aggregated
version. Thus, they have the same cost as the previous ones. %
%

%
However, there are special cases where the computation is quicker, as shown in
the next example.

\begin{example} Consider a call to $\grp_{\pa{s}}(G,p)$ where $p=\langle
    a,X,Y\rangle$. In this case, we can query \nodemgr with $a$ and return at
    most one tuple with the cardinality stored in $M_a$, which has a cost of O(log(|L|)).
\end{example}

If $\omega$ has length two or $p$ contains a repeated variable, then we also
need to access one or more binary tables, similarly as before.

\begin{proposition} Let $G=(V,E,L,\phi_V,\phi_E)$. The time complexity of
    $\grp_\omega(G,p)$ ranges between $O(log(|L|))$ and $O(log(|L|) + |E|)$ depending on $p$ and
$\omega$. \end{proposition}

\fakeparagraph{$\mathbf{Primitive\;f_{17}\;(\cnt)}$} This primitive returns
the cardinality of the output of $f_5,\ldots,f_{16}$. Therefore, it
can be simply implemented by iterating over the results returned by these
functions. However, there there are cases when we can avoid this iteration.
Some of such cases are the ones below:\squishlist

\item If the input is $\edg_\omega(G,p)$ and $p$ contains no constant nor repeated
    variables. In this case the output is $|E|$.

\item If the input is $\edg_\omega(G,p)$ and $p$ contains only one constant $c$
    and no repeated variables. In this case the cardinality is stored in
    $M_c$.

\item If the input is $\grp_\omega(G,p)$, $\isprefix(\omega,\omega')=true$, and
    $p$ contains at most one constant and no repeated variables, then the output
    can be obtained either by consulting \nodemgr or the metadata of one of the
    byte streams.

\squishend

\noindent Otherwise, we also need to access one binary table to compute
the results, which, in the worst case, takes $O(|E|)$.

\begin{proposition} Let $G=(V,E,L,\phi_V,\phi_E)$. The time complexity of
executing $\cnt(\cdot)$ ranges between $O(log(|L|))$ and $O(log(|L|)+|E|)$.  \end{proposition}

\fakeparagraph{$\mathbf{Primitives\;f_{18},\ldots,f_{23}\;(\pos_*)}$} In order
to efficiently support these primitives, we need to provide a fast random access
to the edges. Given a generic $\pos_\omega(G,p,i)$, we distinguish four cases:

\squishlist

\item \textbf{C1} If $p$ contains repeated variables, then we iterate over the
results and return the $i^{th}$ edge;

\item \textbf{C2} If $p$ contains only one constant, then the search space is
    restricted to a single binary table. In this case, the computation depends
    on how the content of the table is serialized on the byte stream. If it
    allows random access to the rows, then the cost reduces to $O(log(|L|))$,
    i.e., query \nodemgr. Otherwise we also need to iterate through the table
    and count until the $i^{th}$ row;

\item \textbf{C3} If $p$ contains more than one constant, then we need to search
    through the table for the right interval, and then scan until we retrieve
    the $i^{th}$ row;

\item \textbf{C4} Finally, if $p$ does not contain any constants or repeated
    variables, we must consider all edges stored in one byte stream. In this
    case, we first search for the binary table that contains the $i^{th}$ edge.
    This operation requires a scan of the metadata associated to the byte
    stream, which can take up to $O(|L|)$. Then, the complexity depends on
    whether the physical storage of the table allows a random access, as in
    \textbf{C2} and \textbf{C3}. Since a scan over the metadata takes  $O(|L|)$,
    this last case represents the worst-case in terms of complexity as it sums
    to $O(|L|+|E|)$. Note that in this case, simply going through all edges is
    faster as it takes $O(|E|)$. However, in practice tables have more than one
    row so we can advance more quickly despite the higher worst-case complexity.
    \squishend

\begin{proposition} Let $G=(V,E,L,\phi_V,\phi_E)$. The time complexity of
executing $\pos_\omega(G,p,i)$ ranges between $O(log(|L|))$ and $O(|L|+|E|)$.\end{proposition}

\subsection{Bulk Loading and Updates}
\label{sec:loadingdb}

\begin{figure}[t] \centering
    \includegraphics[width=\linewidth]{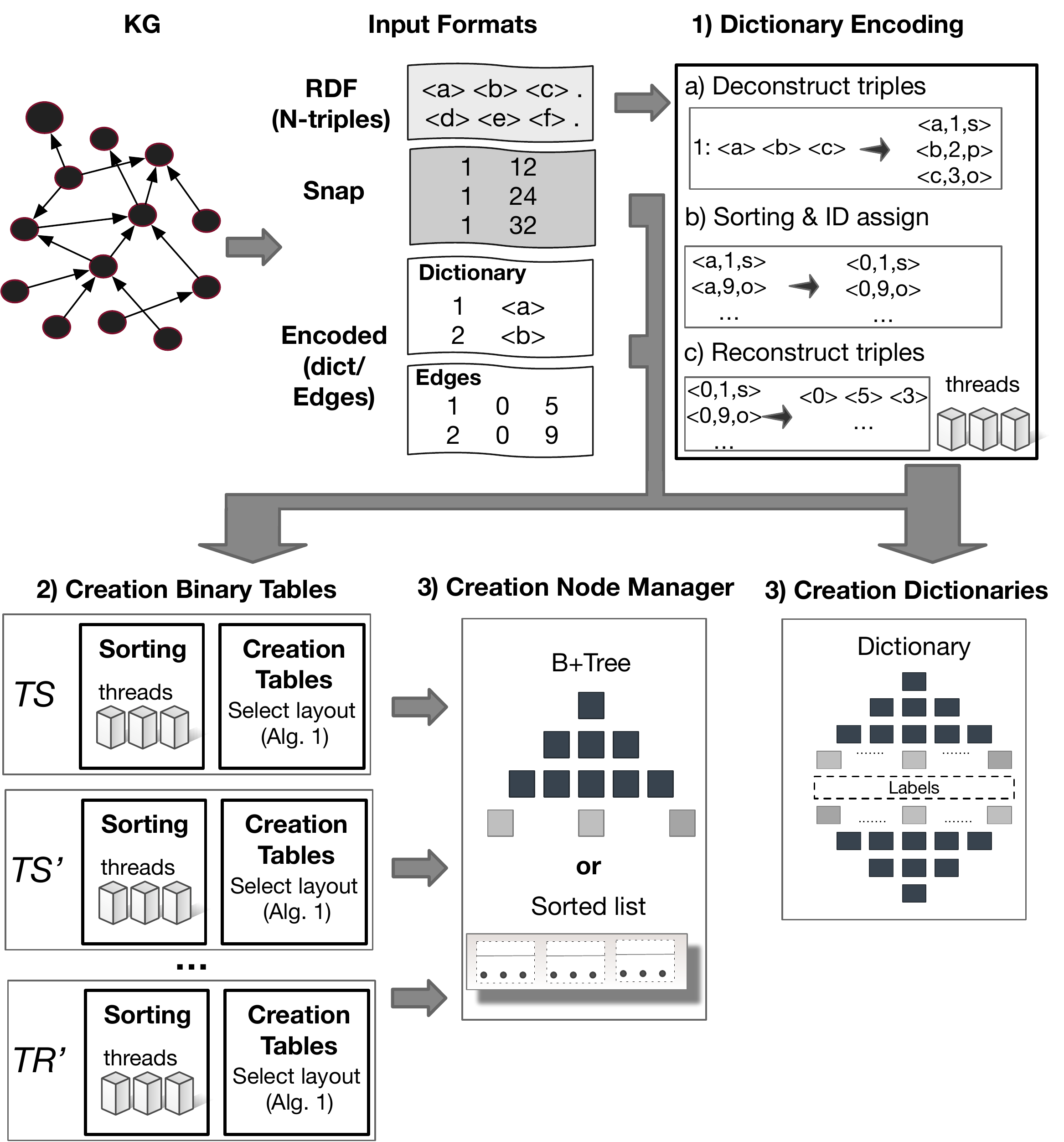} \caption{
Bulk loading in \sys} \label{fig:loading} \end{figure}

\fakeparagraph{Bulk Loading} Loading a large KG can be a lengthy process,
especially if the resources are constrained. In \sys, we developed a loading
routine which exploits the multi-core architecture and maximizes the (limited)
I/O bandwidth.

The main operations are shown in Figure~\ref{fig:loading}. Our implementation
can receive the input KG in multiple formats. Currently, we considered the
N-Triples format (popular in the Semantic Web) and the SNAP
format~\cite{snapnets} (used for generic graphs). The first operation is
encoding the graph, i.e., assigning unique IDs to the entities and relation
labels. For this task, we adapted the MapReduce technique presented
at~\cite{rdfcompr} to work in a multi-core environment. This technique first
deconstructs the triples, then assigns unique IDs to all the terms, and finally
reconstruct the triples. If the graph is already encoded, then our procedure
skips the encoding and proceeds to the second operation of the loading, the
creation of the database.

The creation of the binary tables requires that the triples are pre-sorted
according to a given ordering. We use a disk-based parallel merge sort
algorithm for this purpose. The tables are serialized one-by-one selecting
the most efficient layout for each of them. After all the tables are created, the
loading procedure will create the \nodemgr and the B+Trees with the
dictionaries. The encoding and sorting procedures are parallelized using
threads, which might need to communicate with the secondary storage. Modern
architectures can have >64 cores, but such a number of threads can easily
saturate the disk bandwidth and cause serious slowdowns. To avoid this problem,
we have two types of threads: Processing threads, which perform computation like
sorting, and I/O threads, which only read and write from disk. In this way, we
can control the maximum number of concurrent accesses to the disks.

\fakeparagraph{Updates} To avoid a complete re-loading of the entire KG after
each change, our implementation supports incremental updates. Our procedure is
built following the well-known advice by Jim Gray~\cite{gray1981transaction}
that discourages in-place updates, and it is inspired by the idea of
differential indexing~\cite{rdf3x}, which proposes to create additional indices
and perform a lazy merging with the main database when the number of indices
becomes too high.

Our procedure first encodes the update, which can be either an addition or
removal, and then stores it in a smaller ``delta'' database with its own
\nodemgr and byte streams. Multiple updates will be stored in multiple
databases, which are timestamped to remember the order of updates. Also, updates
create an extra dictionary if they introduce new terms. Whenever the primitives are
executed, the content of the updates is combined with the main KG so that
the execution returns an updated view of the graph.

In contrast to differential indexing, our merging does not copy the
updates in the main database, but only groups them in two
updates, one for the additions and one for the removals. This is to avoid the
process of rebuilding binary tables with possibly different layouts. If the size
of the merged updates becomes too large, then we proceed with a full reload of
the entire database.

\section{Adaptive Storage Layout}
\label{sec:dynamic}

The binary tables can be serialized in different ways. For instance, we can
store them row-by-row or column-by-column. Using a single serialization strategy
for the entire KG is inefficient because the tables can be very different from
each other, so one strategy may be efficient with one table but
inefficient with another. Our approach addresses this inefficiency by choosing
the best serialization strategy for each table depending on its size and
content.

For example, consider two tables $T_1$ and $T_2$. Table $T_1$ contains all the
edges with label ``isA'', while $T_2$ contains all the edges with label
``isbnValue''. These two tables are not only different in terms of sizes, but
also in the number of duplicated values. In fact, the second column of $T_1$ is
likely to contain many more duplicate values than the second column of $T_2$
because there are (typically) many more instances than classes while
``isbnValue'' is a functional property, which means that every entity in the
first column is associated with a unique ISBN code. In this case, it makes sense
to serialize $T_1$ in a column-by-column fashion so that we can apply
run-length-encoding (RLE)~\cite{rle}, a well-known compression scheme of
repeated values, to save space when storing the second column. This type of
compression would be ineffective with $T_2$ since there each value appears only
once. Therefore, $T_2$ can be stored row-by-row.

In our approach, we consider three different serialization strategies, which we
call \emph{serialization layouts} (or simply layouts) and employ an ad-hoc
procedure to select, for each binary table, the best layout among these three.

\subsection{Serialization Layouts}
\label{sec:desclayouts}

We refer to the three layouts that we consider as \emph{row}, \emph{column}, and
\emph{cluster} layouts respectively. The first layout stores the content
row-by-row, the second column-by-column, while the third uses an intermediate
representation.

\fakeparagraph{Row layout} Let $T=\langle \langle t'_1,t''_1\rangle, \ldots,
\langle t'_n,t''_n\rangle \rangle$ be a binary table that contains $n$ sorted
pairs of elements. With this layout, the pairs are stored one after the other.
In terms of space consumption, this layout is optimal if the two columns do not
contain any duplicated value. Moreover, if each row takes a fixed number of
bytes, then it is possible to perform binary search or perform a random access
to a subset of rows. The disadvantage is that with this layout all values are
explicitly written on the stream while the other layouts allow us to compress
duplicate values.

\fakeparagraph{Column layout} With this layout, the elements in $T$ are
serialized as $\langle t'_1,\ldots,t'_n\rangle,\langle
t''_1,\ldots,t''_n\rangle$. The space consumption required by this layout is
equal to the previous one but with the difference that here we can use RLE to
reduce the space of $\langle t'_1,\ldots,t'_n\rangle$. In fact, if
$t'_1=t'_2=\ldots=t'_n$, then we can simply write $t'_1\times n$. Also this
layout allows binary search and a random access to the table. However, it is
slightly less efficient than the row layout for full scans because here one row
is not stored at contiguous locations, and the system needs to ``jump'' between
columns in order to return the entire pair. On the other hand, this layout is more
suitable than the row layout for aggregate reads (required, for instance, for
executing $grp$ primitives) because in this case we only need to read the
content of one column which is stored at contiguous locations.

\fakeparagraph{Cluster layout} Let $g_{t}=\langle \langle t, t''_k \rangle,
\ldots , \langle t,t''_l \rangle \rangle$ be the longest sub-sequence of pairs
in $T$ which share the first term $t$. With this layout, all groups are first
ordered in the sequence $\langle g_{t_1},\ldots,g_{t_i},g_{t_{i+1}},\ldots,
g_{t_{m}} \rangle$ such that $t_i\leq t_{i+1}$ for all $1 \leq i < m$. Then,
they are serialized one-by-one. Each group $g_t$ is serialized by first
writing $t$, then $|g_t|$, and finally the list $t''_k,\ldots,t''_l$.
This layout needs less space than the row layout if the groups contain
multiple elements. Otherwise, it uses more space because it also stores the size
of the groups, and this takes an extra $\lceil log_2 n \rceil$ bits. Another
disadvantage is that with this layout binary search is only possible within one
group.

\subsection{Dynamic Layout Selection}
\label{sec:dynamicselection}

\begin{algorithm}[tb] \small \caption{$\selectlayout(T)$}
    \label{alg:decider} \DontPrintSemicolon
    $U \defeq \{ u \mid \langle u,v \rangle \in T\}$\\
    \If{$n\leq \tau$ and $|U|\leq \upsilon$}{
        $m_1 \defeq 0, m_2 \defeq 0, m_3 \defeq 0$\\
        \ForEach{$ u \in U$}{
            $Z \defeq \{ v \mid \langle u,v \rangle\in T\}$\\
            \lIf{$u>m_1$}{$m_1 \defeq u$}
            \lIf{$|Z|>m_3$}{$m_3 \defeq |Z|$}
            \ForEach{$z \in Z$}{
                \lIf{$z>m_2$}{$m_2 \defeq z$}
            }
        }
        $t_{c} \defeq |U| * (\sizeof{m_1} + \sizeof{m_3}) + |T| * \sizeof{m_2}$\\
        $t_{r} \defeq |T| * (\sizeof{m_1} + \sizeof{m_2})$\\
        \If{$t_{r}\leq t_{c}$}{
            \Return{$\langle \pa{ROW}, \sizeof{m_1}, \sizeof{m_2}, 0 \rangle$}
            }\lElse{
                \Return{$\langle \pa{CLUSTER},\sizeof{m_1}, \sizeof{m_2}, \sizeof{m_3} \rangle$}
        }
    }\lElse{\Return{$\langle \pa{COLUMN}, 5, 5, 0 \rangle$}}
\end{algorithm}

The procedure for selecting the best layout for each table is reported in
Algorithm~\ref{alg:decider}. Its goal is to select the layout which leads to the
best compression without excessively compromising the performance. In our
implementation, Algorithm~\ref{alg:decider} is applied by default, but the user
can disable it and use one layout for all tables.

The procedure receives as input a binary table $T$ with $n$ rows and returns a
tuple that specifies the layout that should be chosen. It proceeds as follows.
First, it makes a distinction between tables that have less than $\tau$ rows
(default value of $\tau$ is 1M) and contain less than $\upsilon$ unique elements
in the first column from tables that do not (line 2). We make this distinction
because 1) if the number of rows is too high then searching for the most optimal
layout becomes expensive and 2) if the number of unique pairs is too high, then
the cluster layout should not be used due to the lack of support of binary
search. With small tables, this is not a problem because it is well known that
in these cases linear search is faster than binary search due to a better usage
of the CPU's cache memory. The value for $\upsilon$ is automatically determined
with a small routine that performs some micro-benchmarks to identify the
threshold after which binary search becomes faster. In our experiments, this
value ranged between 16 and 64 elements.

If the table satisfies the condition of line 2, then the algorithm selects
either the \pa{ROW} or the \pa{CLUSTER} layout. The \pa{COLUMN} layout is not
considered because its main benefit against the other two is a better
compression (e.g., RLE) but this is anyway limited if the table is small. The
procedure scans the table and keeps track of the largest numbers and groups used
in the table ($m_1,m_2,m_3$). Then, the function invokes the subroutine
$\sizeof{\cdot}$ to retrieve the number of bytes needed to store these numbers.
It uses this information to compute the total number of bytes that would be
needed to store the table with the \pa{ROW} and \pa{CLUSTER} layout respectively
(variables $t_r$ and $t_c$). Then, it selects the layout that leads to maximum
compression.

If the condition in line 2 fails, then either the \pa{ROW} or the
\pa{COLUMN} layout can be selected. An exact computation would be too
expensive given the size of the table. Therefore, we always choose \pa{COLUMN}
since the other one cannot be compressed with RLE.

Next to choosing the best layout, Algorithm~\ref{alg:decider} also returns the
maximum number of bytes needed to store the values in the two fields in the
table ($m_1$ and $m_2$) and (optionally) also for storing the cluster size
($m_3$, this last value is only needed for \pa{CLUSTER}). The reason for doing
so is that it would be wasteful to use four- or eight-byte integers to store
small IDs. In the worst case, we assume that all IDs in both fields can be
stored with five bytes, which means it can store up to $2^{40}-1$ terms. We
decided to use byte-wise compression rather than bit-wise compression because
the latter does not appear to be worthwhile~\cite{rdf3x}. Note that more complex
compression schemes could also be used (e.g., VByte~\cite{vbyte}) but this
should be seen as future work.

The tuple returned by $\selectlayout$ contains the information necessary to
properly read the content of the table from the byte stream. The first field is
the chosen layout while the other fields are the number of bytes that should be
used to store the entries of the table. We store this tuple both in \nodemgr (in
one of the $m_*$ fields) and at the beginning of the byte stream.

\subsection{Table Pruning}
\label{sec:table_pruning}

With Algorithm~\ref{alg:decider}, the system adapts to the KG while storing
a single table. We discuss two other forms of compression that consider multiple
tables and decide whether some tables should be skipped or stored in aggregated
form.

\fakeparagraph{On-the-fly reconstruction (OFR)} Every table in one stream $T_x$
maps to another table in $T'_x$ where the first column is swapped with the
second column. If the tables are sufficiently small, one of them can be
re-constructed on-the-fly from the other whenever needed. While this operation
introduces some computational overhead, the saving in terms of space may justify
it. Furthermore, the overhead can be limited to the first access by
serializing the table on disk after the first re-construction.

We refer to this strategy as \emph{on-the-fly reconstruction (OFR)}. If the user
selects it at loading time, the system will not store any binary table in
$T'_x$ which has less than $\eta$ rows, $\eta$ being a value passed by the user
(default value is 20, determined after microbenchmarking).

\fakeparagraph{Aggregate Indexing} Finally, we can construct aggregate indices
to further reduce the storage space. The usage of aggregate indices is not novel
for KG storage~\cite{hexastore}. Here, we limit their usage to the tables
in $T'_r$ if they lead to a storage space reduction.

To illustrate the main idea, consider a generic table $t$ that contains the set
of tuples $F'_r(isA)$. This table stores all the $\langle object, subject
\rangle$ pairs of the triples with the predicate $isA$. Since there are
typically many more instances than classes, the first column of $t$ (the
classes) will contain many duplicate values. If we range-partition $t$ with the
first field, then we can identify a copy of the values in the second field of
$t$ in the partitions of tables in $T'_d$ where the first term is $isA$. With
this technique, we avoid storing the same sequence of values twice but instead
store a pointer to the partition in the other table.

\section{Evaluation}
\label{sec:evaluation}

\begin{table}[t]
  \scriptsize
  \centering
  \begin{tabular}{lrrr|lrrr}
     & \textbf{Type} & \textbf{\#Edges} & \textbf{\#Nodes} & &\textbf{Type} & \textbf{\#Edges} & \textbf{\#Nodes} \\
    \hline
    \pa{LUBM} & KG & Var. & Var. & \pa{YAGO2S} & KG & 76M & 37M \\
    \pa{DBPedia} & KG & 1B & 233M & \pa{Google} & Dir.& 5.1M & 875k \\
    \pa{Wikidata} & KG & 1.1B & 299M & \pa{Twitter} & Dir. & 1.7M & 81k \\
    \pa{Uniprot} & KG & 168M & 177M & \pa{Astro} & Undi. & 198k & 18k \\
    \pa{BTC2012} & KG & 1B & 367M & & & & \\
  \end{tabular}
  \caption{Details about the used datasets}
  \label{tab:characteristics}
\end{table}

\sys is developed in C++, is freely available, and works under Windows, Linux,
MacOS. \sys is also released in the form of a Docker image. The user can
interact via command line, web interface, or HTTP requests according to
the SPARQL standard.

\begin{figure*}
    \centering
    \scriptsize
\begin{subfigure}[T]{0.45\linewidth}
    \centering
    \includegraphics[width=\linewidth]{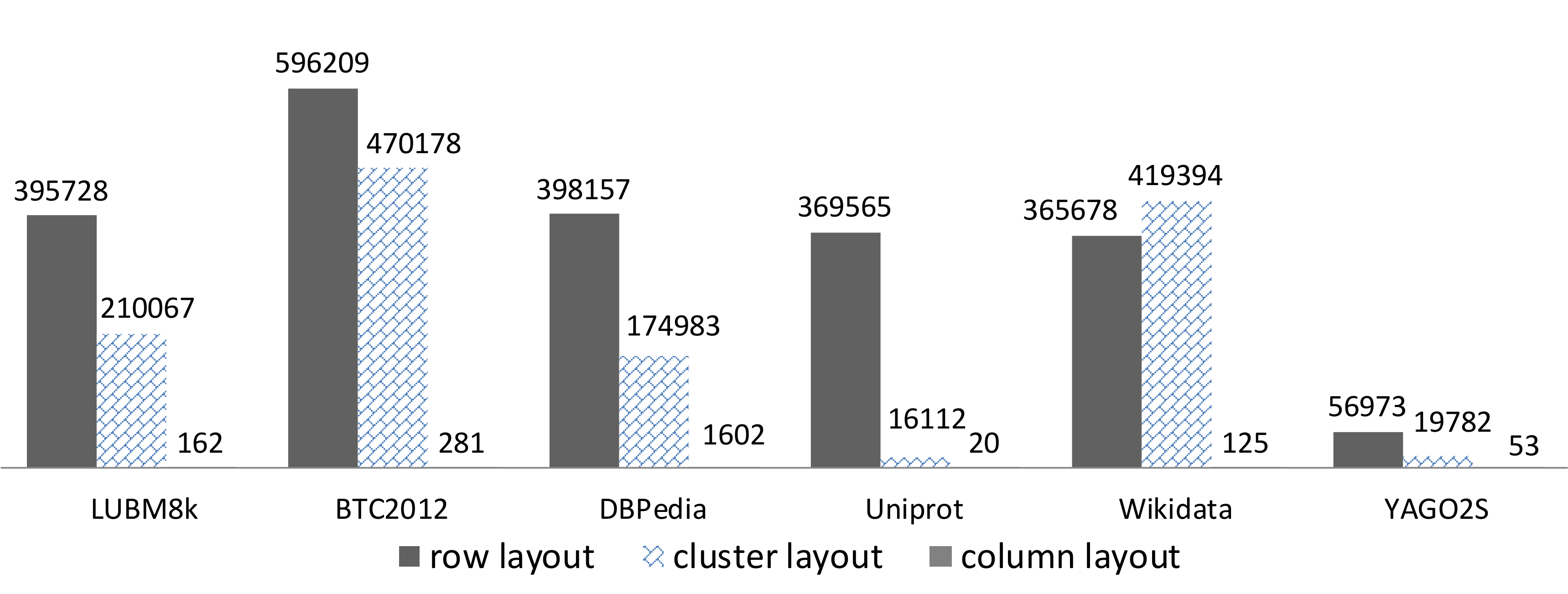}
    \caption{Number of tables (in 1k) of each type with various KGs}\label{fig:ntables}
\end{subfigure}%
\begin{subfigure}[T]{0.5\linewidth}
        \centering
            \begin{tabular}{l|ccccc}
                    & \multicolumn{5}{c}{\bf Graph triple patterns} \\
                    & 0 & 1 & 2 & 3 & 4 \\
                    Default & 1.28s & 0.07s & \bf0.15$\mu$s &
                    \bf 0.14$\mu$s & 0.18$\mu$s\\
                    With \pa{OFR} & 1.76s & 0.07s & 0.38$\mu$s & 0.38$\mu$s &
                    0.30$\mu$s \\
                    With \pa{AGGR} & 1.30s & 0.07s & 0.16$\mu$s & \bf 0.14$\mu$s & 0.19$\mu$s \\
                    Only \pa{ROW} & 1.25s & 0.07s & 0.16$\mu$s & 0.16$\mu$s & \bf 0.15$\mu$s\\
                    Only \pa{COLUMN} & 1.82s & 0.07s & 0.22$s\mu$s & 0.18$\mu$s & 0.27$\mu$s \\
                    \cline{2-6}
                    RDF3X & \bf 0.70s & \bf 0.06s & 22.84$\mu$s & 26.53$\mu$s &
                    18.55$\mu$s \\
            \end{tabular}
            \caption{Median runtimes (best ones are in bold)}\label{tab:lookup_runtimes}
            \begin{tabular}{cccc}
                      Default & With \pa{OFR} & With \pa{AGGR} & With \pa{OFR} \\
                        3.9GB & 2.7GB         & 3.4GB          & 2.5GB \\
                        \hline
                      \multicolumn{4}{l}{RDF3X: 5.1GB\;\;\;\;TripleBit:
                      3.3GB\;\;\;\;\virtuoso: 6.3GB}\\
            \end{tabular}
            \caption{Size of the database with \sys
            with/without optimizations}\label{tab:sizedatabases}
    \end{subfigure}
\caption{Statistics using various layouts/configurations and runtimes of triple
pattern lookups}
\end{figure*}

\fakeparagraph{Integration with other systems} Since our system offers low-level
primitives, we integrated it with the following other engines with simple
wrappers to evaluate our engine in
multiple scenarios:

\squishlist

\item \textbf{RDF3X~\cite{rdf3x}.} RDF3X is one of the fastest and most
    well-known SPARQL engines. We replaced its storage layer with ours so
    that we can reuse its SPARQL
    operators and query optimizations.

\item \textbf{SNAP~\cite{snap}.} Stanford Network Analysis Platform (SNAP) is a
    high-performance open-source library to execute over 100 different graph
    algorithms. As with RDF3X, we removed the SNAP storage layer and added an
    interface to our own engine.

\item \textbf{VLog~\cite{vlog}.} VLog is one
    of most scalable datalog reasoners. We implemented an interface allowing VLog
    to reason using our system as underlying database.  \squishend

\begin{tr}
We also implemented a native procedure to answer basic graph patterns (BGPs) that
applies greedy query optimization based on cardinalities, and uses either merge
joins or index loop joins if the first cannot be used.
\end{tr}
\begin{paper}
We also implemented a native procedure to answer basic graph patterns (BGPs) that
applies greedy query optimization based on cardinalities, and uses either merge
or index loop joins.
\end{paper}

\fakeparagraph{Testbed} We used a Linux machine (kernel 3.10, GCC 6.4, page size 4k) with dual Intel E5-2630v3 eight-core
CPUs of 2.4 GHz, 64 GB of memory and two 4TB SATA hard disks in RAID-0 mode. The
commercial value is well below \$5K. We compared against RDF3X
and SNAP with their native storages, TripleBit~\cite{triplebit}, a in-memory
state-of-the-art RDF database (in contrast to RDF3X which uses disks), and
\virtuoso, a widely used commercial SPARQL engine\footnote{We hide
    the real name as it is a commercial product, as usual in
database research.}. As inputs, we considered a selection of real-world and
artifical KGs, and other non-KG graphs from SNAP~\cite{snapnets} (see
Table~\ref{tab:characteristics} for statistics).

\squishlist
\item \textbf{KGs.} \pa{LUBM}~\cite{lubm}, a well-known artificial benchmark
    that creates KGs of arbitrary sizes. The KG is in the domain of
    universities and each university
    contributes ca. 100k new triples. Henceforth, we write $LUBMX$ to
    indicate a KG with $X$ universities, e.g. LUBM10 contains 1M triples; \pa{DBPedia}~\cite{dbpedia},
    YAGO2S~\cite{yago} and \pa{Wikidata}~\cite{wikidata}, three widely used KGs
    with encyclopedic knowledge;
    \pa{Uniprot}~\cite{uniprot}, a KG that contains biomedical knowledge, and
    \pa{BTC2012}~\cite{btc2012}, a collection of crawled interlinked KGs.

\item \textbf{Other graphs.} We considered the graphs: \pa{Google}, a Web graph from
    Google, \pa{Twitter}, which contains a social circle, and \pa{Astro}, a
    collaboration network in Physics.

\squishend
\sys was configured to use
the B+Tree for \nodemgr and table pruning was disabled, unless otherwise stated.

\subsection{Lookups}

\begin{tr}
    \begin{table}[t]
        \centering
        \scriptsize
        \begin{tabular}{ l | c | r | r }
            \bf Type Patt/ & \bf Example & \multicolumn{1}{c |}{\bf N.} &
            \multicolumn{1}{c}{\bf Avg. \#} \\
            \bf Ordering(s) & \bf Pattern & \multicolumn{1}{c|}{\bf Queries} &
            \multicolumn{1}{c}{\bf Results}\\
            \hline
            0/\pa{all} & X Y Z & 1 & 75,999,246 \\
            1/\pa{srd}-\pa{sdr} & X $*$ $*$ & 1 & 8,617,963 \\
            1/\pa{drs}-\pa{dsr} & $*$ $*$ X & 1 &  29,835,479\\
            1/\pa{rds}-\pa{rsd} & $*$ X $*$ & 1 & 99 \\
            2/\pa{srd}-\pa{sdr} & a X Y & 8,617,963 & 8 \\
            2/\pa{drs}-\pa{dsr} & X Y a & 29,835,479 & 2 \\
            2/\pa{rds}-\pa{rsd} & X a Y & 99 & 767,669 \\
            3/\pa{srd}-\pa{sdr} & a X $*$ / a $*$ X & 8,617,963 & 4/8 \\
            3/\pa{drs}-\pa{dsr} & $*$ X a / X $*$ a & 29,835,479 & 1/2 \\
            3/\pa{rds}-\pa{rsd} & $*$ a X / X a $*$ & 99  & 369,011/423,335 \\
            4/\pa{srd}-\pa{rsd} & a b X& 41,910,232 &  1\\
            4/\pa{drs}-\pa{rds} & a X b & 36,532,121 & 2 \\
            4/\pa{sdr}-\pa{dsr} & X b a & 69,564,969 & 1 \\
        \end{tabular}

        \caption{Type of patterns (1,2,3,4) on several orderings on YAGO2S. $X,Y,Z$
        are variables, $a,b$ are constants, $*$ means the column is ignored.}

        \label{tab:scans1}
    \end{table}
\end{tr}

During the loading procedure, \sys applies Algorithm~\ref{alg:decider} to
determine the best layout for each table. Figure~\ref{fig:ntables} shows the
number of tables of each type for the KGs. The vast
majority of tables is stored either with the \pa{ROW} or \pa{CLUSTER} layout.
Only a few tables are stored with the \pa{COLUMN} layout. These are mostly the
ones in the \TR and \TRI byte streams.  It is interesting to note that the
number of tables varies differently among different KGs. For instance, the
number of row tables is twice the number of cluster tables with LUBM. In
contrast, with Wikidata there are more cluster tables than row ones. These
differences show to what extent \sys adapted its physical storage
to the structure of the KG.

One key operation of our system is to retrieve answers of simple triple patterns.
First, we generated all possible triple patterns that return non-empty answers from
\pa{YAGO2S}. We considered five types of patterns. Patterns of type 0 are full
scans; of type 2 contain one constants and two variables (e.g., $X,type,Y$),
while of type 4 contain two constants and one variable (e.g., $X,type,person$).
These patterns are answered with $\edg_*$. Patterns of types 1 request an
aggregated version of a full scan (e.g., retrieve all subjects) while patterns
of type 3 request an aggregation where the pattern contains one constant (e.g.,
return all objects of the predicate \emph{type}). These two patterns are
answered with $\grp_*$.

\begin{tr}
    The number, types of patterns, and average number of answers per type is
    reported in Table~\ref{tab:scans1}. The first column reports the type of
    pattern and the orderings we can apply when we retrieve it. The second
    column reports an example pattern of this type. The third column
    contains the number of different queries that we can construct of this type. The
    fourth column reports the average number of answers that we get if we
    execute queries of this type.

    For example, the first row describes the pattern of type 0, which is a full
    scan. For this type of pattern, we can retrieve the answers with all the
    orderings in $\cR$. There is only one possible query of this type (column 3)
    and if we execute it then we obtain about 76M answers (column 4). Patterns
    of type 1 correspond to full aggregated scans. An example pattern
    of this type is shown in the second row. If this query is
    executed, the system will return the list of all subjects with the count of
    triples that share each subject. With this input, this query will return
    about 8M results (i.e., the number of subjects). We can construct a similar
    query if we consider the variables in the second or third position. Details
    for these two cases are reported in the third and fourth rows.

    Patterns of type 2 have one constant and two variables. Like
    before the constant can appear in three positions. Note that in this case we
    can construct many more queries by using different constants. For instance,
    we can construct 8.6M queries if the constant appears as subject, and 99 if
    it appears as predicate. Similarly, Table~\ref{tab:scans1} reports such
    details also for queries of type 3 and 4. By testing our system on all these
    types of patterns, we are effectively evaluating the performance over all
    possible queries of these types which would return non-empty answers.

\end{tr}

We used the primitives to retrieve the answers for these patterns with various
configurations of our system, and compared against \pa{RDF3X}, which was
the system with fastest runtimes. The
median warm runtimes of all executions are reported in
Figure~\ref{tab:lookup_runtimes}.

The row ``Default'' reports the results with the adaptive storage selected by
Algorithm~\ref{alg:decider} but without table pruning. The rows ``With
\pa{OFR}'' and ``With \pa{AGGR}'' use Algorithm~\ref{alg:decider} and the two
techniques for table pruning discussed in Section~\ref{sec:table_pruning}
respectively. The rows ``Only \pa{ROW} (\pa{COLUMN})'' use only the \pa{ROW} and
\pa{COLUMN} layouts (the \pa{CLUSTER} is not competitive alone due to the lack
of binary search). From the table, we see that if the two pruning strategies are
enabled, then the runtimes increase, especially with \pa{OFR}. This was expected
since these two techniques trade speed for space. Their benefit is that they
reduce the size of the database, as shown in Figure~\ref{tab:sizedatabases}. In
particular, \pa{OFR} is very effective, and they can reduce the size by 35\%.
Therefore, they should only be used if space is critical. The \pa{ROW} layout
returns competitive performance if used alone but then the database size
is about 9\% larger due to the suboptimal compression.
Figure~\ref{tab:sizedatabases} also reports the size of the databases with the
other systems as reference. Note that the reported numbers for \sys do not
include the size of the dictionary (764MB). This size should be added to the
reported numbers for a fair comparison with the other systems' databases.

A comparison against \pa{RDF3X} shows that the latter is faster with full scans
(patterns of type 0) because our approach has to visit more tables stored with
different configurations. However, our approach has comparable performance with
the second pattern type and performs significantly better when the scan is
limited to a single table, with, in the best case, improvements of more than two
orders of magnitude (pattern 3). Note that in contexts like SPARQL query
answering, patterns that contain at least one constant are much more frequent
than full scans (e.g., see Table 2 of~\cite{yasgui}).

\begin{tr}
\begin{table*}[tb]
    \scriptsize
    \centering
    \begin{tabular}{ l  l  c | c  c  c  c  c | c  c  c  c  c}
                & \bf Q. & \bf N. Query & \multicolumn{5}{c |}{\bf Cold runtime
    (ms.)} &
    \multicolumn{5}{c}{\bf Warm runtime (ms.)} \\
                \cline{4-13}
                & & \bf Answers &
        \bf TN & \bf TR & \bf R3X & \bf TripleBit & \bf \virtuoso &
        \bf TN & \bf TR & \bf R3X & \bf TripleBit & \bf \virtuoso \\
        \hline
        \parbox[t]{2mm}{\multirow{5}{*}{\rotatebox[origin=c]{90}{\bf LUBM1B}}} &
        1 &  10  &
        136.165 & \bf 50.839 &  186.127 & 61.78  & 956.34  &
        \bf 0.107 &  0.164 & 1.331 &  11.76  &  0.39 \\
                                    & 2 &  10  &
        307.625 &  3,182 &  \bf 174.040 & \cellcolor{red} x  & 10,620  &
        \bf 0.181 &  1,067 & 3.528 & \cellcolor{red} x  &  0.33 \\
                                    & 3 &  0 &
        807,870 &  6,468 &  712,365 &  \bf 0.31  & 24,538  &
        806,194 &  2,387 & 701,738 &  \bf 0.04  &  9,365 \\
                & 4 &  2,528 &
        643,072 &  24,848 &  738,473 & \bf 20,308  & 49,816  &
        611,083 &  19,435 & 709,541 &  20,298  & \bf 9,244 \\
                & 5 &  351,919  &
        \bf 9,801 &  25,694 &  105,386 & \cellcolor{red} 15,352  &  171,850  &
        \bf 4,704 &  14,452 & 22,468 & \cellcolor{red} 12,212  &  150,407 \\
        \hline

        \parbox[t]{2mm}{\multirow{5}{*}{\rotatebox[origin=c]{90}{\bf Uniprot}}} &
        1 & 853 &
        469.759 &  643.739 &  2,969 &  \bf 141.76  & 8,532 &
        \bf 1.245&  8.676 & 15.058 &  2.347 & 17.145 \\
                                   & 2 & 32 &
        371.790 &  1,475 &  267.184 & \bf 64.00  & 2,767 &
        \bf 0.168 &  52.000 & 4.144 &  0.545 & 1.121 \\
                                    & 3 &  10,715,646  &
        20,557 &  9,660 &  132,966 & \bf 8,224  & 189,962 &
        17,729 &  7,318 &  64,923 & \bf 6,532 & 96,063 \\
               & 4 &  5,564  &
        1,355 &  3,469 &  1,474 & \bf 352.70  & 4,112  &
        \bf 17.025 & 438.864 &  50.898 &  211.71 & 58.498 \\
                                    & 5 & 0 &
        \bf 78.583 &  892.596 &  168.806 & \cellcolor{red} x & 770.707  &
        2.759 & 298.881 & 2.489 &  \cellcolor{red} x & \bf 1.226 \\

        \hline
        \parbox[t]{2mm}{\multirow{5}{*}{\rotatebox[origin=c]{90}{\bf DBPEDIA}}} &
        1 & 449  &
        \bf 537.754 &  885.549 &  687.848 &  \cellcolor{red} 46.93  &  2,706 &
        \bf 0.632 &  80.511 &  21.157 & \cellcolor{red} 3.14  &  25.86 \\
                                    & 2 &  600  &
        249.147 &  306.902 &  354.489 & \bf 127.17  &  1,675  &
        \bf 0.075 &  0.118 &  3.533 &  0.58  &  4.47 \\
                                    & 3 &  270  &
        \bf 472.855 &  562.649 &  756.241 & \cellcolor{red} 71.22  & 2,091  &
        \bf 0.436 &  7.761 &  9.356 & \cellcolor{red} 0.92  &  3.22 \\
                                    & 4 &  68  &
        \bf 869.719 &  1,445 &  1,128 & \cellcolor{red} 116.60  & 6,233 &
        \bf 0.371 &  124.391 &  9.600 & \cellcolor{red} 1.42  &  4.32 \\
                                    & 5 &  1,643  &
        \bf 433.677 &  811.409 &  4,705 & \cellcolor{red} 355.93 &  10,055 &
        \bf 5.158 &  24.207 &  31.842 & \cellcolor{red} 8.89  & 47.46 \\
        \hline
        \parbox[t]{2mm}{\multirow{5}{*}{\rotatebox[origin=c]{90}{\bf BTC2012}}} &
        1 &  0 &
        \bf 23.828 &  24.260 &  297.849 &  127.25  &  N.A.  &
        0.069 &  \bf 0.050 &  4.903 &  0.99 &  N.A. \\
                                    & 2 &  1  &
        355.780 &  607.617 &  185.877 & \bf 65.48  & N.A. &
        \bf 0.197 &  24.911 & 6.279 &  0.55  &  N.A. \\
                                    & 3 &  1  &
        415.230 &  1,515 &  506.125 & \bf 244.29  & N.A.  &
        \bf 0.257 &  116.572 & 18.896  &  1.80  &  N.A. \\
                                    & 4 &  664  &
        \bf 1,340 &  2,914 &  4,773 & \cellcolor{red} 1,000 &   N.A. &
        \bf 24.693 &  1,049 &  290.655 & \cellcolor{red} 75.98  &  N.A. \\
                                     & 5 &  5,996  &
        6,120 &  19,525 & \bf 4,446 & \cellcolor{red} 3,404  &  N.A.  &
        \bf 5.493 & 9,410 &  528.668 & \cellcolor{red} 18.95  & N.A. \\
        \hline
        \parbox[t]{2mm}{\multirow{5}{*}{\rotatebox[origin=c]{90}{\bf WIKIDATA}}} &
        1 & 43 &
        174.608 &  248.589 &  706.03 & \bf 29.72  & N.A. &
        \bf 0.183 &  0.386 &  2.278 &  0.505  &  N.A.\\
                                    & 2 & 55 &
        154.305 &  189.611 &  851.00 & \bf 32.94  &  N.A. &
        0.066 &  \bf 0.063 &  1.081 & 0.107  & N.A.\\
                                    & 3 & 1,583 &
        578.537  &  873.176  & \bf 327.25 &  522.03  &  N.A.  &
        \bf 4.095 & 140.315 & 52.731 & 57.054 & N.A.\\
                                    & 4 & 682 &
        355.267 &  742.303  &  5,461  & \bf 232.02  &  N.A. &
        24.110  &  27.627 &  179.36  & \bf 15.219  &  N.A.\\
                & 5 & 1,975,090 &
        2,013 &  \bf 1,995  &  75,810  & \cellcolor{red} 2,946  &  N.A.&
        817.614 & \bf 728.641 & 6,399  & \cellcolor{red} 1,533  &  N.A.\\
    \end{tabular}

    \caption{Average runtimes of SPARQL queries. Column ``TN'' reports the
        runtime of our approach with the native SPARQL implementation while
        ``TR'' is the runtime with the RDF3X SPARQL engine. LUBM8k is a
        generated database with about 1B RDF triples. Red background
        means that TripleBit returned wrong
        answers ('x' means it crashed); ``N.A'' means that the experiment was not possible due to
failure at loading time.}
    \label{tab:sparql}
\end{table*}
\end{tr}

\begin{paper}

\begin{figure*}[t]
    \centering
    \scriptsize
    \begin{subfigure}[T]{0.45\linewidth}
        \centering
        \begin{tabular}{ l  l  c | c  c  c  c  c}
                & \bf Q. & \bf N. Query &
        \multicolumn{5}{c}{\bf Avg warm runtime (ms.)} \\
                \cline{4-8}
                & & \bf Answers &
        \bf TN & \bf TR & \bf R3X & \bf TripleBit & \bf \virtuoso \\
        \hline
        \parbox[t]{2mm}{\multirow{5}{*}{\rotatebox[origin=c]{90}{\bf LUBM8k}}} &
        1 &  10  &
        \bf 0.107 &  0.164 & 1.331 &  11.76  &  0.39 \\
                                    & 2 &  10  &
        \bf 0.181 &  1,067 & 3.528 & \cellcolor{red} x  &  0.33 \\
                                    & 3 &  0 &
        806,194 &  2,387 & 701,738 &  \bf 0.04  &  9,365 \\
                & 4 &  2,528 &
        611,083 &  19,435 & 709,541 &  20,298  & \bf 9,244 \\
                & 5 &  351,919  &
        \bf 4,704 &  14,452 & 22,468 & \cellcolor{red} 12,212  &  150,407 \\

        \hline

        \parbox[t]{2mm}{\multirow{5}{*}{\rotatebox[origin=c]{90}{\bf Uniprot}}} &
        1 & 853 &
        \bf 1.245&  8.676 & 15.058 &  2.347 & 17.145 \\
                                   & 2 & 32 &
        \bf 0.168 &  52.000 & 4.144 &  0.545 & 1.121 \\
                                    & 3 &  10,715,646  &
        17,729 &  7,318 &  64,923 & \bf 6,532 & 96,063 \\
               & 4 &  5,564  &
        \bf 17.025 & 438.864 &  50.898 &  211.71 & 58.498 \\
                                    & 5 & 0 &
        2.759 & 298.881 & 2.489 &  \cellcolor{red} x & \bf 1.226 \\

        \hline
        \parbox[t]{2mm}{\multirow{5}{*}{\rotatebox[origin=c]{90}{\bf DBPEDIA}}} &
        1 & 449  &

        \bf 0.632 &  80.511 &  21.157 & \cellcolor{red} 3.14  &  25.86 \\
                                    & 2 &  600  &
        \bf 0.075 &  0.118 &  3.533 &  0.58  &  4.47 \\
                                    & 3 &  270  &
        \bf 0.436 &  7.761 &  9.356 & \cellcolor{red} 0.92  &  3.22 \\
                                    & 4 &  68  &
        \bf 0.371 &  124.391 &  9.600 & \cellcolor{red} 1.42  &  4.32 \\
                                    & 5 &  1,643  &
        \bf 5.158 &  24.207 &  31.842 & \cellcolor{red} 8.89  & 47.46 \\
        \hline
        \parbox[t]{2mm}{\multirow{5}{*}{\rotatebox[origin=c]{90}{\bf BTC2012}}} &
        1 &  0 &
        0.069 &  \bf 0.050 &  4.903 &  0.99 &  N.A. \\
                                    & 2 &  1  &
        \bf 0.197 &  24.911 & 6.279 &  0.55  &  N.A. \\
                                    & 3 &  1  &
        \bf 0.257 &  116.572 & 18.896  &  1.80  &  N.A. \\
                                    & 4 &  664  &
        \bf 24.693 &  1,049 &  290.655 & \cellcolor{red} 75.98  &  N.A. \\
                                     & 5 &  5,996  &
        \bf 5.493 & 9,410 &  528.668 & \cellcolor{red} 18.95  & N.A. \\
        \hline
        \parbox[t]{2mm}{\multirow{5}{*}{\rotatebox[origin=c]{90}{\bf WIKIDATA}}} &
        1 & 43 &
        \bf 0.183 &  0.386 &  2.278 &  0.505  &  N.A.\\
                                    & 2 & 55 &
        0.066 &  \bf 0.063 &  1.081 & 0.107  & N.A.\\
                                    & 3 & 1,583 &
        \bf 4.095 & 140.315 & 52.731 & 57.054 & N.A.\\
                                    & 4 & 682 &
        24.110  &  27.627 &  179.36  & \bf 15.219  &  N.A.\\
                & 5 & 1,975,090 &
        817.614 & \bf 728.641 & 6,399  & \cellcolor{red} 1,533  &  N.A.\\
        \end{tabular}
    \caption{Runtime SPARQL queries. ``TN'' is the
        runtime of our SPARQL engine while ``TR''
        uses our storage and the RDF3X SPARQL engine. Red background
        means that TripleBit returned wrong
        answers ('x' means it crashed);
        ``N.A.'' means that there was a failure at loading time}\label{tab:runtimesparql}
    \end{subfigure}\;\;%
    \begin{subfigure}[T]{0.55\linewidth}
        \begin{subfigure}[T]{\linewidth}
            \centering
            \begin{tabular}{ l  c  c  c  c  c  c}
 & \multicolumn{2}{c }{\bf ASTRO} & \multicolumn{2}{c }{\bf GOOGLE} & \multicolumn{2}{c}{\bf TWITTER} \\
\multicolumn{1}{l }{\bf Task} & \multicolumn{1}{c }{\bf Snap} &
\multicolumn{1}{c }{\bf Ours} & \multicolumn{1}{c }{\bf Snap} &
\multicolumn{1}{c }{\bf Ours} & \multicolumn{1}{c }{\bf Snap} &
\multicolumn{1}{c}{\bf Ours} \\
        \cline{1-7}
	HITS & 431 & \bf 89 & 9252 & \bf 3557 & 2399 & \bf 588 \\
	BFS & 81993 & \bf 62241 & \bf 1604037 & 1709823 & \bf 215704 & 243740 \\
	Triangles & 69 & \bf 31 & 1353 & \bf 526 & 607 & \bf 105 \\
	Random Walks & \bf 25 & 34 & \bf 30 & 41 & \bf 26 & 32 \\
	MaxWCC & 22 & \bf 15 & 594 & \bf 351 & 132 & \bf 65 \\
	MaxSCC & 47 & \bf 29 & 1177 & \bf 712 & 228 & \bf 148 \\
	Diameter & 11767 & \bf 5233 & \bf 168211 & 243669 & 56132 & \bf 42581 \\
	PageRank & 515 & \bf 319 & 14616 & \bf 7771 & 4482 & \bf 1738 \\
	ClustCoef & \bf 375 & 417 & 8519 & \bf 7114 & 5886 & \bf 4178 \\
	mod & 7 & \bf 6 & \bf 5 & 23 & \bf 8 & 13 \\
\end{tabular}

            \caption{Runtime of graph analytics algorithms (ms)}\label{tab:snap}
        \end{subfigure}
        \begin{subfigure}[T]{\linewidth}
                \begin{subfigure}[T]{0.6\linewidth}
                    \begin{subfigure}[T]{\linewidth}
                        \centering

\begin{tabular}{lcc}
    \multicolumn{3}{l}{\bf Datalog reasoning using LUBM1k (130M triples)} \\
    \bf Ruleset from~\cite{rdfox} & \bf VLog+Ours & \bf VLog \\
        LUBM-L & \bf 17.3s & 25.6s \\
        LUBM-LE & \bf 31m & 34m \\

        \multicolumn{3}{l}{\textbf{Runtime training 10 epochs with TransE and
        YAGO}} \\
        \multicolumn{3}{l}{\tiny Params:
        bathsize=100,learningrate=0.001,dims=50,adagrad,margin=1} \\
        \multicolumn{3}{l}{Ours: \textbf{8.6s}\;\;\;\;\;\;\;OpenKE~\cite{openke}: 18.72s} \\
\end{tabular}

                        \caption{Runtime of reasoning and learning}\label{tab:reasoning}
                    \end{subfigure}
                    \begin{subfigure}[T]{\linewidth}
                        \centering
                        \includegraphics[width=\linewidth,trim={1em 1em 1em 10em},clip]{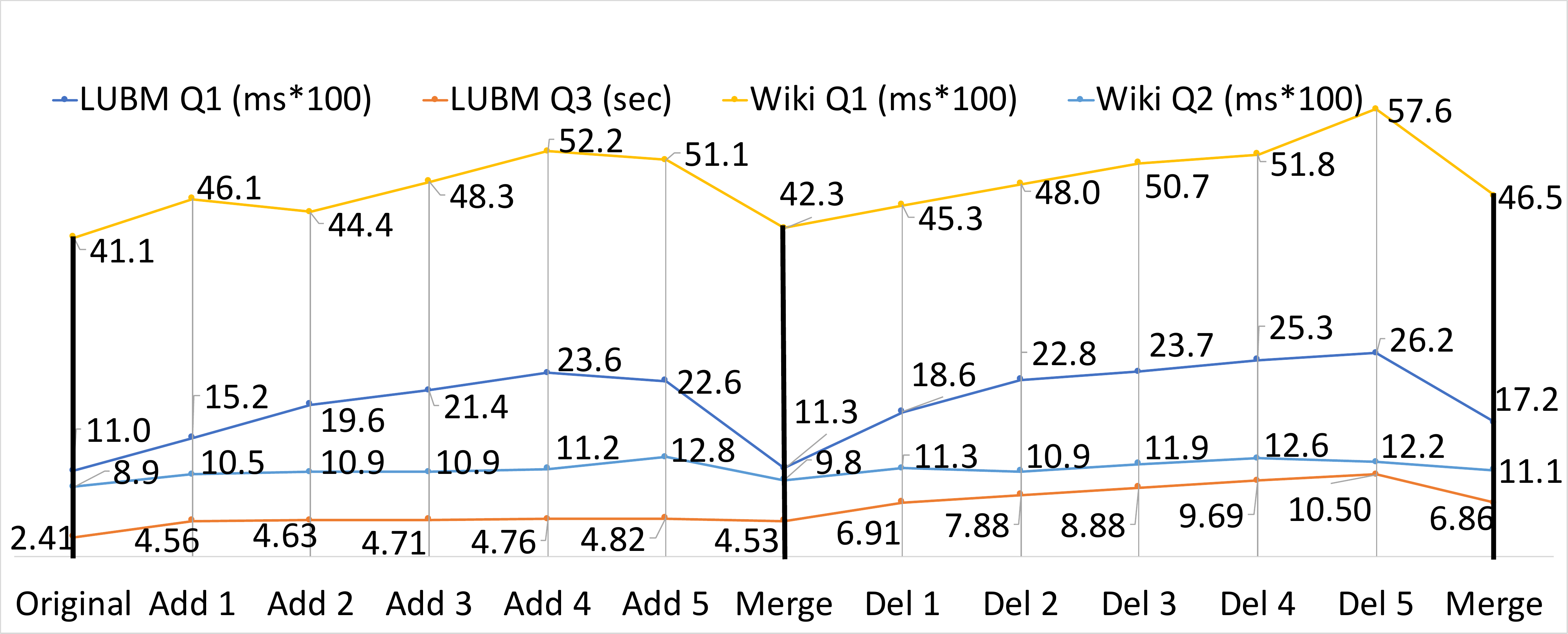}
                        \caption{Warm runtimes on Wikidata and LUBM8k after adding/removing 1M triples}\label{fig:queryupdates}
                    \end{subfigure}
                \end{subfigure}\;\;%
                \begin{subfigure}[T]{0.35\linewidth}
                    \centering


\begin{tabular}{p{4.7em}|p{1.3em}p{1.3em}p{1em}}
    \textbf{Universities} & \textbf{Q1)} & \textbf{Q2} &
    \textbf{Q3} \\
    \bf (\# facts) & \bf (ms) & \bf (ms) & \\
    10k (1.3B) & 0.05 & 0.09 & 25m \\
    20k (2.6B) & 0.05 & 0.09 & 52m \\
    40k (5B) & 0.05 & 0.09 & 1h50m \\
    80k (10B) & 0.05 & 0.09 & 3h52m \\
    160k (21B)& 0.05 & 0.09 & >8h \\
    800k (100B) & 0.05 & 0.09 & >8h \\
                  & \textbf{Q4} & \textbf{Q5} & \\
    10k (1.3B) &  11m & 6s\\
    20k (2.6B) &  41m & 12s\\
    40k (5B) &  1h42s & 25s\\
    80k (10B) &  3h1m & 56s\\
    160k (21B)&  6h49m & 1m51s\\
    800k (100B) & >8h & 12m \\
\end{tabular}

                    \caption{Runtime LUBM Q1-Q5 and KGs which size ranges
                    between 1-100B triples}\label{tab:scalability}
                \end{subfigure}
        \end{subfigure}
    \end{subfigure}
    \caption{Performance of our storage engine on several tasks, with larger KGs and updates}
\end{figure*}
\end{paper}

\subsection{SPARQL}

\begin{paper}
Figure~\ref{tab:runtimesparql} shows the average of five warm runtimes with our system
and with other state-of-the-art engines. For \pa{LUBM8k}, \pa{DBPedia}, \pa{Uniprot}, and \pa{BTC2012}, we
considered some of the queries used to evaluate TripleBit~\cite{triplebit}. For
Wikidata, we designed example queries of various complexity looking at
published examples. All queries are available
at~\cite{tr}. Unfortunately, we could not load \pa{Wikidata} and \pa{BTC2012} with
\virtuoso due to errors during the loading phase.
\end{paper}

\begin{tr}
Table~\ref{tab:sparql} reports the average of five cold and warm runtimes with
our system and with other state-of-the-art engines. For \pa{LUBM}, \pa{DBPedia},
\pa{Uniprot}, and \pa{BTC2012}, we considered queries used to evaluate previous
systems~\cite{triplebit}. For Wikidata, we designed five example queries of
various complexity looking at published examples. The queries are reported in
Appendix~\ref{sec:sparql_queries}. Unfortunately, we could not load
\pa{Wikidata} and \pa{BTC2012} with \virtuoso due to raised exceptions during
the loading phase.

We can make a few observations from the obtained results.
First, a direct comparison against \pa{TripleBit} is problematic
because sometimes \pa{TripleBit} crashed or returned wrong results (checked
after manual inspection). Looking at the other systems, we observe that
our approach returned the best cold runtimes for 20 out of 25 queries, counting
in both the executions with our native SPARQL engine and the integration with
RDF3X. If we compare the warm runtimes, our system is faster 20 out of 25
times. Furthermore, we observe that \sys/N is faster than \sys/R mostly with
selective queries that require only a few joins. Otherwise the second is faster.
The reason is that \pa{RDF3X} uses a sophisticated query optimizer that builds
multiple plans in a bottom-up fashion. This procedure is costly if applied to
simple queries, but it pays off for more complex ones because it can detect a
better execution plan.
\end{tr}

\begin{paper}
We can make a few observations from the obtained results.
    First, a direct comparison against \pa{TripleBit}
    is problematic because sometimes \pa{TripleBit} crashed or returned wrong
    results (checked after manual inspection). Looking at the other systems,
    we observe that our system is faster 20 out of 25 times. Furthermore,
    we observe that TN (our native SPARQL engine) is faster than TR
    (our storage+RDF3X) mostly with selective queries that require fewer joins
    (e.g., LUBM Q1,Q2, DBPedia Q1, Q2, Q3, Wikidata Q1, Q3). The reason
    is that \pa{RDF3X} uses a sophisticated query optimizer that considers more
    plans in a bottom-up fashion. This query optimization procedure is costly
    and pays off only with more complex queries.
\end{paper}

\subsection{Graph Analytics, Reasoning, Learning}

\begin{tr}
\begin{table}[t]
    \centering
    \scriptsize
    \begin{tabular}{ l  c  c  c  c  c  c}
 & \multicolumn{2}{c }{\bf ASTRO} & \multicolumn{2}{c }{\bf GOOGLE} & \multicolumn{2}{c}{\bf TWITTER} \\
\multicolumn{1}{l }{\bf Task} & \multicolumn{1}{c }{\bf Snap} &
\multicolumn{1}{c }{\bf Ours} & \multicolumn{1}{c }{\bf Snap} &
\multicolumn{1}{c }{\bf Ours} & \multicolumn{1}{c }{\bf Snap} &
\multicolumn{1}{c}{\bf Ours} \\
        \cline{1-7}
	HITS & 431 & \bf 89 & 9252 & \bf 3557 & 2399 & \bf 588 \\
	BFS & 81993 & \bf 62241 & \bf 1604037 & 1709823 & \bf 215704 & 243740 \\
	Triangles & 69 & \bf 31 & 1353 & \bf 526 & 607 & \bf 105 \\
	Random Walks & \bf 25 & 34 & \bf 30 & 41 & \bf 26 & 32 \\
	MaxWCC & 22 & \bf 15 & 594 & \bf 351 & 132 & \bf 65 \\
	MaxSCC & 47 & \bf 29 & 1177 & \bf 712 & 228 & \bf 148 \\
	Diameter & 11767 & \bf 5233 & \bf 168211 & 243669 & 56132 & \bf 42581 \\
	PageRank & 515 & \bf 319 & 14616 & \bf 7771 & 4482 & \bf 1738 \\
	ClustCoef & \bf 375 & 417 & 8519 & \bf 7114 & 5886 & \bf 4178 \\
	mod & 7 & \bf 6 & \bf 5 & 23 & \bf 8 & 13 \\
\end{tabular}

    \caption{Runtime of various graph analytics algorithms}
    \label{tab:snap}
\end{table}
\end{tr}

\begin{tr}
\begin{table}[t]
    \centering
    \footnotesize
\begin{tabular}{lcc}
    \multicolumn{3}{l}{\bf Datalog reasoning using LUBM1k (130M triples)} \\
    \bf Ruleset from~\cite{rdfox} & \bf VLog+Ours & \bf VLog \\
        LUBM-L & \bf 17.3s & 25.6s \\
        LUBM-LE & \bf 31m & 34m \\

        \multicolumn{3}{l}{\textbf{Runtime training 10 epochs with TransE and
        YAGO}} \\
        \multicolumn{3}{l}{Params:
        bathsize=100,learningrate=0.001,dims=50,adagrad,margin=1} \\
        \multicolumn{3}{l}{Ours: \textbf{8.6s}\;\;\;\;\;\;\;OpenKE~\cite{openke}: 18.72s} \\
\end{tabular}

\caption{Runtime of reasoning and learning}\label{tab:reasoning}
\end{table}
\end{tr}

\fakeparagraph{Graph analytics} Algorithms for graph analytics are used for path
analysis (e.g., find the shortest paths), community analysis (e.g., triangle counting), or to compute centrality metrics (e.g., PageRank).
They use frequently the primitives $pos_*$ and $count$ to traverse the graph or
to obtain the nodes' degree. For these experiments, we used the sorted list as
\pa{NODEMGR} since these algorithms are node-centric.

We selected ten well-known algorithms: \emph{HITS} and \emph{PageRank} compute
centrality metrics; \emph{Breadth First Search (BFS)} performs a search;
\emph{MOD} computes the modularity of the network, which is used for community
detection; \emph{Triangle Counting} counts all triangles; \emph{Random Walks}
extracts random paths; \emph{MaxWCC} and \emph{MaxSCC} compute the largest weak
and strong connected components respectively; \emph{Diameter} computes the
diameter of the graph while \emph{ClustCoeff} computes the clustering
coefficient.

We executed these algorithms using the original SNAP library and in combination
with our engine. Note that the implementation of the algorithms is the same;
only the storage changes. \FigOrTab~\ref{tab:snap} reports the runtimes. From it, we
see that our engine is faster in most cases. It is only with random walks that
our approach is slower. From these results, we conclude that also with this type
of computation our approach leads to competitive runtimes.

\fakeparagraph{Reasoning and Learning} We also tested the performance of our
system for rule-based reasoning. In this task, rules are used to materialize all
possible derivations from the KG. First, we computed reasoning considering \sys
and VLog, using LUBM and two popular rulesets (LUBM-L and LUBM-LE)~\cite{rdfox,vlog}. Then, we repeated the process with the native storage of
VLog. The runtime, reported in \FigOrTab~\ref{tab:reasoning}, shows that our
engine leads to an improvement of the performance (48\% faster in the best case).

Finally, we considered statistical relational learning as another class of
problems that could benefit from our engine.
\begin{tr}
These techniques associate each entity and relation label in the KG to a
numerical vector (called embedding) and then learn optimal values for the
embeddings so that truth values of some unseen triples
can be computed via algebraic operations on the vectors.

\end{tr}
We implemented
TransE~\cite{transe}, one of the most popular techniques of this kind, on top of
\sys and compared the runtime of training vs. the one produced by
OpenKE~\cite{openke}, a state-of-the-art library.  \FigOrTab~\ref{tab:reasoning}
reports the runtime to train a model using as input a subset of YAGO which was
used in other works~\cite{subgraphs}. The results indicate competitive
performance also in this case.

\subsection{Scalability, updates, and bulk loading}
\label{sec:loading}

\begin{tr}
    \begin{table}[t]
        \centering
        \footnotesize
        
\begin{tabular}{ l | ccccc}
    \textbf{Universities} & \textbf{Q1)} & \textbf{Q2} &
    \textbf{Q3} & \textbf{Q4} & \textbf{Q5}\\
    \bf (\# facts) & \bf (ms) & \bf (ms) & & \\
    10k (1.3B) & 0.05 & 0.09 & 25m & 11m & 6s\\
    20k (2.6B) & 0.05 & 0.09 & 52m & 41m & 12s\\
    40k (5B) & 0.05 & 0.09 & 1h50m & 1h42s & 25s\\
    80k (10B) & 0.05 & 0.09 & 3h52m & 3h1m & 56s\\
    160k (21B)& 0.05 & 0.09 & >8h & 6h49m & 1m51s\\
    800k (100B) & 0.05 & 0.09 & >8h & >8h & 12m \\
\end{tabular}

        \caption{Runtime LUBM Q1-Q5 and KGs which size
                ranges between 1B and 100B triples}
        \label{tab:scalability}
    \end{table}
\end{tr}

\begin{paper}%
We executed the five LUBM queries using our native SPARQL procedure on KGs of
different sizes (between 1B-100B triples). We used another machine with 256GB of
RAM for these experiments (which also costs $<$\$5K) due to lack of disk space.
The warm runtimes are shown in \FigOrTab~\ref{tab:scalability}. The runtime of
the first two queries remains constant. This was expected since their
selectivity does not decrease as the size of the KG increases. In contrast, the
runtime of the other queries increases as the KG becomes larger.
\end{paper}

\begin{tr}
We executed the five LUBM queries using our native SPARQL procedure on KGs of
different sizes (between 1B-100B triples). We used another machine with 256GB of
RAM for these experiments (which also costs $<$\$5K) due to lack of disk space.
The warm runtimes are shown in \FigOrTab~\ref{tab:scalability}. The runtime of
the first two queries remains constant. This was expected since their
selectivity does not decrease as the size of the KG increases. In contrast, the
runtime of the other queries increases as the KG becomes larger.

\begin{figure}[t]
    \centering
\includegraphics[width=\linewidth,trim={1em
1em 1em 1em},clip]{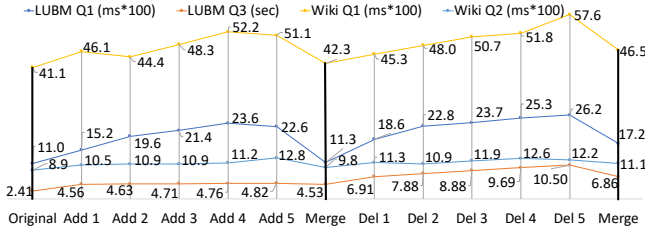}
\caption{Warm runtimes on Wikidata and LUBM8k after adding/removing updates with 1M
        triples each}
\label{fig:queryupdates}
\end{figure}
\end{tr}

\begin{figure}[t]
    \centering
    \scriptsize
    \begin{subfigure}[T]{0.4\linewidth}
        \centering
         \begin{tabular}{p{1em}cc}
                 \bf Op & \bf Wikidata & \bf LUBM8K \\
                 ADD  & 308s & 175s\\
                 ADD  & 386s & 230s\\
                 ADD  & 404s & 242s\\
                 ADD  & 477s & 261s\\
                 ADD  & 431s & 260s\\
                 Merge & 200s & 114s\\
                 DEL  & 399s & 222s\\
                 DEL  & 465s & 278s\\
                 DEL  & 501s & 319s\\
                 DEL  & 531s & 342s\\
                 DEL  & 566s & 369s\\
                 Merge & 291s & 181s\\
        \end{tabular}
         \caption{Runtime of updates}\label{tab:runtimeupdates}
 \end{subfigure}\;\;\;%
    \begin{subfigure}[T]{0.6\linewidth}
        \includegraphics[width=\linewidth]{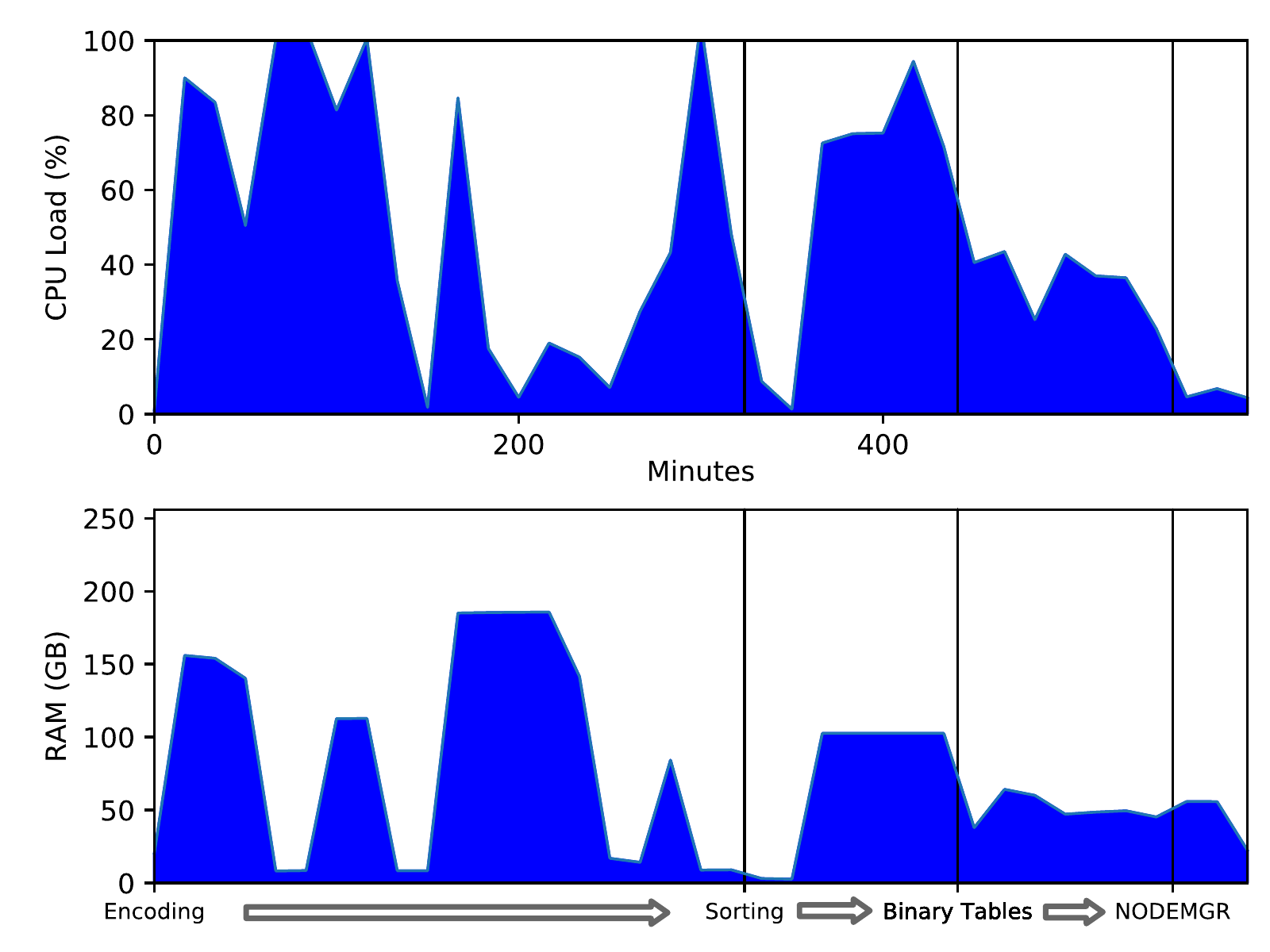}
        \caption{CPU/RAM usage loading LUBM80k}\label{fig:cputrace}
    \end{subfigure}
    \begin{subfigure}[T]{\linewidth}
        \begin{tabular}{lc | lc | lc}
            \bf System & \bf Runtime & \bf System & \bf Runtime & \bf System & \bf Runtime\\
            Ours (seq) & 20min & RDF3X (seq) & 24min & TripleBit (par) & 9min\\
            Ours (par) & 6min  & \virtuoso (par) & 1h9min  & & \\
        \end{tabular}
        \caption{Loading runtime of LUBM1k (130M triples)}\label{tab:comparisonloading}
    \end{subfigure}
    \caption{Loading and update runtimes}
    \label{fig:scalability}
\end{figure}

Figure~\ref{fig:queryupdates} shows the runtime of four SPARQL queries after we
added five new sets of triples to the KG, merged them, removed five
other sets of triples, and merged again.
\begin{tr}
    Each set of added triples does not contain triples contained in
    previous updates. Similarly, each set of removed triples contains only triples in the original
    KG and not in previous updates.
\end{tr}
We selected the queries so that the
content of the updates is considered. We observe that the runtime increases
(because more deltas are considered) and that it drops after they are merged in
a single update. Figure~\ref{tab:runtimeupdates} reports the runtime to process
five additions of ca. 1M novel triples, one merge, five removals of ca. 1M
existing triples, and another merge. As we can see, with both datasets the
runtime is much smaller than re-creating the database from scratch (>1h). The
runtime with LUBM8k is faster than with Wikidata because the updates with the
latter KG contained 4X more new entities.

In Figure~\ref{fig:cputrace}, we show the trace of the resource consumption
during the loading of LUBM80k (10B triples). We plot the CPU (100\% means all
physical cores are used) and RAM usage. From it, we see that most of the
runtime is taken to dictionary encoding, sorting the edges, and to create the
binary tables.

In general, \sys has competitive loading times.
Figure~\ref{tab:comparisonloading} shows the loading time of ours and other
systems on LUBM1k. With larger KGs, RDF3X becomes significantly slower than ours
(e.g., it takes ca. 7 hours to load LUBM8k on our smaller machine while \sys
needs 1 hour and 18 minutes) due to lack of parallelism. TripleBit is an
in-memory database and thus it cannot scale to some of our largest inputs. In
some of our largest experiments, \sys could load LUBM400k (50B triples) in about
48 hours which is a size that other systems cannot handle. If the graph is
already encoded, then loading is faster. We loaded the Hyperlink
Graph~\cite{hypergraph}, a graph with 128B edges, in about 13 hours (with the
larger machine) and the database required 1.4TB of space.

\section{Related Work}
\label{sec:related}

In this section, we describe the most relevant works to our problem. For a
broader introduction to graph and RDF processing, we redirect to existing
surveys~\cite{distsurvey,survey1,survey2,survey3,survey4,survey5,survey6}.
Current approaches can be classified either as \emph{native} (i.e., designed for
this task) or \emph{non-native} (adapt pre-existing technology). Native engines
have better performance~\cite{ibm}, but less
functionalities~\cite{ibm,fan_case_2015}. Our approach belongs to the first
category.

Research on native systems has focused on advanced indexing structures. The most
popular approach is to extensively materialize a dedicated index for each
permutation. This was initially proposed by YARS~\cite{yars2}, and further
explored in RDF3X~\cite{hprd,redland,triplet,owlim,rdf3x}. Also
Hexastore~\cite{hexastore} proposes a six-way permutation-based indexing, but
implemented it using hierarchical in-memory Java hash maps. Instead, we use
on-disk data structures and therefore can scale to larger inputs. Recently,
other types of indices, based on 2D or 3D bit matrices~\cite{bitmat,triplebit},
hash-maps~\cite{rdfox}, or data structures used for graph matching
approaches~\cite{gstore,taming} have been proposed. If compared with these
works, our approach uses a novel layout of data structures and uses multiple
layouts to store the subgraphs.

Non-native approaches offload the indexing to external engines (mostly DBMS).
Here, the challenge is to find efficient partitioning/replication criteria to
exploit the multi-table nature of relational engines. Existing partitioning
criteria group the triples either by
predicates~\cite{4store,swstore,jena,rstar,charsets,peterwww,pham_exploiting_2016},
clusters of predicates~\cite{oracle2005}, or by using other entity-based
splitting criteria~\cite{ibm}. The various partitioning schemes are designed to
create few tables to meet the constraints of relational
engines~\cite{swan}. Our approach differs because we group the edges at a much
higher granularity generating a number of binary tables that is too large for
such engines.

Some popular commercial systems for graph processing are Virtuoso~\cite{virtuoso},
BlazeGraph~\cite{blazegraph}, Titan~\cite{titan}, Neo4J~\cite{neo4j}
Sparksee~\cite{sparksee}, and InfiniteGraph~\cite{infinitegraph}. We compared
\sys against such a leading commercial system and observed that ours
has very competitive performance; other comparisons are presented
in~\cite{swan,schemaless}. In general, a direct comparison is challenging
because these systems provide end-to-end solutions tailored for specific tasks
while we offer general-purpose low-level APIs.

Finally, many works have focused on distributed graph
processing~\cite{4store,triad,distrspark,otzu1,s2rdf,
accelerating,scaling,distributed,sparql_flink,vertexcentric}. We do not view
these approaches as competitors since they operate on different hardware
architectures. Instead, we view ours as a potential complement that can be
employed by them to speed up distributed processing.

\begin{tr}
In our approach, we use numerical IDs to store the terms. This form of
compression has been the subject of some studies. First, some systems use the
Hash-code of the strings as
IDs~\cite{harris_3store:_2003,harris_sparql_2005,4store}. Most systems, however,
use counters to assign new IDs~\cite{sesame,harth1,yars2,rdf3x,compression1}. It
has been shown in~\cite{kognac} that assigning some IDs rather than others can
improve the query answering due to data locality. It is straightforward to
include these procedures in our system. Finally, some approaches focused on
compressing RDF collections~\cite{martinez-prieto_querying_2012} and on the
management of the strings~\cite{compact,uriencoding,singh_efficient_2018}. We
adopted a conventional approach to store such strings.  Replacing our dictionary
with these proposals is an interesting direction for future work.
\end{tr}

\section{Conclusion}

We proposed a novel centralized architecture for the low-level storage of very
large KGs which provides both node- and edge-centric access to the KG. One of the
main novelties of our approach is that it exhaustively decomposes the storage of
the KGs in many binary tables, serializing them in multiple byte streams to
facilitate inter-table scanning, akin to permutation-based approaches. Another
main novelty is that the storage effectively adapts to the KG by choosing a
different layout for each table depending on the graph topology. Our empirical
evaluation in multiple scenarios shows that our approach offers competitive
performance and that it can load very large graphs without expensive hardware.

Future work is necessary to apply or adapt our architecture for additional
scenarios. In particular, we believe that our system can be used to support
Triple Pattern Fragments~\cite{fragment}, an emerging paradigm to query RDF
datasets, and GraphQL~\cite{graphql}, a more complex graph query language.
Finally, it is also interesting to study whether the integration of additional
compression techniques, like locality-based dictionary encoding~\cite{kognac} or
HDT~\cite{hdt}, can further improve the runtime and/or reduce the storage space.

\fakeparagraph{Acknowledgments} We would like to thank (in alphabetical order)
Peter Boncz, Martin Kersten, Stefan Manegold, and Gerhard Weikum for discussing
and providing comments to improve this work. This project was partly funded by
the NWO research programme 400.17.605 (VWData) and NWO VENI project 639.021.335.

\begin{tr}
\appendix
\section{SPARQL queries}
\label{sec:sparql_queries}

\subsection{LUBM}

\footnotesize
\begin{verbatim}
PREFIX rdf: <http://www.w3.org/1999/02/
             22-rdf-syntax-ns#>
PREFIX ub:  <http://www.lehigh.edu/~zhp2/2004/
             0401/univ-bench.owl#>
Q1:
SELECT ?x WHERE {
?x ub:subOrganizationOf
   <http://www.Department0.University0.edu> .
?x rdf:type ub:ResearchGroup . }
Q2:
SELECT ?x WHERE {
?x ub:worksFor <http://www.Department0.University0.edu> .
?x rdf:type ub:FullProfessor . ?x ub:name ?y1 .
?x ub:emailAddress ?y2 . ?x ub:telephone ?y3 . }
Q3:
SELECT ?x ?y ?z WHERE {
?y rdf:type ub:University . ?z ub:subOrganizationOf ?y .
?z rdf:type ub:Department . ?x ub:memberOf ?z .
?x ub:undergraduateDegreeFrom ?y .
?x rdf:type ub:UndergraduateStudent. }
Q4:
SELECT ?x ?y ?z WHERE {
?y rdf:type ub:University . ?z ub:subOrganizationOf ?y .
?z rdf:type ub:Department . ?x ub:memberOf ?z .
?x rdf:type ub:GraduateStudent .
?x ub:undergraduateDegreeFrom ?y . }
Q5:
SELECT ?x ?y ?z WHERE {
?y rdf:type ub:FullProfessor . ?y ub:teacherOf ?z .
?z rdf:type ub:Course . ?x ub:advisor ?y .
?x ub:takesCourse ?z . }
\end{verbatim}

\subsection{DBPedia}

\footnotesize
\begin{verbatim}
PREFIX foaf: <http://xmlns.com/foaf/0.1/>
PREFIX dbo: <http://dbpedia.org/ontology/>
PREFIX db: <http://dbpedia.org/resource/>
PREFIX purl: <http://purl.org/dc/terms/>
PREFIX rdfs: <http://www.w3.org/2000/01/rdf-schema#>
Q1:
SELECT ?manufacturer ?name ?car
WHERE {
?car purl:subject db:Category:Luxury_vehicles .
?car foaf:name ?name .
?car dbo:manufacturer ?man .
?man foaf:name ?manufacturer . }
Q2:
SELECT ?film WHERE {
?film purl:subject db:Category:French_films . }
Q3:
SELECT ?title WHERE {
?game purl:subject db:Category:First-person_shooters .
?game foaf:name ?title . }
Q4:
SELECT ?name ?birth ?description ?person WHERE {
?person dbo:birthPlace db:Berlin .
?person purl:subject db:Category:German_musicians .
?person dbo:birthDate ?birth .
?person foaf:name ?name .
?person rdfs:comment ?description . }
Q5:
SELECT ?name ?birth ?death ?person WHERE {
?person dbo:birthPlace db:Berlin .
?person dbo:birthDate ?birth .
?person foaf:name ?name .
?person dbo:deathDate ?death .}
\end{verbatim}

\subsection{BTC2012}

\footnotesize
\begin{verbatim}
PREFIX geo: <http://www.geonames.org/ontology#>
PREFIX rdf: <http://www.w3.org/1999/02/
             22-rdf-syntax-ns#>
PREFIX dbpedia: <http://dbpedia.org/property/>
PREFIX dbpediares: <http://dbpedia.org/resource/>
PREFIX pos: <http://www.w3.org/2003/01/geo/wgs84_pos#>
PREFIX owl: <http://www.w3.org/2002/07/owl#>
Q1:
SELECT ?lat ?long WHERE {
?a ?x "Bro-C'hall" .
?a geo:inCountry <http://www.geonames.org/countries/
                  #FR> .
?a pos:lat ?lat . ?a pos:long ?long . }
Q2:
SELECT ?t ?lat ?long WHERE {
?a dbpedia:region
   dbres:List_of_World_Heritage_Sites_in_Europe .
?a dbpedia:title ?t . ?a pos:lat ?lat .
?a pos:long ?long .
?a dbpedia:link <http://whc.unesco.org/en/list/728> . }
Q3:
SELECT ?d WHERE {
?a dbpedia:senators ?c . ?a dbpedia:name ?d .
?c dbpedia:profession dbpediares:Politician .
?a owl:sameAs ?b .
?b geo:inCountry <http://www.geonames.org/countries/
                  #US> .}
Q4:
SELECT ?a ?b ?lat ?long WHERE {
?a dbpedia:spouse ?b .
?a rdf:type <http://dbpedia.org/ontology/Person> .
?b rdf:type <http://dbpedia.org/ontology/Person> .
?a dbpedia:placeOfBirth ?c . ?b dbpedia:placeOfBirth ?c .
?c owl:sameAs ?c2 . ?c2 pos:lat ?lat .
?c2 pos:long ?long . }
Q5:
SELECT ?a ?y WHERE {
?a rdf:type
   <http://dbpedia.org/class/yago/Politician110451263> .
?a dbpedia:years ?y.
?a <http://xmlns.com/foaf/0.1/name> ?n.
?b ?bn ?n.
?b rdf:type <http://dbpedia.org/ontology/OfficeHolder> . }
\end{verbatim}
\subsection{Uniprot}

\footnotesize
\begin{verbatim}
PREFIX r: <http://www.w3.org/1999/02/22-rdf-syntax-ns#>
PREFIX rs: <http://www.w3.org/2000/01/rdf-schema#>
PREFIX u: <http://purl.uniprot.org/core/>
Q1:
SELECT ?a ?vo WHERE {
?a u:encodedBy ?vo .
?a rs:seeAlso <http://purl.uniprot.org/eggnog/COG0787> .
?a rs:seeAlso <http://purl.uniprot.org/pfam/PF00842>.
?a rs:seeAlso <http://purl.uniprot.org/prints/PR00992>. }
Q2:
SELECT ?a ?vo WHERE {
?a u:annotation ?vo .
?a rs:seeAlso <http://purl.uniprot.org/interpro/IPR000842> .
?a rs:seeAlso <http://purl.uniprot.org/geneid/945772> .
?a u:citation <http://purl.uniprot.org/citations/9298646> . }
Q3:
SELECT ?p ?a WHERE {
?p u:annotation ?a .
?p r:type u:Protein .
?a r:type u:Transmembrane_Annotation . }
Q4:
SELECT ?p ?a WHERE {
?p u:annotation ?a .
?p r:type u:Protein .
?p u:organism <http://purl.uniprot.org/taxonomy/9606> .
?a r:type core/Disease_Annotation> . }
Q5:
SELECT ?a ?b ?ab WHERE {
?b u:modified "2008-07-22"^^
              <http://www.w3.org/2001/XMLSchema#date> .
?b r:type u:Protein . ?a u:replaces ?ab .
?ab u:replacedBy ?b . }
\end{verbatim}

\subsection{Wikidata}

\footnotesize
\begin{verbatim}
PREFIX wd: <http://www.wikidata.org/entity/>
PREFIX wdt: <http://www.wikidata.org/prop/direct/>
Q1:
SELECT ?h ?cause WHERE {
?h wdt:P39 wd:Q11696 .
?h wdt:P509 ?cause . }
Q2:
SELECT ?cat WHERE {
?cat  wdt:P31 wd:Q146 . }
Q3:
select ?gender WHERE {
?human wdt:P31 wd:Q5 .
?human wdt:P21 ?gender .
?human wdt:P106 wd:Q901 . }
Q4:
SELECT ?u ?state WHERE {
?u wdt:P31 wd:Q3918 .
?u wdt:P131 ?state .
?state wdt:P31 wd:Q35657 . }
Q5:
SELECT ?u ?x  WHERE {
?u wdt:P31 ?x .
?u wdt:P569 ?date }
\end{verbatim}

\end{tr}

\bibliographystyle{ACM-Reference-Format}
\bibliography{references}


\begin{thebibliography}{99}


\ifx \showCODEN    \undefined \def \showCODEN     #1{\unskip}     \fi
\ifx \showDOI      \undefined \def \showDOI       #1{#1}\fi
\ifx \showISBNx    \undefined \def \showISBNx     #1{\unskip}     \fi
\ifx \showISBNxiii \undefined \def \showISBNxiii  #1{\unskip}     \fi
\ifx \showISSN     \undefined \def \showISSN      #1{\unskip}     \fi
\ifx \showLCCN     \undefined \def \showLCCN      #1{\unskip}     \fi
\ifx \shownote     \undefined \def \shownote      #1{#1}          \fi
\ifx \showarticletitle \undefined \def \showarticletitle #1{#1}   \fi
\ifx \showURL      \undefined \def \showURL       {\relax}        \fi
\providecommand\bibfield[2]{#2}
\providecommand\bibinfo[2]{#2}
\providecommand\natexlab[1]{#1}
\providecommand\showeprint[2][]{arXiv:#2}

\bibitem[\protect\citeauthoryear{Abadi, Madden, and Ferreira}{Abadi
  et~al\mbox{.}}{2006}]%
        {rle}
\bibfield{author}{\bibinfo{person}{Daniel Abadi}, \bibinfo{person}{Samuel
  Madden}, {and} \bibinfo{person}{Miguel Ferreira}.}
  \bibinfo{year}{2006}\natexlab{}.
\newblock \showarticletitle{Integrating Compression and Execution in
  Column-Oriented Database Systems}. In \bibinfo{booktitle}{\emph{Proceedings
  of the 2006 ACM SIGMOD International Conference on Management of Data}}
  (Chicago, IL, USA) \emph{(\bibinfo{series}{SIGMOD ’06})}.
  \bibinfo{publisher}{Association for Computing Machinery},
  \bibinfo{address}{New York, NY, USA}, \bibinfo{pages}{671–682}.
\newblock
\showISBNx{1595934340}
\urldef\tempurl%
\url{https://doi.org/10.1145/1142473.1142548}
\showDOI{\tempurl}


\bibitem[\protect\citeauthoryear{Abadi, Marcus, Madden, and Hollenbach}{Abadi
  et~al\mbox{.}}{2009}]%
        {swstore}
\bibfield{author}{\bibinfo{person}{Daniel~J. Abadi}, \bibinfo{person}{Adam
  Marcus}, \bibinfo{person}{Samuel~R. Madden}, {and} \bibinfo{person}{Kate
  Hollenbach}.} \bibinfo{year}{2009}\natexlab{}.
\newblock \showarticletitle{{SW}-{Store}: a vertically partitioned {DBMS} for
  {Semantic} {Web} data management}.
\newblock \bibinfo{journal}{\emph{The VLDB Journal}} \bibinfo{volume}{18},
  \bibinfo{number}{2} (\bibinfo{year}{2009}), \bibinfo{pages}{385--406}.
\newblock


\bibitem[\protect\citeauthoryear{Abdelaziz, Harbi, Khayyat, and
  Kalnis}{Abdelaziz et~al\mbox{.}}{2017a}]%
        {survey5}
\bibfield{author}{\bibinfo{person}{Ibrahim Abdelaziz}, \bibinfo{person}{Razen
  Harbi}, \bibinfo{person}{Zuhair Khayyat}, {and} \bibinfo{person}{Panos
  Kalnis}.} \bibinfo{year}{2017}\natexlab{a}.
\newblock \showarticletitle{A Survey and Experimental Comparison of Distributed
  SPARQL Engines for Very Large RDF Data}.
\newblock \bibinfo{journal}{\emph{Proceedings of the VLDB Endowment}}
  \bibinfo{volume}{10}, \bibinfo{number}{13} (\bibinfo{date}{Sept.}
  \bibinfo{year}{2017}), \bibinfo{pages}{2049–2060}.
\newblock
\showISSN{2150-8097}
\urldef\tempurl%
\url{https://doi.org/10.14778/3151106.3151109}
\showDOI{\tempurl}


\bibitem[\protect\citeauthoryear{Abdelaziz, Harbi, Salihoglu, and
  Kalnis}{Abdelaziz et~al\mbox{.}}{2017b}]%
        {vertexcentric}
\bibfield{author}{\bibinfo{person}{I. Abdelaziz}, \bibinfo{person}{R. Harbi},
  \bibinfo{person}{S. Salihoglu}, {and} \bibinfo{person}{P. Kalnis}.}
  \bibinfo{year}{2017}\natexlab{b}.
\newblock \showarticletitle{Combining vertex-centric graph processing with
  {SPARQL} for large-scale {RDF} data analytics}.
\newblock \bibinfo{journal}{\emph{IEEE Transactions on Parallel and Distributed
  Systems}} \bibinfo{volume}{28}, \bibinfo{number}{12} (\bibinfo{date}{Dec.}
  \bibinfo{year}{2017}), \bibinfo{pages}{3374--3388}.
\newblock
\showISSN{1045-9219}


\bibitem[\protect\citeauthoryear{Alu{\c{c}}, {\"{O}}zsu, Daudjee, and
  Hartig}{Alu{\c{c}} et~al\mbox{.}}{2015}]%
        {schemaless}
\bibfield{author}{\bibinfo{person}{G{\"{u}}nes Alu{\c{c}}},
  \bibinfo{person}{M.~Tamer {\"{O}}zsu}, \bibinfo{person}{Khuzaima Daudjee},
  {and} \bibinfo{person}{Olaf Hartig}.} \bibinfo{year}{2015}\natexlab{}.
\newblock \showarticletitle{Executing queries over schemaless {RDF} databases}.
  In \bibinfo{booktitle}{\emph{31st {IEEE} International Conference on Data
  Engineering, {ICDE} 2015, Seoul, South Korea, April 13-17, 2015}}.
  \bibinfo{publisher}{{IEEE} Computer Society}, \bibinfo{address}{Seoul, South
  Korea}, \bibinfo{pages}{807--818}.
\newblock
\urldef\tempurl%
\url{https://doi.org/10.1109/ICDE.2015.7113335}
\showDOI{\tempurl}


\bibitem[\protect\citeauthoryear{Amann, Curé, and Naacke}{Amann
  et~al\mbox{.}}{2018}]%
        {distrspark}
\bibfield{author}{\bibinfo{person}{Bernd Amann}, \bibinfo{person}{Olivier
  Curé}, {and} \bibinfo{person}{Hubert Naacke}.}
  \bibinfo{year}{2018}\natexlab{}.
\newblock \bibinfo{booktitle}{\emph{{Distributed SPARQL Query Processing: a
  Case Study with Apache Spark}}}.
\newblock \bibinfo{publisher}{John Wiley \& Sons, Ltd}, Chapter~2,
  \bibinfo{pages}{21--55}.
\newblock
\showISBNx{9781119528227}
\urldef\tempurl%
\url{https://doi.org/10.1002/9781119528227.ch2}
\showDOI{\tempurl}


\bibitem[\protect\citeauthoryear{Antoniou, Batsakis, Mutharaju, Pan, Qi,
  Tachmazidis, Urbani, and Zhou}{Antoniou et~al\mbox{.}}{2018}]%
        {reasoningsurvey}
\bibfield{author}{\bibinfo{person}{Grigoris Antoniou}, \bibinfo{person}{Sotiris
  Batsakis}, \bibinfo{person}{Raghava Mutharaju}, \bibinfo{person}{Jeff~Z.
  Pan}, \bibinfo{person}{Guilin Qi}, \bibinfo{person}{Ilias Tachmazidis},
  \bibinfo{person}{Jacopo Urbani}, {and} \bibinfo{person}{Zhangquan Zhou}.}
  \bibinfo{year}{2018}\natexlab{}.
\newblock \showarticletitle{A survey of large-scale reasoning on the {Web} of
  data}.
\newblock \bibinfo{journal}{\emph{The Knowledge Engineering Review}}
  \bibinfo{volume}{33} (\bibinfo{year}{2018}), \bibinfo{pages}{1--43}.
\newblock
\showISSN{0269-8889, 1469-8005}


\bibitem[\protect\citeauthoryear{Atre, Chaoji, Zaki, and Hendler}{Atre
  et~al\mbox{.}}{2010}]%
        {bitmat}
\bibfield{author}{\bibinfo{person}{Medha Atre}, \bibinfo{person}{Vineet
  Chaoji}, \bibinfo{person}{Mohammed~J. Zaki}, {and} \bibinfo{person}{James~A.
  Hendler}.} \bibinfo{year}{2010}\natexlab{}.
\newblock \showarticletitle{{Matrix "Bit" Loaded: A Scalable Lightweight Join
  Query Processor for RDF Data}}. In \bibinfo{booktitle}{\emph{Proceedings of
  the 19th International Conference on World Wide Web}} (Raleigh, North
  Carolina, USA) \emph{(\bibinfo{series}{WWW ’10})}.
  \bibinfo{publisher}{Association for Computing Machinery},
  \bibinfo{address}{New York, NY, USA}, \bibinfo{pages}{41–50}.
\newblock
\showISBNx{9781605587998}
\urldef\tempurl%
\url{https://doi.org/10.1145/1772690.1772696}
\showDOI{\tempurl}


\bibitem[\protect\citeauthoryear{Azzam, Kirrane, and Polleres}{Azzam
  et~al\mbox{.}}{2018}]%
        {sparql_flink}
\bibfield{author}{\bibinfo{person}{A. Azzam}, \bibinfo{person}{S. Kirrane},
  {and} \bibinfo{person}{A. Polleres}.} \bibinfo{year}{2018}\natexlab{}.
\newblock \showarticletitle{{Towards Making Distributed RDF Processing
  FLINKer}}. In \bibinfo{booktitle}{\emph{2018 4th International Conference on
  Big Data Innovations and Applications (Innovate-Data)}}.
  \bibinfo{publisher}{IEEE Computer Society}, \bibinfo{address}{Los Alamitos,
  CA, USA}, \bibinfo{pages}{9--16}.
\newblock
\urldef\tempurl%
\url{https://doi.org/10.1109/Innovate-Data.2018.00009}
\showDOI{\tempurl}


\bibitem[\protect\citeauthoryear{Baolin and Bo}{Baolin and Bo}{2007}]%
        {hprd}
\bibfield{author}{\bibinfo{person}{Liu Baolin} {and} \bibinfo{person}{Hu Bo}.}
  \bibinfo{year}{2007}\natexlab{}.
\newblock \showarticletitle{{HPRD}: {A} {High} {Performance} {RDF} {Database}}.
  In \bibinfo{booktitle}{\emph{IFIP International Conference on Network and
  {Parallel} {Computing}}} \emph{(\bibinfo{series}{Lecture {Notes} in
  {Computer} {Science}})}, \bibfield{editor}{\bibinfo{person}{Keqiu Li},
  \bibinfo{person}{Chris Jesshope}, \bibinfo{person}{Hai Jin}, {and}
  \bibinfo{person}{Jean-Luc Gaudiot}} (Eds.). \bibinfo{publisher}{Springer
  Berlin Heidelberg}, \bibinfo{address}{Dalian, China},
  \bibinfo{pages}{364--374}.
\newblock
\showISBNx{978-3-540-74784-0}


\bibitem[\protect\citeauthoryear{Bazoobandi, de~Rooij, Urbani, ten Teije, van
  Harmelen, and Bal}{Bazoobandi et~al\mbox{.}}{2015}]%
        {compact}
\bibfield{author}{\bibinfo{person}{Hamid~R. Bazoobandi},
  \bibinfo{person}{Steven de Rooij}, \bibinfo{person}{Jacopo Urbani},
  \bibinfo{person}{Annette ten Teije}, \bibinfo{person}{Frank van Harmelen},
  {and} \bibinfo{person}{Henri Bal}.} \bibinfo{year}{2015}\natexlab{}.
\newblock \showarticletitle{A {Compact} {In}-memory {Dictionary} for {RDF}
  {Data}}. In \bibinfo{booktitle}{\emph{The Semantic Web. Latest Advances and
  New Domains}}. \bibinfo{publisher}{Springer-Verlag New York, Inc.},
  \bibinfo{address}{Portoroz, Slovenia}, \bibinfo{pages}{205--220}.
\newblock


\bibitem[\protect\citeauthoryear{Beckett}{Beckett}{2001}]%
        {redland}
\bibfield{author}{\bibinfo{person}{David Beckett}.}
  \bibinfo{year}{2001}\natexlab{}.
\newblock \showarticletitle{The Design and Implementation of the Redland RDF
  Application Framework}. In \bibinfo{booktitle}{\emph{Proceedings of the 10th
  International Conference on World Wide Web}} (Hong Kong)
  \emph{(\bibinfo{series}{WWW ’01})}. \bibinfo{publisher}{Association for
  Computing Machinery}, \bibinfo{address}{New York, NY, USA},
  \bibinfo{pages}{449–456}.
\newblock
\showISBNx{1581133480}
\urldef\tempurl%
\url{https://doi.org/10.1145/371920.372099}
\showDOI{\tempurl}


\bibitem[\protect\citeauthoryear{Bishop, Kiryakov, Ognyanoff, Peikov, Tashev,
  and Velkov}{Bishop et~al\mbox{.}}{2011}]%
        {owlim}
\bibfield{author}{\bibinfo{person}{Barry Bishop}, \bibinfo{person}{Atanas
  Kiryakov}, \bibinfo{person}{Damyan Ognyanoff}, \bibinfo{person}{Ivan Peikov},
  \bibinfo{person}{Zdravko Tashev}, {and} \bibinfo{person}{Ruslan Velkov}.}
  \bibinfo{year}{2011}\natexlab{}.
\newblock \showarticletitle{{OWLIM}: {A} family of scalable semantic
  repositories}.
\newblock \bibinfo{journal}{\emph{Semantic Web}} \bibinfo{volume}{2},
  \bibinfo{number}{1} (\bibinfo{year}{2011}), \bibinfo{pages}{33--42}.
\newblock


\bibitem[\protect\citeauthoryear{Bizer, Lehmann, Kobilarov, Auer, Becker,
  Cyganiak, and Hellmann}{Bizer et~al\mbox{.}}{2009}]%
        {dbpedia}
\bibfield{author}{\bibinfo{person}{Christian Bizer}, \bibinfo{person}{Jens
  Lehmann}, \bibinfo{person}{Georgi Kobilarov}, \bibinfo{person}{Sören Auer},
  \bibinfo{person}{Christian Becker}, \bibinfo{person}{Richard Cyganiak}, {and}
  \bibinfo{person}{Sebastian Hellmann}.} \bibinfo{year}{2009}\natexlab{}.
\newblock \showarticletitle{{DBpedia}-{A} crystallization point for the {Web}
  of {Data}}.
\newblock \bibinfo{journal}{\emph{Web Semantics: science, services and agents
  on the world wide web}} \bibinfo{volume}{7}, \bibinfo{number}{3}
  (\bibinfo{year}{2009}), \bibinfo{pages}{154--165}.
\newblock


\bibitem[\protect\citeauthoryear{Blanco, Cambazoglu, Mika, and Torzec}{Blanco
  et~al\mbox{.}}{2013}]%
        {websearch2}
\bibfield{author}{\bibinfo{person}{Roi Blanco}, \bibinfo{person}{Berkant~Barla
  Cambazoglu}, \bibinfo{person}{Peter Mika}, {and} \bibinfo{person}{Nicolas
  Torzec}.} \bibinfo{year}{2013}\natexlab{}.
\newblock \showarticletitle{Entity Recommendations in Web Search}. In
  \bibinfo{booktitle}{\emph{The Semantic Web -- ISWC 2013}}.
  \bibinfo{publisher}{Springer Berlin Heidelberg}, \bibinfo{address}{Berlin,
  Heidelberg}, \bibinfo{pages}{33--48}.
\newblock


\bibitem[\protect\citeauthoryear{Bordes, Usunier, Garcia-Duran, Weston, and
  Yakhnenko}{Bordes et~al\mbox{.}}{2013}]%
        {transe}
\bibfield{author}{\bibinfo{person}{Antoine Bordes}, \bibinfo{person}{Nicolas
  Usunier}, \bibinfo{person}{Alberto Garcia-Duran}, \bibinfo{person}{Jason
  Weston}, {and} \bibinfo{person}{Oksana Yakhnenko}.}
  \bibinfo{year}{2013}\natexlab{}.
\newblock \showarticletitle{Translating {Embeddings} for {Modeling}
  {Multi}-relational {Data}}.
\newblock In \bibinfo{booktitle}{\emph{Advances in {Neural} {Information}
  {Processing} {Systems} 26}}, \bibfield{editor}{\bibinfo{person}{C.~J.~C.
  Burges}, \bibinfo{person}{L.~Bottou}, \bibinfo{person}{M.~Welling},
  \bibinfo{person}{Z.~Ghahramani}, {and} \bibinfo{person}{K.~Q. Weinberger}}
  (Eds.). \bibinfo{publisher}{Curran Associates, Inc.}, \bibinfo{address}{Lake
  Tahoe, Nevada, USA}, \bibinfo{pages}{2787--2795}.
\newblock


\bibitem[\protect\citeauthoryear{Bornea, Dolby, Kementsietsidis, Srinivas,
  Dantressangle, Udrea, and Bhattacharjee}{Bornea et~al\mbox{.}}{2013}]%
        {ibm}
\bibfield{author}{\bibinfo{person}{Mihaela~A. Bornea}, \bibinfo{person}{Julian
  Dolby}, \bibinfo{person}{Anastasios Kementsietsidis},
  \bibinfo{person}{Kavitha Srinivas}, \bibinfo{person}{Patrick Dantressangle},
  \bibinfo{person}{Octavian Udrea}, {and} \bibinfo{person}{Bishwaranjan
  Bhattacharjee}.} \bibinfo{year}{2013}\natexlab{}.
\newblock \showarticletitle{Building an {E}fficient {RDF} {S}tore over a
  {R}elational {D}atabase}. In \bibinfo{booktitle}{\emph{SIGMOD '13:
  Proceedings of the 2013 ACM SIGMOD International Conference on Management of
  Data}}. \bibinfo{publisher}{ACM}, \bibinfo{address}{New York, NY, USA},
  \bibinfo{pages}{121--132}.
\newblock


\bibitem[\protect\citeauthoryear{Broekstra, Kampman, and
  Van~Harmelen}{Broekstra et~al\mbox{.}}{2002}]%
        {sesame}
\bibfield{author}{\bibinfo{person}{Jeen Broekstra}, \bibinfo{person}{Arjohn
  Kampman}, {and} \bibinfo{person}{Frank Van~Harmelen}.}
  \bibinfo{year}{2002}\natexlab{}.
\newblock \showarticletitle{Sesame: {A} {G}eneric {A}rchitecture for {S}toring
  and {Q}uerying {RDF} and {RDF} {S}chema}. In \bibinfo{booktitle}{\emph{1st
  International Semantic Web Conference}}. \bibinfo{publisher}{Springer},
  \bibinfo{address}{Sardinia, Italia}, \bibinfo{pages}{54--68}.
\newblock


\bibitem[\protect\citeauthoryear{Callahan, Cruz-Toledo, Ansell, and
  Dumontier}{Callahan et~al\mbox{.}}{2013}]%
        {bio2rdf}
\bibfield{author}{\bibinfo{person}{Alison Callahan}, \bibinfo{person}{Jos\'{e}
  Cruz-Toledo}, \bibinfo{person}{Peter Ansell}, {and} \bibinfo{person}{Michel
  Dumontier}.} \bibinfo{year}{2013}\natexlab{}.
\newblock \showarticletitle{Bio2RDF {R}elease 2: {I}mproved {C}overage,
  {I}nteroperability and {P}rovenance of {L}ife {S}cience {L}inked {D}ata}. In
  \bibinfo{booktitle}{\emph{10th Extended {Semantic} {Web} {Conference}}}.
  \bibinfo{publisher}{Springer}, \bibinfo{address}{Montperlier, France},
  \bibinfo{pages}{200--212}.
\newblock


\bibitem[\protect\citeauthoryear{Ching, Edunov, Kabiljo, Logothetis, and
  Muthukrishnan}{Ching et~al\mbox{.}}{2015}]%
        {trillion}
\bibfield{author}{\bibinfo{person}{Avery Ching}, \bibinfo{person}{Sergey
  Edunov}, \bibinfo{person}{Maja Kabiljo}, \bibinfo{person}{Dionysios
  Logothetis}, {and} \bibinfo{person}{Sambavi Muthukrishnan}.}
  \bibinfo{year}{2015}\natexlab{}.
\newblock \showarticletitle{One Trillion Edges: Graph Processing at
  Facebook-Scale}.
\newblock \bibinfo{journal}{\emph{Proceedings of the VLDB Endowment}}
  \bibinfo{volume}{8}, \bibinfo{number}{12} (\bibinfo{date}{Aug.}
  \bibinfo{year}{2015}), \bibinfo{pages}{1804–1815}.
\newblock
\showISSN{2150-8097}
\urldef\tempurl%
\url{https://doi.org/10.14778/2824032.2824077}
\showDOI{\tempurl}


\bibitem[\protect\citeauthoryear{Chong, Das, Eadon, and Srinivasan}{Chong
  et~al\mbox{.}}{2005}]%
        {oracle2005}
\bibfield{author}{\bibinfo{person}{Eugene~Inseok Chong},
  \bibinfo{person}{Souripriya Das}, \bibinfo{person}{George Eadon}, {and}
  \bibinfo{person}{Jagannathan Srinivasan}.} \bibinfo{year}{2005}\natexlab{}.
\newblock \showarticletitle{{An Efficient SQL-Based RDF Querying Scheme}}. In
  \bibinfo{booktitle}{\emph{Proceedings of the 31st International Conference on
  Very Large Data Bases}} \emph{(\bibinfo{series}{VLDB ’05})}.
  \bibinfo{publisher}{VLDB Endowment}, \bibinfo{address}{Trondheim, Norway},
  \bibinfo{pages}{1216–1227}.
\newblock
\showISBNx{1595931546}


\bibitem[\protect\citeauthoryear{{DATASTAX, Inc.}}{{DATASTAX, Inc.}}{2019}]%
        {titan}
\bibfield{author}{\bibinfo{person}{{DATASTAX, Inc.}}}
  \bibinfo{year}{2019}\natexlab{}.
\newblock \bibinfo{title}{Titan: {Distributed} {Graph} {Database}}.
\newblock
\newblock
\urldef\tempurl%
\url{http://titan.thinkaurelius.com/}
\showURL{%
\tempurl}


\bibitem[\protect\citeauthoryear{Fan, Raj, and Patel}{Fan
  et~al\mbox{.}}{2015}]%
        {fan_case_2015}
\bibfield{author}{\bibinfo{person}{Jing Fan}, \bibinfo{person}{Adalbert
  Gerald~Soosai Raj}, {and} \bibinfo{person}{Jignesh~M. Patel}.}
  \bibinfo{year}{2015}\natexlab{}.
\newblock \showarticletitle{The {Case} {Against} {Specialized} {Graph}
  {Analytics} {Engines}.}. In \bibinfo{booktitle}{\emph{The 7th Biennial
  Conference on Innovative Data Systems Research {CIDR 2015}}}.
  \bibinfo{publisher}{www.cidrdb.org}, \bibinfo{address}{Asilomar, California,
  USA}.
\newblock


\bibitem[\protect\citeauthoryear{Faye, Cur\'{e}, and Blin}{Faye
  et~al\mbox{.}}{2011}]%
        {survey1}
\bibfield{author}{\bibinfo{person}{David~C. Faye}, \bibinfo{person}{Olivier
  Cur\'{e}}, {and} \bibinfo{person}{Guillaume Blin}.}
  \bibinfo{year}{2011}\natexlab{}.
\newblock \showarticletitle{A survey of {RDF} storage approaches}.
\newblock \bibinfo{journal}{\emph{Revue Africaine de la Recherche en
  Informatique et Math\'{e}matiques Appliqu\'{e}es}}  \bibinfo{volume}{15}
  (\bibinfo{year}{2011}), \bibinfo{pages}{11--35}.
\newblock


\bibitem[\protect\citeauthoryear{Fern\'andez, Mart\'inez-Prieto, Guti\'errez,
  Polleres, and Arias}{Fern\'andez et~al\mbox{.}}{2013}]%
        {hdt}
\bibfield{author}{\bibinfo{person}{Javier~D. Fern\'andez},
  \bibinfo{person}{Miguel~A. Mart\'inez-Prieto}, \bibinfo{person}{Claudio
  Guti\'errez}, \bibinfo{person}{Axel Polleres}, {and} \bibinfo{person}{Mario
  Arias}.} \bibinfo{year}{2013}\natexlab{}.
\newblock \showarticletitle{Binary {RDF} representation for publication and
  exchange ({HDT})}.
\newblock \bibinfo{journal}{\emph{Journal of Web Semantics}}
  \bibinfo{volume}{19} (\bibinfo{date}{March} \bibinfo{year}{2013}),
  \bibinfo{pages}{22--41}.
\newblock
\showISSN{1570-8268}


\bibitem[\protect\citeauthoryear{Fletcher and Beck}{Fletcher and Beck}{2009}]%
        {triplet}
\bibfield{author}{\bibinfo{person}{George~H.L. Fletcher} {and}
  \bibinfo{person}{Peter~W. Beck}.} \bibinfo{year}{2009}\natexlab{}.
\newblock \showarticletitle{{Scalable Indexing of RDF Graphs for Efficient Join
  Processing}}. In \bibinfo{booktitle}{\emph{Proceedings of the 18th ACM
  Conference on Information and Knowledge Management}} (Hong Kong)
  \emph{(\bibinfo{series}{CIKM ’09})}. \bibinfo{publisher}{Association for
  Computing Machinery}, \bibinfo{address}{New York, NY, USA},
  \bibinfo{pages}{1513–1516}.
\newblock
\showISBNx{9781605585123}
\urldef\tempurl%
\url{https://doi.org/10.1145/1645953.1646159}
\showDOI{\tempurl}


\bibitem[\protect\citeauthoryear{Gonzalez, Low, Gu, Bickson, and
  Guestrin}{Gonzalez et~al\mbox{.}}{2012}]%
        {powergraph}
\bibfield{author}{\bibinfo{person}{Joseph~E. Gonzalez},
  \bibinfo{person}{Yucheng Low}, \bibinfo{person}{Haijie Gu},
  \bibinfo{person}{Danny Bickson}, {and} \bibinfo{person}{Carlos Guestrin}.}
  \bibinfo{year}{2012}\natexlab{}.
\newblock \showarticletitle{{PowerGraph: Distributed Graph-Parallel Computation
  on Natural Graphs}}. In \bibinfo{booktitle}{\emph{10th {USENIX} Symposium on
  Operating Systems Design and Implementation ({OSDI} 12)}}.
  \bibinfo{publisher}{{USENIX}}, \bibinfo{address}{Hollywood, CA},
  \bibinfo{pages}{17--30}.
\newblock
\showISBNx{978-1-931971-96-6}


\bibitem[\protect\citeauthoryear{Gonzalez, Xin, Dave, Crankshaw, Franklin, and
  Stoica}{Gonzalez et~al\mbox{.}}{2014}]%
        {graphx}
\bibfield{author}{\bibinfo{person}{Joseph~E. Gonzalez},
  \bibinfo{person}{Reynold~S. Xin}, \bibinfo{person}{Ankur Dave},
  \bibinfo{person}{Daniel Crankshaw}, \bibinfo{person}{Michael~J. Franklin},
  {and} \bibinfo{person}{Ion Stoica}.} \bibinfo{year}{2014}\natexlab{}.
\newblock \showarticletitle{{GraphX}: {Graph} {Processing} in a {Distributed}
  {Dataflow} {Framework}}. In \bibinfo{booktitle}{\emph{11th {USENIX} Symposium
  on Operating Systems Design and Implementation ({OSDI} 14)}}.
  \bibinfo{publisher}{{USENIX} Association}, \bibinfo{address}{Broomfield, CO},
  \bibinfo{pages}{599--613}.
\newblock
\showISBNx{978-1-931971-16-4}


\bibitem[\protect\citeauthoryear{Gray}{Gray}{1981}]%
        {gray1981transaction}
\bibfield{author}{\bibinfo{person}{Jim Gray}.} \bibinfo{year}{1981}\natexlab{}.
\newblock \showarticletitle{{The Transaction Concept: Virtues and Limitations
  (Invited Paper)}}. In \bibinfo{booktitle}{\emph{Very Large Data Bases, 7th
  International Conference, September 9-11, 1981, Cannes, France,
  Proceedings}}. \bibinfo{publisher}{VLDB Endowment}, \bibinfo{address}{Cannes,
  France}, \bibinfo{pages}{144--154}.
\newblock


\bibitem[\protect\citeauthoryear{Greaves and Mika}{Greaves and Mika}{2008}]%
        {websearch3}
\bibfield{author}{\bibinfo{person}{Mark Greaves} {and} \bibinfo{person}{Peter
  Mika}.} \bibinfo{year}{2008}\natexlab{}.
\newblock \showarticletitle{Semantic {Web} and {Web} 2.0}.
\newblock \bibinfo{journal}{\emph{Web Semantics: Science, Services and Agents
  on the World Wide Web}} \bibinfo{volume}{6}, \bibinfo{number}{1}
  (\bibinfo{year}{2008}), \bibinfo{pages}{1--3}.
\newblock


\bibitem[\protect\citeauthoryear{Guha, McCool, and Miller}{Guha
  et~al\mbox{.}}{2003}]%
        {guha_semantic_2003}
\bibfield{author}{\bibinfo{person}{R. Guha}, \bibinfo{person}{Rob McCool},
  {and} \bibinfo{person}{Eric Miller}.} \bibinfo{year}{2003}\natexlab{}.
\newblock \showarticletitle{{Semantic Search}}. In
  \bibinfo{booktitle}{\emph{Proceedings of the 12th International Conference on
  World Wide Web}} (Budapest, Hungary) \emph{(\bibinfo{series}{WWW ’03})}.
  \bibinfo{publisher}{Association for Computing Machinery},
  \bibinfo{address}{New York, NY, USA}, \bibinfo{pages}{700–709}.
\newblock
\showISBNx{1581136803}
\urldef\tempurl%
\url{https://doi.org/10.1145/775152.775250}
\showDOI{\tempurl}


\bibitem[\protect\citeauthoryear{Guo, Pan, and Heflin}{Guo
  et~al\mbox{.}}{2005}]%
        {lubm}
\bibfield{author}{\bibinfo{person}{Yuanbo Guo}, \bibinfo{person}{Zhengxiang
  Pan}, {and} \bibinfo{person}{Jeff Heflin}.} \bibinfo{year}{2005}\natexlab{}.
\newblock \showarticletitle{{LUBM}: {A} benchmark for {OWL} knowledge base
  systems}.
\newblock \bibinfo{journal}{\emph{Web Semantics: Science, Services and Agents
  on the World Wide Web}} \bibinfo{volume}{3}, \bibinfo{number}{2}
  (\bibinfo{year}{2005}), \bibinfo{pages}{158--182}.
\newblock


\bibitem[\protect\citeauthoryear{Gurajada, Seufert, Miliaraki, and
  Theobald}{Gurajada et~al\mbox{.}}{2014}]%
        {triad}
\bibfield{author}{\bibinfo{person}{Sairam Gurajada}, \bibinfo{person}{Stephan
  Seufert}, \bibinfo{person}{Iris Miliaraki}, {and} \bibinfo{person}{Martin
  Theobald}.} \bibinfo{year}{2014}\natexlab{}.
\newblock \showarticletitle{{TriAD: A Distributed Shared-Nothing RDF Engine
  Based on Asynchronous Message Passing}}. In
  \bibinfo{booktitle}{\emph{Proceedings of the 2014 ACM SIGMOD International
  Conference on Management of Data}} (Snowbird, Utah, USA)
  \emph{(\bibinfo{series}{SIGMOD ’14})}. \bibinfo{publisher}{Association for
  Computing Machinery}, \bibinfo{address}{New York, NY, USA},
  \bibinfo{pages}{289–300}.
\newblock
\showISBNx{9781450323765}
\urldef\tempurl%
\url{https://doi.org/10.1145/2588555.2610511}
\showDOI{\tempurl}


\bibitem[\protect\citeauthoryear{Han, Daudjee, Ammar, \"{O}zsu, Wang, and
  Jin}{Han et~al\mbox{.}}{2014}]%
        {pregel_comp}
\bibfield{author}{\bibinfo{person}{Minyang Han}, \bibinfo{person}{Khuzaima
  Daudjee}, \bibinfo{person}{Khaled Ammar}, \bibinfo{person}{M.~Tamer
  \"{O}zsu}, \bibinfo{person}{Xingfang Wang}, {and} \bibinfo{person}{Tianqi
  Jin}.} \bibinfo{year}{2014}\natexlab{}.
\newblock \showarticletitle{{An Experimental Comparison of Pregel-like Graph
  Processing Systems}}.
\newblock \bibinfo{journal}{\emph{Proceedings of the VLDB Endowment}}
  \bibinfo{volume}{7}, \bibinfo{number}{12} (\bibinfo{year}{2014}),
  \bibinfo{pages}{1047–1058}.
\newblock
\showISSN{2150-8097}
\urldef\tempurl%
\url{https://doi.org/10.14778/2732977.2732980}
\showDOI{\tempurl}


\bibitem[\protect\citeauthoryear{Han, Cao, Lv, Lin, Liu, Sun, and Li}{Han
  et~al\mbox{.}}{2018}]%
        {openke}
\bibfield{author}{\bibinfo{person}{Xu Han}, \bibinfo{person}{Shulin Cao},
  \bibinfo{person}{Xin Lv}, \bibinfo{person}{Yankai Lin},
  \bibinfo{person}{Zhiyuan Liu}, \bibinfo{person}{Maosong Sun}, {and}
  \bibinfo{person}{Juanzi Li}.} \bibinfo{year}{2018}\natexlab{}.
\newblock \showarticletitle{{OpenKE: An Open Toolkit for Knowledge Embedding}}.
  In \bibinfo{booktitle}{\emph{Proceedings of the 2018 Conference on Empirical
  Methods in Natural Language Processing, {EMNLP} 2018: System Demonstrations,
  Brussels, Belgium, October 31 - November 4, 2018}}.
  \bibinfo{publisher}{Association for Computational Linguistics},
  \bibinfo{address}{Brussels, Belgium}, \bibinfo{pages}{139--144}.
\newblock


\bibitem[\protect\citeauthoryear{Harbi, Abdelaziz, Kalnis, Mamoulis, Ebrahim,
  and Sahli}{Harbi et~al\mbox{.}}{2016}]%
        {accelerating}
\bibfield{author}{\bibinfo{person}{Razen Harbi}, \bibinfo{person}{Ibrahim
  Abdelaziz}, \bibinfo{person}{Panos Kalnis}, \bibinfo{person}{Nikos Mamoulis},
  \bibinfo{person}{Yasser Ebrahim}, {and} \bibinfo{person}{Majed Sahli}.}
  \bibinfo{year}{2016}\natexlab{}.
\newblock \showarticletitle{Accelerating {SPARQL} queries by exploiting
  hash-based locality and adaptive partitioning}.
\newblock \bibinfo{journal}{\emph{The VLDB Journal}} \bibinfo{volume}{25},
  \bibinfo{number}{3} (\bibinfo{year}{2016}), \bibinfo{pages}{355--380}.
\newblock


\bibitem[\protect\citeauthoryear{Harris and Gibbins}{Harris and
  Gibbins}{2003}]%
        {harris_3store:_2003}
\bibfield{author}{\bibinfo{person}{Stephen Harris} {and}
  \bibinfo{person}{Nicholas Gibbins}.} \bibinfo{year}{2003}\natexlab{}.
\newblock \showarticletitle{3store: {E}fficient {B}ulk {RDF} {S}torage}. In
  \bibinfo{booktitle}{\emph{1st International Workshop on Practical and
  Scalable Semantic Systems (PSSS'03)}}. \bibinfo{publisher}{ePrints Soton},
  \bibinfo{address}{Sanibel Island, FL, USA}, \bibinfo{pages}{1--15}.
\newblock


\bibitem[\protect\citeauthoryear{Harris, Lamb, and Shadbolt}{Harris
  et~al\mbox{.}}{2009}]%
        {4store}
\bibfield{author}{\bibinfo{person}{Steve Harris}, \bibinfo{person}{Nick Lamb},
  {and} \bibinfo{person}{Nigel Shadbolt}.} \bibinfo{year}{2009}\natexlab{}.
\newblock \showarticletitle{4store: {The} Design and Implementation of a
  Clustered {RDF} Store}. In \bibinfo{booktitle}{\emph{5th {International}
  {Workshop} on {Scalable} {Semantic} {Web} {Knowledge} {Base} {Systems}
  ({SSWS}2009)}}. \bibinfo{publisher}{CEUR Workshop Proceedings},
  \bibinfo{address}{Washington, DC, USA}, \bibinfo{pages}{94--109}.
\newblock


\bibitem[\protect\citeauthoryear{Harris, Seaborne, and Prud'hommeaux}{Harris
  et~al\mbox{.}}{2013}]%
        {sparql}
\bibfield{author}{\bibinfo{person}{Steve Harris}, \bibinfo{person}{Andy
  Seaborne}, {and} \bibinfo{person}{Eric Prud'hommeaux}.}
  \bibinfo{year}{2013}\natexlab{}.
\newblock \bibinfo{title}{{SPARQL 1.1 Query Language}}.
\newblock
\newblock
\urldef\tempurl%
\url{http://www.w3.org/TR/sparql11-query}
\showURL{%
\tempurl}


\bibitem[\protect\citeauthoryear{Harris and Shadbolt}{Harris and
  Shadbolt}{2005}]%
        {harris_sparql_2005}
\bibfield{author}{\bibinfo{person}{Stephen Harris} {and} \bibinfo{person}{Nigel
  Shadbolt}.} \bibinfo{year}{2005}\natexlab{}.
\newblock \showarticletitle{{SPARQL} query processing with conventional
  relational database systems}. In \bibinfo{booktitle}{\emph{International
  {Conference} on {Web} {Information} {Systems} {Engineering}}}.
  \bibinfo{publisher}{Springer}, \bibinfo{address}{New York, NY, USA},
  \bibinfo{pages}{235--244}.
\newblock


\bibitem[\protect\citeauthoryear{Harth}{Harth}{2012}]%
        {btc2012}
\bibfield{author}{\bibinfo{person}{Andreas Harth}.}
  \bibinfo{year}{2012}\natexlab{}.
\newblock \bibinfo{title}{{Billion Triples Challenge} data set}.
\newblock
\newblock
\urldef\tempurl%
\url{http://km.aifb.kit.edu/projects/btc-2012/}
\showURL{%
\tempurl}


\bibitem[\protect\citeauthoryear{Harth and Decker}{Harth and Decker}{2005}]%
        {harth1}
\bibfield{author}{\bibinfo{person}{Andreas Harth} {and} \bibinfo{person}{Stefan
  Decker}.} \bibinfo{year}{2005}\natexlab{}.
\newblock \showarticletitle{Optimized Index Structures for Querying RDF from
  the Web}. In \bibinfo{booktitle}{\emph{Proceedings of the Third Latin
  American Web Congress}} \emph{(\bibinfo{series}{LA-WEB '05})}.
  \bibinfo{publisher}{IEEE Computer Society}, \bibinfo{address}{Washington, DC,
  USA}, \bibinfo{pages}{71--80}.
\newblock
\showISBNx{0-7695-2471-0}


\bibitem[\protect\citeauthoryear{Harth, Umbrich, Hogan, and Decker}{Harth
  et~al\mbox{.}}{2007}]%
        {yars2}
\bibfield{author}{\bibinfo{person}{Andreas Harth}, \bibinfo{person}{J\"{u}rgen
  Umbrich}, \bibinfo{person}{Aidan Hogan}, {and} \bibinfo{person}{Stefan
  Decker}.} \bibinfo{year}{2007}\natexlab{}.
\newblock \showarticletitle{{YARS}2: {A} {Federated} {Repository} for
  {Querying} {Graph} {Structured} {Data} from the {Web}}. In
  \bibinfo{booktitle}{\emph{The 6th International Semantic Web Conference}}
  \emph{(\bibinfo{series}{Lecture {Notes} in {Computer} {Science}})}.
  \bibinfo{publisher}{Springer Berlin Heidelberg}, \bibinfo{address}{Busan,
  South Korea}, \bibinfo{pages}{211--224}.
\newblock
\showISBNx{978-3-540-76298-0}


\bibitem[\protect\citeauthoryear{Hartig and P\'{e}rez}{Hartig and
  P\'{e}rez}{2018}]%
        {graphql}
\bibfield{author}{\bibinfo{person}{Olaf Hartig} {and} \bibinfo{person}{Jorge
  P\'{e}rez}.} \bibinfo{year}{2018}\natexlab{}.
\newblock \showarticletitle{{Semantics and Complexity of GraphQL}}. In
  \bibinfo{booktitle}{\emph{Proceedings of the 2018 World Wide Web Conference}}
  (Lyon, France) \emph{(\bibinfo{series}{WWW ’18})}.
  \bibinfo{publisher}{International World Wide Web Conferences Steering
  Committee}, \bibinfo{address}{Republic and Canton of Geneva, CHE},
  \bibinfo{pages}{1155–1164}.
\newblock
\showISBNx{9781450356398}
\urldef\tempurl%
\url{https://doi.org/10.1145/3178876.3186014}
\showDOI{\tempurl}


\bibitem[\protect\citeauthoryear{Hayes}{Hayes}{2004}]%
        {rdf}
\bibfield{author}{\bibinfo{person}{Patrick Hayes}.}
  \bibinfo{year}{2004}\natexlab{}.
\newblock \bibinfo{title}{{RDF} Semantics, {W3C} Recommendation}.
\newblock
\newblock
\urldef\tempurl%
\url{http://www.w3.org/TR/rdf-mt/}
\showURL{%
\tempurl}


\bibitem[\protect\citeauthoryear{Kasneci, Suchanek, Ifrim, Ramanath, and
  Weikum}{Kasneci et~al\mbox{.}}{2008}]%
        {naga}
\bibfield{author}{\bibinfo{person}{Gjergji Kasneci}, \bibinfo{person}{Fabian~M.
  Suchanek}, \bibinfo{person}{Georgiana Ifrim}, \bibinfo{person}{Maya
  Ramanath}, {and} \bibinfo{person}{Gerhard Weikum}.}
  \bibinfo{year}{2008}\natexlab{}.
\newblock \showarticletitle{{NAGA}: {Searching} and Ranking Knowledge}. In
  \bibinfo{booktitle}{\emph{24th International Conference on Data {Engineering}
  ({ICDE})}}. \bibinfo{publisher}{IEEE}, \bibinfo{address}{Cancun, Mexico},
  \bibinfo{pages}{953--962}.
\newblock


\bibitem[\protect\citeauthoryear{Kim, Shin, Han, Hong, and Chafi}{Kim
  et~al\mbox{.}}{2015}]%
        {taming}
\bibfield{author}{\bibinfo{person}{Jinha Kim}, \bibinfo{person}{Hyungyu Shin},
  \bibinfo{person}{Wook-Shin Han}, \bibinfo{person}{Sungpack Hong}, {and}
  \bibinfo{person}{Hassan Chafi}.} \bibinfo{year}{2015}\natexlab{}.
\newblock \showarticletitle{Taming {Subgraph} {Isomorphism} for {RDF} {Query}
  {Processing}}.
\newblock \bibinfo{journal}{\emph{Proceedings of the VLDB Endowment}}
  \bibinfo{volume}{8}, \bibinfo{number}{11} (\bibinfo{year}{2015}),
  \bibinfo{pages}{1238--1249}.
\newblock
\showISSN{2150-8097}


\bibitem[\protect\citeauthoryear{Lee and Liu}{Lee and Liu}{2013}]%
        {scaling}
\bibfield{author}{\bibinfo{person}{Kisung Lee} {and} \bibinfo{person}{Ling
  Liu}.} \bibinfo{year}{2013}\natexlab{}.
\newblock \showarticletitle{Scaling queries over big {RDF} graphs with semantic
  hash partitioning}.
\newblock \bibinfo{journal}{\emph{Proceedings of the VLDB Endowment}}
  \bibinfo{volume}{6}, \bibinfo{number}{14} (\bibinfo{year}{2013}),
  \bibinfo{pages}{1894--1905}.
\newblock


\bibitem[\protect\citeauthoryear{Leskovec and Krevl}{Leskovec and
  Krevl}{2014}]%
        {snapnets}
\bibfield{author}{\bibinfo{person}{Jure Leskovec} {and} \bibinfo{person}{Andrej
  Krevl}.} \bibinfo{year}{2014}\natexlab{}.
\newblock \bibinfo{title}{{SNAP Datasets}: {Stanford} Large Network Dataset
  Collection}.
\newblock \bibinfo{howpublished}{\url{http://snap.stanford.edu/data}}.
\newblock


\bibitem[\protect\citeauthoryear{Leskovec and Sosi\v{c}}{Leskovec and
  Sosi\v{c}}{2016}]%
        {snap}
\bibfield{author}{\bibinfo{person}{Jure Leskovec} {and} \bibinfo{person}{Rok
  Sosi\v{c}}.} \bibinfo{year}{2016}\natexlab{}.
\newblock \showarticletitle{Snap: {A} general-purpose network analysis and
  graph-mining library}.
\newblock \bibinfo{journal}{\emph{ACM Transactions on Intelligent Systems and
  Technology (TIST)}} \bibinfo{volume}{8}, \bibinfo{number}{1}
  (\bibinfo{year}{2016}), \bibinfo{pages}{1}.
\newblock


\bibitem[\protect\citeauthoryear{Ma, Su, Pan, Zhang, and Liu}{Ma
  et~al\mbox{.}}{2004}]%
        {rstar}
\bibfield{author}{\bibinfo{person}{Li Ma}, \bibinfo{person}{Zhong Su},
  \bibinfo{person}{Yue Pan}, \bibinfo{person}{Li Zhang}, {and}
  \bibinfo{person}{Tao Liu}.} \bibinfo{year}{2004}\natexlab{}.
\newblock \showarticletitle{{RStar: An RDF Storage and Query System for
  Enterprise Resource Management}}. In \bibinfo{booktitle}{\emph{Proceedings of
  the Thirteenth ACM International Conference on Information and Knowledge
  Management}} (Washington, D.C., USA) \emph{(\bibinfo{series}{CIKM ’04})}.
  \bibinfo{publisher}{Association for Computing Machinery},
  \bibinfo{address}{New York, NY, USA}, \bibinfo{pages}{484–491}.
\newblock
\showISBNx{1581138741}
\urldef\tempurl%
\url{https://doi.org/10.1145/1031171.1031264}
\showDOI{\tempurl}


\bibitem[\protect\citeauthoryear{Malewicz, Austern, Bik, Dehnert, Horn, Leiser,
  and Czajkowski}{Malewicz et~al\mbox{.}}{2010}]%
        {pregel}
\bibfield{author}{\bibinfo{person}{Grzegorz Malewicz},
  \bibinfo{person}{Matthew~H. Austern}, \bibinfo{person}{Aart~J.C Bik},
  \bibinfo{person}{James~C. Dehnert}, \bibinfo{person}{Ilan Horn},
  \bibinfo{person}{Naty Leiser}, {and} \bibinfo{person}{Grzegorz Czajkowski}.}
  \bibinfo{year}{2010}\natexlab{}.
\newblock \showarticletitle{Pregel: A System for Large-Scale Graph Processing}.
  In \bibinfo{booktitle}{\emph{Proceedings of the 2010 ACM SIGMOD International
  Conference on Management of Data}} (Indianapolis, Indiana, USA)
  \emph{(\bibinfo{series}{SIGMOD ’10})}. \bibinfo{publisher}{Association for
  Computing Machinery}, \bibinfo{address}{New York, NY, USA},
  \bibinfo{pages}{135–146}.
\newblock
\showISBNx{9781450300322}
\urldef\tempurl%
\url{https://doi.org/10.1145/1807167.1807184}
\showDOI{\tempurl}


\bibitem[\protect\citeauthoryear{Mart\'{\i}nez-Prieto, Fern\'{a}ndez, and
  C\'{a}novas}{Mart\'{\i}nez-Prieto et~al\mbox{.}}{2012a}]%
        {compression1}
\bibfield{author}{\bibinfo{person}{Miguel~A. Mart\'{\i}nez-Prieto},
  \bibinfo{person}{Javier~D. Fern\'{a}ndez}, {and} \bibinfo{person}{Rodrigo
  C\'{a}novas}.} \bibinfo{year}{2012}\natexlab{a}.
\newblock \showarticletitle{Compression of {RDF} Dictionaries}. In
  \bibinfo{booktitle}{\emph{Proceedings of the 27th {Annual} {ACM} {Symposium}
  on {Applied} {Computing}}}. \bibinfo{publisher}{ACM},
  \bibinfo{address}{Trento, Italy}, \bibinfo{pages}{340--347}.
\newblock


\bibitem[\protect\citeauthoryear{Mart\'{\i}nez-Prieto, Fern\'{a}ndez, and
  C\'{a}novas}{Mart\'{\i}nez-Prieto et~al\mbox{.}}{2012b}]%
        {martinez-prieto_querying_2012}
\bibfield{author}{\bibinfo{person}{Miguel~A. Mart\'{\i}nez-Prieto},
  \bibinfo{person}{Javier~D. Fern\'{a}ndez}, {and} \bibinfo{person}{Rodrigo
  C\'{a}novas}.} \bibinfo{year}{2012}\natexlab{b}.
\newblock \showarticletitle{Querying {RDF} dictionaries in compressed space}.
\newblock \bibinfo{journal}{\emph{SIGAPP Appl. Comput. Rev.}}
  \bibinfo{volume}{12}, \bibinfo{number}{2} (\bibinfo{date}{June}
  \bibinfo{year}{2012}), \bibinfo{pages}{64--77}.
\newblock
\showISSN{1559-6915}


\bibitem[\protect\citeauthoryear{Mavlyutov, Wylot, and Cudre-Mauroux}{Mavlyutov
  et~al\mbox{.}}{2015}]%
        {uriencoding}
\bibfield{author}{\bibinfo{person}{Ruslan Mavlyutov}, \bibinfo{person}{Marcin
  Wylot}, {and} \bibinfo{person}{Philippe Cudre-Mauroux}.}
  \bibinfo{year}{2015}\natexlab{}.
\newblock \showarticletitle{A Comparison of Data Structures to Manage URIs on
  the Web of Data}. In \bibinfo{booktitle}{\emph{The Semantic Web. Latest
  Advances and New Domains}}. \bibinfo{publisher}{Springer-Verlag New York,
  Inc.}, \bibinfo{address}{Portoroz, Slovenia}, \bibinfo{pages}{137--151}.
\newblock
\showISBNx{978-3-319-18817-1}


\bibitem[\protect\citeauthoryear{McBride}{McBride}{2001}]%
        {jena}
\bibfield{author}{\bibinfo{person}{Brian McBride}.}
  \bibinfo{year}{2001}\natexlab{}.
\newblock \showarticletitle{Jena: {Implementing} the {RDF} Model and Syntax
  Specification}. In \bibinfo{booktitle}{\emph{SemWeb'01 Proceedings of the
  {Second} {International} {Conference} on {Semantic} {Web}-{Volume} 40}}.
  \bibinfo{publisher}{CEUR-WS. org}, \bibinfo{address}{Hong Kong},
  \bibinfo{pages}{23--28}.
\newblock


\bibitem[\protect\citeauthoryear{McCune, Weninger, and Madey}{McCune
  et~al\mbox{.}}{2015}]%
        {distsurvey}
\bibfield{author}{\bibinfo{person}{Robert~Ryan McCune}, \bibinfo{person}{Tim
  Weninger}, {and} \bibinfo{person}{Greg Madey}.}
  \bibinfo{year}{2015}\natexlab{}.
\newblock \showarticletitle{Thinking like a vertex: a survey of vertex-centric
  frameworks for large-scale distributed graph processing}.
\newblock \bibinfo{journal}{\emph{ACM Computing Surveys (CSUR)}}
  \bibinfo{volume}{48}, \bibinfo{number}{2} (\bibinfo{year}{2015}),
  \bibinfo{pages}{25}.
\newblock


\bibitem[\protect\citeauthoryear{Meusel, Vigna, Lehmberg, and Bizer}{Meusel
  et~al\mbox{.}}{2015}]%
        {hypergraph}
\bibfield{author}{\bibinfo{person}{Robert Meusel}, \bibinfo{person}{Sebastiano
  Vigna}, \bibinfo{person}{Oliver Lehmberg}, {and} \bibinfo{person}{Christian
  Bizer}.} \bibinfo{year}{2015}\natexlab{}.
\newblock \showarticletitle{The Graph Structure in the Web -- Analyzed on
  Different Aggregation Levels}.
\newblock \bibinfo{journal}{\emph{The Journal of Web Science}}
  \bibinfo{volume}{1}, \bibinfo{number}{1} (\bibinfo{year}{2015}),
  \bibinfo{pages}{33--47}.
\newblock


\bibitem[\protect\citeauthoryear{Modoni, Sacco, and Terkaj}{Modoni
  et~al\mbox{.}}{2014}]%
        {survey2}
\bibfield{author}{\bibinfo{person}{G.~E. Modoni}, \bibinfo{person}{M. Sacco},
  {and} \bibinfo{person}{W. Terkaj}.} \bibinfo{year}{2014}\natexlab{}.
\newblock \showarticletitle{A {S}urvey of {RDF} {S}tore {S}olutions}. In
  \bibinfo{booktitle}{\emph{2014 {International} {Conference} on {Engineering},
  {Technology} and {Innovation} ({ICE})}}. \bibinfo{publisher}{IEEE},
  \bibinfo{address}{Bergamo, Italy}, \bibinfo{pages}{1--7}.
\newblock


\bibitem[\protect\citeauthoryear{Motik, Nenov, Piro, Horrocks, and
  Olteanu}{Motik et~al\mbox{.}}{2014}]%
        {rdfox}
\bibfield{author}{\bibinfo{person}{Boris Motik}, \bibinfo{person}{Yavor Nenov},
  \bibinfo{person}{Robert Piro}, \bibinfo{person}{Ian Horrocks}, {and}
  \bibinfo{person}{Dan Olteanu}.} \bibinfo{year}{2014}\natexlab{}.
\newblock \showarticletitle{{Parallel Materialisation of Datalog Programs in
  Centralised, Main-Memory {RDF} Systems}}. In
  \bibinfo{booktitle}{\emph{Proceedings of the Twenty-Eighth {AAAI} Conference
  on Artificial Intelligence, July 27 -31, 2014, Qu{\'{e}}bec City,
  Qu{\'{e}}bec, Canada}}. \bibinfo{publisher}{AAAI Press},
  \bibinfo{address}{Qu{\'{e}}bec, Canada}, \bibinfo{pages}{129--137}.
\newblock


\bibitem[\protect\citeauthoryear{{Neo4j, Inc.}}{{Neo4j, Inc.}}{2019}]%
        {neo4j}
\bibfield{author}{\bibinfo{person}{{Neo4j, Inc.}}}
  \bibinfo{year}{2019}\natexlab{}.
\newblock \bibinfo{title}{Neo4j {Graph} {Platform}}.
\newblock
\newblock
\urldef\tempurl%
\url{https://neo4j.com/}
\showURL{%
\tempurl}


\bibitem[\protect\citeauthoryear{Neumann and Moerkotte}{Neumann and
  Moerkotte}{2011}]%
        {charsets}
\bibfield{author}{\bibinfo{person}{Thomas Neumann} {and} \bibinfo{person}{Guido
  Moerkotte}.} \bibinfo{year}{2011}\natexlab{}.
\newblock \showarticletitle{Characteristic sets: {Accurate} Cardinality
  Estimation for {RDF} Queries with Multiple Joins}. In
  \bibinfo{booktitle}{\emph{27th {International} {Conference} on {Data}
  {Engineering} ({ICDE})}}. \bibinfo{publisher}{IEEE},
  \bibinfo{address}{Hannover, Germany}, \bibinfo{pages}{984--994}.
\newblock


\bibitem[\protect\citeauthoryear{Neumann and Weikum}{Neumann and
  Weikum}{2010}]%
        {rdf3x}
\bibfield{author}{\bibinfo{person}{Thomas Neumann} {and}
  \bibinfo{person}{Gerhard Weikum}.} \bibinfo{year}{2010}\natexlab{}.
\newblock \showarticletitle{The {RDF}-3X engine for scalable management of
  {RDF} data}.
\newblock \bibinfo{journal}{\emph{The VLDB Journal}} \bibinfo{volume}{19},
  \bibinfo{number}{1} (\bibinfo{year}{2010}), \bibinfo{pages}{91--113}.
\newblock


\bibitem[\protect\citeauthoryear{Nickel, Murphy, Tresp, and Gabrilovich}{Nickel
  et~al\mbox{.}}{2015}]%
        {kgemb_survey}
\bibfield{author}{\bibinfo{person}{Maximilian Nickel}, \bibinfo{person}{Kevin
  Murphy}, \bibinfo{person}{Volker Tresp}, {and} \bibinfo{person}{Evgeniy
  Gabrilovich}.} \bibinfo{year}{2015}\natexlab{}.
\newblock \showarticletitle{A review of relational machine learning for
  knowledge graphs}.
\newblock \bibinfo{journal}{\emph{Proc. IEEE}} \bibinfo{volume}{104},
  \bibinfo{number}{1} (\bibinfo{year}{2015}), \bibinfo{pages}{11--33}.
\newblock


\bibitem[\protect\citeauthoryear{Noy, Shah, Whetzel, Dai, Dorf, Griffith,
  Jonquet, Rubin, Storey, and Chute}{Noy et~al\mbox{.}}{2009}]%
        {bioportal}
\bibfield{author}{\bibinfo{person}{Natalya~F. Noy}, \bibinfo{person}{Nigam~H.
  Shah}, \bibinfo{person}{Patricia~L. Whetzel}, \bibinfo{person}{Benjamin Dai},
  \bibinfo{person}{Michael Dorf}, \bibinfo{person}{Nicholas Griffith},
  \bibinfo{person}{Clement Jonquet}, \bibinfo{person}{Daniel~L. Rubin},
  \bibinfo{person}{Margaret-Anne Storey}, {and} \bibinfo{person}{Christopher~G.
  Chute}.} \bibinfo{year}{2009}\natexlab{}.
\newblock \showarticletitle{{BioPortal}: ontologies and integrated data
  resources at the click of a mouse}.
\newblock \bibinfo{journal}{\emph{Nucleic acids research}}
  \bibinfo{volume}{37}, \bibinfo{number}{suppl\_2} (\bibinfo{year}{2009}),
  \bibinfo{pages}{W170--W173}.
\newblock


\bibitem[\protect\citeauthoryear{{Objectivity Inc.}}{{Objectivity
  Inc.}}{2019}]%
        {infinitegraph}
\bibfield{author}{\bibinfo{person}{{Objectivity Inc.}}}
  \bibinfo{year}{2019}\natexlab{}.
\newblock \bibinfo{title}{{InfiniteGraph}}.
\newblock
\newblock
\urldef\tempurl%
\url{https://www.objectivity.com/products/infinitegraph/}
\showURL{%
\tempurl}


\bibitem[\protect\citeauthoryear{{OpenLink Software}}{{OpenLink
  Software}}{2019}]%
        {virtuoso}
\bibfield{author}{\bibinfo{person}{{OpenLink Software}}.}
  \bibinfo{year}{2019}\natexlab{}.
\newblock \bibinfo{title}{Virtuoso {RDF} {Engine}}.
\newblock
\newblock
\urldef\tempurl%
\url{https://virtuoso.openlinksw.com/}
\showURL{%
\tempurl}


\bibitem[\protect\citeauthoryear{\"Ozsu}{\"Ozsu}{2016}]%
        {survey4}
\bibfield{author}{\bibinfo{person}{M.~Tamer \"Ozsu}.}
  \bibinfo{year}{2016}\natexlab{}.
\newblock \showarticletitle{A survey of {RDF} data management systems}.
\newblock \bibinfo{journal}{\emph{Frontiers of Computer Science}}
  \bibinfo{volume}{10}, \bibinfo{number}{3} (\bibinfo{year}{2016}),
  \bibinfo{pages}{418--432}.
\newblock
\showISSN{2095-2236}


\bibitem[\protect\citeauthoryear{Pal and Urbani}{Pal and Urbani}{2017}]%
        {subgraphs}
\bibfield{author}{\bibinfo{person}{Soumajit Pal} {and} \bibinfo{person}{Jacopo
  Urbani}.} \bibinfo{year}{2017}\natexlab{}.
\newblock \showarticletitle{Enhancing {Knowledge} {Graph} {Completion} {By}
  {Embedding} {Correlations}}. In \bibinfo{booktitle}{\emph{Proceedings of the
  2017 {ACM} on {Conference} on {Information} and {Knowledge} {Management},
  {CIKM} 2017, {Singapore}, {November} 06 - 10, 2017}}.
  \bibinfo{publisher}{ACM}, \bibinfo{address}{New York, NY, USA},
  \bibinfo{pages}{2247--2250}.
\newblock


\bibitem[\protect\citeauthoryear{Peng, Zou, \"Ozsu, Chen, and Zhao}{Peng
  et~al\mbox{.}}{2016}]%
        {otzu1}
\bibfield{author}{\bibinfo{person}{Peng Peng}, \bibinfo{person}{Lei Zou},
  \bibinfo{person}{M.~Tamer \"Ozsu}, \bibinfo{person}{Lei Chen}, {and}
  \bibinfo{person}{Dongyan Zhao}.} \bibinfo{year}{2016}\natexlab{}.
\newblock \showarticletitle{Processing {SPARQL} queries over distributed {RDF}
  graphs}.
\newblock \bibinfo{journal}{\emph{The VLDB Journal}} \bibinfo{volume}{25},
  \bibinfo{number}{2} (\bibinfo{year}{2016}), \bibinfo{pages}{243--268}.
\newblock
\showISSN{1066-8888}


\bibitem[\protect\citeauthoryear{Perez, Sosi\v{c}, Banerjee, Puttagunta,
  Raison, Shah, and Leskovec}{Perez et~al\mbox{.}}{2015}]%
        {ringo}
\bibfield{author}{\bibinfo{person}{Yonathan Perez}, \bibinfo{person}{Rok
  Sosi\v{c}}, \bibinfo{person}{Arijit Banerjee}, \bibinfo{person}{Rohan
  Puttagunta}, \bibinfo{person}{Martin Raison}, \bibinfo{person}{Pararth Shah},
  {and} \bibinfo{person}{Jure Leskovec}.} \bibinfo{year}{2015}\natexlab{}.
\newblock \showarticletitle{{Ringo: Interactive Graph Analytics on Big-Memory
  Machines}}. In \bibinfo{booktitle}{\emph{Proceedings of the 2015 ACM SIGMOD
  International Conference on Management of Data}} (Melbourne, Victoria,
  Australia) \emph{(\bibinfo{series}{SIGMOD ’15})}.
  \bibinfo{publisher}{Association for Computing Machinery},
  \bibinfo{address}{New York, NY, USA}, \bibinfo{pages}{1105–1110}.
\newblock
\showISBNx{9781450327589}
\urldef\tempurl%
\url{https://doi.org/10.1145/2723372.2735369}
\showDOI{\tempurl}


\bibitem[\protect\citeauthoryear{Pham and Boncz}{Pham and Boncz}{2016}]%
        {pham_exploiting_2016}
\bibfield{author}{\bibinfo{person}{Minh-Duc Pham} {and} \bibinfo{person}{Peter
  Boncz}.} \bibinfo{year}{2016}\natexlab{}.
\newblock \showarticletitle{Exploiting {Emergent} {Schemas} to {Make} {RDF}
  {Systems} {More} {Efficient}}. In \bibinfo{booktitle}{\emph{The 15th
  International {Semantic} {Web} Conference – {ISWC} 2016}}
  \emph{(\bibinfo{series}{Lecture {Notes} in {Computer} {Science}})}.
  \bibinfo{publisher}{Springer International Publishing},
  \bibinfo{address}{Kobe, Japan}, \bibinfo{pages}{463--479}.
\newblock
\showISBNx{978-3-319-46523-4}


\bibitem[\protect\citeauthoryear{Pham, Passing, Erling, and Boncz}{Pham
  et~al\mbox{.}}{2015}]%
        {peterwww}
\bibfield{author}{\bibinfo{person}{Minh-Duc Pham}, \bibinfo{person}{Linnea
  Passing}, \bibinfo{person}{Orri Erling}, {and} \bibinfo{person}{Peter
  Boncz}.} \bibinfo{year}{2015}\natexlab{}.
\newblock \showarticletitle{Deriving an {Emergent} {Relational} {Schema} from
  {RDF} {Data}}. In \bibinfo{booktitle}{\emph{Proceedings of the 24th
  {International} {Conference} on {World} {Wide} {Web}}}
  \emph{(\bibinfo{series}{{WWW} '15})}. \bibinfo{publisher}{International World
  Wide Web Conferences Steering Committee}, \bibinfo{address}{Republic and
  Canton of Geneva, Switzerland}, \bibinfo{pages}{864--874}.
\newblock
\showISBNx{978-1-4503-3469-3}


\bibitem[\protect\citeauthoryear{Qin, Yu, Chang, Cheng, Zhang, and Lin}{Qin
  et~al\mbox{.}}{2014}]%
        {graph_mapreduce}
\bibfield{author}{\bibinfo{person}{Lu Qin}, \bibinfo{person}{Jeffrey~Xu Yu},
  \bibinfo{person}{Lijun Chang}, \bibinfo{person}{Hong Cheng},
  \bibinfo{person}{Chengqi Zhang}, {and} \bibinfo{person}{Xuemin Lin}.}
  \bibinfo{year}{2014}\natexlab{}.
\newblock \showarticletitle{{Scalable Big Graph Processing in MapReduce}}. In
  \bibinfo{booktitle}{\emph{Proceedings of the 2014 ACM SIGMOD International
  Conference on Management of Data}} (Snowbird, Utah, USA)
  \emph{(\bibinfo{series}{SIGMOD ’14})}. \bibinfo{publisher}{Association for
  Computing Machinery}, \bibinfo{address}{New York, NY, USA},
  \bibinfo{pages}{827–838}.
\newblock
\showISBNx{9781450323765}
\urldef\tempurl%
\url{https://doi.org/10.1145/2588555.2593661}
\showDOI{\tempurl}


\bibitem[\protect\citeauthoryear{Redaschi and Consortium}{Redaschi and
  Consortium}{2009}]%
        {uniprot}
\bibfield{author}{\bibinfo{person}{Nicole Redaschi} {and}
  \bibinfo{person}{UniProt Consortium}.} \bibinfo{year}{2009}\natexlab{}.
\newblock \showarticletitle{{UniProt} in {RDF}: {Tackling} {Data} {Integration}
  and {Distributed} {Annotation} with the {Semantic} {Web}}.
\newblock \bibinfo{journal}{\emph{Nature Precedings}} (\bibinfo{year}{2009}).
\newblock
\showISSN{1756-0357}


\bibitem[\protect\citeauthoryear{Rietveld and Hoekstra}{Rietveld and
  Hoekstra}{2014}]%
        {yasgui}
\bibfield{author}{\bibinfo{person}{Laurens Rietveld} {and}
  \bibinfo{person}{Rinke Hoekstra}.} \bibinfo{year}{2014}\natexlab{}.
\newblock \showarticletitle{{YASGUI}: {Feeling} the {Pulse} of {Linked}
  {Data}}. In \bibinfo{booktitle}{\emph{Knowledge Engineering and Knowledge
  Management - 19th International Conference, {EKAW} 2014, Link{\"{o}}ping,
  Sweden, November 24-28, 2014. Proceedings}} \emph{(\bibinfo{series}{Lecture
  {Notes} in {Computer} {Science}})}. \bibinfo{publisher}{Springer
  International Publishing}, \bibinfo{address}{Link\"oping, Sweden},
  \bibinfo{pages}{441--452}.
\newblock


\bibitem[\protect\citeauthoryear{Sakr and Al-Naymat}{Sakr and
  Al-Naymat}{2010}]%
        {survey3}
\bibfield{author}{\bibinfo{person}{Sherif Sakr} {and} \bibinfo{person}{Ghazi
  Al-Naymat}.} \bibinfo{year}{2010}\natexlab{}.
\newblock \showarticletitle{Relational Processing of RDF Queries: A Survey}.
\newblock \bibinfo{journal}{\emph{SIGMOD Record}} \bibinfo{volume}{38},
  \bibinfo{number}{4} (\bibinfo{year}{2010}), \bibinfo{pages}{23–28}.
\newblock
\showISSN{0163-5808}
\urldef\tempurl%
\url{https://doi.org/10.1145/1815948.1815953}
\showDOI{\tempurl}


\bibitem[\protect\citeauthoryear{Sch\"atzle, Przyjaciel-Zablocki, Skilevic, and
  Lausen}{Sch\"atzle et~al\mbox{.}}{2016}]%
        {s2rdf}
\bibfield{author}{\bibinfo{person}{Alexander Sch\"atzle},
  \bibinfo{person}{Martin Przyjaciel-Zablocki}, \bibinfo{person}{Simon
  Skilevic}, {and} \bibinfo{person}{Georg Lausen}.}
  \bibinfo{year}{2016}\natexlab{}.
\newblock \showarticletitle{S2RDF: {RDF} {Querying} with {SPARQL} on {Spark}}.
\newblock \bibinfo{journal}{\emph{Proceedings of the VLDB Endowment}}
  \bibinfo{volume}{9}, \bibinfo{number}{10} (\bibinfo{year}{2016}),
  \bibinfo{pages}{804--815}.
\newblock
\showISSN{2150-8097}


\bibitem[\protect\citeauthoryear{Shao, Wang, and Li}{Shao
  et~al\mbox{.}}{2013}]%
        {trinity}
\bibfield{author}{\bibinfo{person}{Bin Shao}, \bibinfo{person}{Haixun Wang},
  {and} \bibinfo{person}{Yatao Li}.} \bibinfo{year}{2013}\natexlab{}.
\newblock \showarticletitle{Trinity: A Distributed Graph Engine on a Memory
  Cloud}. In \bibinfo{booktitle}{\emph{Proceedings of the 2013 ACM SIGMOD
  International Conference on Management of Data}} (New York, New York, USA)
  \emph{(\bibinfo{series}{SIGMOD ’13})}. \bibinfo{publisher}{Association for
  Computing Machinery}, \bibinfo{address}{New York, NY, USA},
  \bibinfo{pages}{505–516}.
\newblock
\showISBNx{9781450320375}
\urldef\tempurl%
\url{https://doi.org/10.1145/2463676.2467799}
\showDOI{\tempurl}


\bibitem[\protect\citeauthoryear{Shen, Wang, and Han}{Shen
  et~al\mbox{.}}{2015}]%
        {shen_entity_2015}
\bibfield{author}{\bibinfo{person}{W. Shen}, \bibinfo{person}{J. Wang}, {and}
  \bibinfo{person}{J. Han}.} \bibinfo{year}{2015}\natexlab{}.
\newblock \showarticletitle{Entity linking with a knowledge base: issues,
  techniques, and solutions}.
\newblock \bibinfo{journal}{\emph{IEEE Transactions on Knowledge and Data
  Engineering}} \bibinfo{volume}{27}, \bibinfo{number}{2}
  (\bibinfo{year}{2015}), \bibinfo{pages}{443--460}.
\newblock
\showISSN{1041-4347}


\bibitem[\protect\citeauthoryear{Sidirourgos, Goncalves, Kersten, Nes, and
  Manegold}{Sidirourgos et~al\mbox{.}}{2008}]%
        {swan}
\bibfield{author}{\bibinfo{person}{Lefteris Sidirourgos},
  \bibinfo{person}{Romulo Goncalves}, \bibinfo{person}{Martin Kersten},
  \bibinfo{person}{Niels Nes}, {and} \bibinfo{person}{Stefan Manegold}.}
  \bibinfo{year}{2008}\natexlab{}.
\newblock \showarticletitle{Column-Store Support for {RDF} Data Management: not
  all swans are white}.
\newblock \bibinfo{journal}{\emph{Proceedings of the VLDB Endowment}}
  \bibinfo{volume}{1}, \bibinfo{number}{2} (\bibinfo{year}{2008}),
  \bibinfo{pages}{1553--1563}.
\newblock


\bibitem[\protect\citeauthoryear{Singh, Upadhyay, and Atre}{Singh
  et~al\mbox{.}}{2018}]%
        {singh_efficient_2018}
\bibfield{author}{\bibinfo{person}{Gurkirat Singh}, \bibinfo{person}{Dhawal
  Upadhyay}, {and} \bibinfo{person}{Medha Atre}.}
  \bibinfo{year}{2018}\natexlab{}.
\newblock \showarticletitle{Efficient {RDF} {Dictionaries} with {B}+ {Trees}}.
  In \bibinfo{booktitle}{\emph{Proceedings of the {ACM} {India} {Joint}
  {International} {Conference} on {Data} {Science} and {Management} of {Data}}}
  \emph{(\bibinfo{series}{{CoDS}-{COMAD} '18})}. \bibinfo{publisher}{ACM},
  \bibinfo{address}{New York, NY, USA}, \bibinfo{pages}{128--136}.
\newblock
\showISBNx{978-1-4503-6341-9}


\bibitem[\protect\citeauthoryear{{Sparsity Technologies}}{{Sparsity
  Technologies}}{2019}]%
        {sparksee}
\bibfield{author}{\bibinfo{person}{{Sparsity Technologies}}.}
  \bibinfo{year}{2019}\natexlab{}.
\newblock \bibinfo{title}{Sparksee}.
\newblock
\newblock
\urldef\tempurl%
\url{http://sparsity-technologies.com/}
\showURL{%
\tempurl}


\bibitem[\protect\citeauthoryear{Suchanek, Kasneci, and Weikum}{Suchanek
  et~al\mbox{.}}{2008}]%
        {yago}
\bibfield{author}{\bibinfo{person}{Fabian~M. Suchanek},
  \bibinfo{person}{Gjergji Kasneci}, {and} \bibinfo{person}{Gerhard Weikum}.}
  \bibinfo{year}{2008}\natexlab{}.
\newblock \showarticletitle{{YAGO}: {A} {Large} {Ontology} from {Wikipedia} and
  {WordNet}}.
\newblock \bibinfo{journal}{\emph{Web Semantics: Science, Services and Agents
  on the World Wide Web}} \bibinfo{volume}{6}, \bibinfo{number}{3}
  (\bibinfo{year}{2008}), \bibinfo{pages}{203--217}.
\newblock
\showISSN{1570-8268}


\bibitem[\protect\citeauthoryear{Systap}{Systap}{2019}]%
        {blazegraph}
\bibfield{author}{\bibinfo{person}{Systap}.} \bibinfo{year}{2019}\natexlab{}.
\newblock \bibinfo{title}{{BlazeGraph}}.
\newblock
\newblock
\urldef\tempurl%
\url{https://blazegraph.com/}
\showURL{%
\tempurl}


\bibitem[\protect\citeauthoryear{Tandon, Melo, Suchanek, and Weikum}{Tandon
  et~al\mbox{.}}{2014}]%
        {webchild}
\bibfield{author}{\bibinfo{person}{Niket Tandon}, \bibinfo{person}{Gerard~de
  Melo}, \bibinfo{person}{Fabian~M. Suchanek}, {and} \bibinfo{person}{Gerhard
  Weikum}.} \bibinfo{year}{2014}\natexlab{}.
\newblock \showarticletitle{{WebChild}: Harvesting and Organizing Commonsense
  Knowledge from the Web}. In \bibinfo{booktitle}{\emph{Seventh {ACM}
  {International} {Conference} on {Web} {Search} and {Data} {Mining}, {WSDM}
  2014}}. \bibinfo{publisher}{ACM}, \bibinfo{address}{New York, NY, USA},
  \bibinfo{pages}{523--532}.
\newblock
\showISBNx{978-1-4503-2351-2}


\bibitem[\protect\citeauthoryear{Tonon, Catasta, Prokofyev, Demartini, Aberer,
  and Cudre-Mauroux}{Tonon et~al\mbox{.}}{2016}]%
        {ranking}
\bibfield{author}{\bibinfo{person}{Alberto Tonon}, \bibinfo{person}{Michele
  Catasta}, \bibinfo{person}{Roman Prokofyev}, \bibinfo{person}{Gianluca
  Demartini}, \bibinfo{person}{Karl Aberer}, {and} \bibinfo{person}{Philippe
  Cudre-Mauroux}.} \bibinfo{year}{2016}\natexlab{}.
\newblock \showarticletitle{Contextualized ranking of entity types based on
  knowledge graphs}.
\newblock \bibinfo{journal}{\emph{Web Semantics: Science, Services and Agents
  on the World Wide Web}}  \bibinfo{volume}{37} (\bibinfo{year}{2016}),
  \bibinfo{pages}{170--183}.
\newblock


\bibitem[\protect\citeauthoryear{Urbani, Dutta, Gurajada, and Weikum}{Urbani
  et~al\mbox{.}}{2016a}]%
        {kognac}
\bibfield{author}{\bibinfo{person}{Jacopo Urbani}, \bibinfo{person}{Sourav
  Dutta}, \bibinfo{person}{Sairam Gurajada}, {and} \bibinfo{person}{Gerhard
  Weikum}.} \bibinfo{year}{2016}\natexlab{a}.
\newblock \showarticletitle{{KOGNAC:} Efficient Encoding of Large Knowledge
  Graphs}. In \bibinfo{booktitle}{\emph{Proceedings of the Twenty-Fifth
  International Joint Conference on Artificial Intelligence, {IJCAI} 2016, New
  York, NY, USA, 9-15 July 2016}}. \bibinfo{publisher}{AAAI Press},
  \bibinfo{address}{New York, NY, USA}, \bibinfo{pages}{3896--3902}.
\newblock


\bibitem[\protect\citeauthoryear{Urbani, Jacobs, and Kr\"otzsch}{Urbani
  et~al\mbox{.}}{2016b}]%
        {vlog}
\bibfield{author}{\bibinfo{person}{Jacopo Urbani}, \bibinfo{person}{Ceriel
  Jacobs}, {and} \bibinfo{person}{Markus Kr\"otzsch}.}
  \bibinfo{year}{2016}\natexlab{b}.
\newblock \showarticletitle{Column-{Oriented} {Datalog} {Materialization} for
  {Large} {Knowledge} {Graphs}}. In \bibinfo{booktitle}{\emph{Proceedings of
  the Thirtieth {AAAI} Conference on Artificial Intelligence}}.
  \bibinfo{publisher}{AAAI Press}, \bibinfo{address}{Phoenix, Arizona, USA},
  \bibinfo{pages}{258--264}.
\newblock


\bibitem[\protect\citeauthoryear{Urbani, Maassen, Drost, Seinstra, and
  Bal}{Urbani et~al\mbox{.}}{2013}]%
        {rdfcompr}
\bibfield{author}{\bibinfo{person}{Jacopo Urbani}, \bibinfo{person}{Jason
  Maassen}, \bibinfo{person}{Niels Drost}, \bibinfo{person}{Frank Seinstra},
  {and} \bibinfo{person}{Henri Bal}.} \bibinfo{year}{2013}\natexlab{}.
\newblock \showarticletitle{Scalable {RDF} data compression with {MapReduce}}.
\newblock \bibinfo{journal}{\emph{Concurrency and Computation: Practice and
  Experience}} \bibinfo{volume}{25}, \bibinfo{number}{1}
  (\bibinfo{year}{2013}), \bibinfo{pages}{24--39}.
\newblock


\bibitem[\protect\citeauthoryear{Verborgh, Vander~Sande, Hartig, Van~Herwegen,
  De~Vocht, De~Meester, Haesendonck, and Colpaert}{Verborgh
  et~al\mbox{.}}{2016}]%
        {fragment}
\bibfield{author}{\bibinfo{person}{Ruben Verborgh}, \bibinfo{person}{Miel
  Vander~Sande}, \bibinfo{person}{Olaf Hartig}, \bibinfo{person}{Joachim
  Van~Herwegen}, \bibinfo{person}{Laurens De~Vocht}, \bibinfo{person}{Ben
  De~Meester}, \bibinfo{person}{Gerald Haesendonck}, {and}
  \bibinfo{person}{Pieter Colpaert}.} \bibinfo{year}{2016}\natexlab{}.
\newblock \showarticletitle{{Triple Pattern Fragments}: a low-cost knowledge
  graph interface for the Web}.
\newblock \bibinfo{journal}{\emph{Web Semantics: Science, Services and Agents
  on the World Wide Web}}  \bibinfo{volume}{37} (\bibinfo{year}{2016}),
  \bibinfo{pages}{184--206}.
\newblock


\bibitem[\protect\citeauthoryear{Vrande\v{c}i\'c and
  Kr\"otzsch}{Vrande\v{c}i\'c and Kr\"otzsch}{2014}]%
        {wikidata}
\bibfield{author}{\bibinfo{person}{Denny Vrande\v{c}i\'c} {and}
  \bibinfo{person}{Markus Kr\"otzsch}.} \bibinfo{year}{2014}\natexlab{}.
\newblock \showarticletitle{Wikidata: a free collaborative knowledge base}.
\newblock \bibinfo{journal}{\emph{Commun. ACM}} \bibinfo{volume}{57},
  \bibinfo{number}{10} (\bibinfo{year}{2014}), \bibinfo{pages}{78--85}.
\newblock


\bibitem[\protect\citeauthoryear{Weiss, Karras, and Bernstein}{Weiss
  et~al\mbox{.}}{2008}]%
        {hexastore}
\bibfield{author}{\bibinfo{person}{Cathrin Weiss}, \bibinfo{person}{Panagiotis
  Karras}, {and} \bibinfo{person}{Abraham Bernstein}.}
  \bibinfo{year}{2008}\natexlab{}.
\newblock \showarticletitle{Hexastore: sextuple indexing for semantic web data
  management}.
\newblock \bibinfo{journal}{\emph{Proceedings of the VLDB Endowment}}
  \bibinfo{volume}{1}, \bibinfo{number}{1} (\bibinfo{year}{2008}),
  \bibinfo{pages}{1008--1019}.
\newblock


\bibitem[\protect\citeauthoryear{Williams and Zobel}{Williams and
  Zobel}{1999}]%
        {vbyte}
\bibfield{author}{\bibinfo{person}{Hugh~E. Williams} {and}
  \bibinfo{person}{Justin Zobel}.} \bibinfo{year}{1999}\natexlab{}.
\newblock \showarticletitle{Compressing {Integers} for {Fast} {File} {Access}}.
\newblock \bibinfo{journal}{\emph{Comput. J.}} \bibinfo{volume}{42},
  \bibinfo{number}{3} (\bibinfo{year}{1999}), \bibinfo{pages}{193--201}.
\newblock
\showISSN{0010-4620}
\urldef\tempurl%
\url{https://doi.org/10.1093/comjnl/42.3.193}
\showDOI{\tempurl}


\bibitem[\protect\citeauthoryear{Wylot, Hauswirth, Cudr{\'{e}}{-}Mauroux, and
  Sakr}{Wylot et~al\mbox{.}}{2018}]%
        {survey6}
\bibfield{author}{\bibinfo{person}{Marcin Wylot}, \bibinfo{person}{Manfred
  Hauswirth}, \bibinfo{person}{Philippe Cudr{\'{e}}{-}Mauroux}, {and}
  \bibinfo{person}{Sherif Sakr}.} \bibinfo{year}{2018}\natexlab{}.
\newblock \showarticletitle{{RDF} Data Storage and Query Processing Schemes:
  {A} Survey}.
\newblock \bibinfo{journal}{\emph{ACM Computing Surveys (CSUR)}}
  \bibinfo{volume}{51}, \bibinfo{number}{4} (\bibinfo{year}{2018}),
  \bibinfo{pages}{84:1--84:36}.
\newblock


\bibitem[\protect\citeauthoryear{Yahya, Barbosa, Berberich, Wang, and
  Weikum}{Yahya et~al\mbox{.}}{2016}]%
        {kgqueryans}
\bibfield{author}{\bibinfo{person}{Mohamed Yahya}, \bibinfo{person}{Denilson
  Barbosa}, \bibinfo{person}{Klaus Berberich}, \bibinfo{person}{Qiuyue Wang},
  {and} \bibinfo{person}{Gerhard Weikum}.} \bibinfo{year}{2016}\natexlab{}.
\newblock \showarticletitle{Relationship {Queries} on {Extended} {Knowledge}
  {Graphs}}. In \bibinfo{booktitle}{\emph{Proceedings of the {Ninth} {ACM}
  {International} {Conference} on {Web} {Search} and {Data} {Mining}}}
  \emph{(\bibinfo{series}{{WSDM} '16})}. \bibinfo{publisher}{ACM},
  \bibinfo{address}{New York, NY, USA}, \bibinfo{pages}{605--614}.
\newblock
\showISBNx{978-1-4503-3716-8}


\bibitem[\protect\citeauthoryear{Yuan, Liu, Wu, Jin, Zhang, and Liu}{Yuan
  et~al\mbox{.}}{2013}]%
        {triplebit}
\bibfield{author}{\bibinfo{person}{Pingpeng Yuan}, \bibinfo{person}{Pu Liu},
  \bibinfo{person}{Buwen Wu}, \bibinfo{person}{Hai Jin}, \bibinfo{person}{Wenya
  Zhang}, {and} \bibinfo{person}{Ling Liu}.} \bibinfo{year}{2013}\natexlab{}.
\newblock \showarticletitle{{TripleBit}: a fast and compact system for large
  scale {RDF} data}.
\newblock \bibinfo{journal}{\emph{Proceedings of the VLDB Endowment}}
  \bibinfo{volume}{6}, \bibinfo{number}{7} (\bibinfo{year}{2013}),
  \bibinfo{pages}{517--528}.
\newblock


\bibitem[\protect\citeauthoryear{Zeng, Yang, Wang, Shao, and Wang}{Zeng
  et~al\mbox{.}}{2013}]%
        {distributed}
\bibfield{author}{\bibinfo{person}{Kai Zeng}, \bibinfo{person}{Jiacheng Yang},
  \bibinfo{person}{Haixun Wang}, \bibinfo{person}{Bin Shao}, {and}
  \bibinfo{person}{Zhongyuan Wang}.} \bibinfo{year}{2013}\natexlab{}.
\newblock \showarticletitle{A distributed graph engine for web scale {RDF}
  data}.
\newblock \bibinfo{journal}{\emph{Proceedings of the VLDB Endowment}}
  \bibinfo{volume}{6}, \bibinfo{number}{4} (\bibinfo{year}{2013}),
  \bibinfo{pages}{265--276}.
\newblock


\bibitem[\protect\citeauthoryear{Zou, \"Ozsu, Chen, Shen, Huang, and Zhao}{Zou
  et~al\mbox{.}}{2014}]%
        {gstore}
\bibfield{author}{\bibinfo{person}{Lei Zou}, \bibinfo{person}{M.~Tamer \"Ozsu},
  \bibinfo{person}{Lei Chen}, \bibinfo{person}{Xuchuan Shen},
  \bibinfo{person}{Ruizhe Huang}, {and} \bibinfo{person}{Dongyan Zhao}.}
  \bibinfo{year}{2014}\natexlab{}.
\newblock \showarticletitle{{gStore}: a graph-based {SPARQL} query engine}.
\newblock \bibinfo{journal}{\emph{The VLDB Journal}} \bibinfo{volume}{23},
  \bibinfo{number}{4} (\bibinfo{year}{2014}), \bibinfo{pages}{565--590}.
\newblock


\end{thebibliography}

\end{document}